\documentclass[aps,prd,notitlepage,onecolumn,superscriptaddress,nofootinbib,a4paper]{revtex4-1}
\usepackage[all]{xy}
\usepackage{graphicx}
\usepackage{mathrsfs}
\usepackage{dsfont}
\usepackage{slashed}
\usepackage{amsmath,bbm,latexsym,amssymb}
\usepackage{epsfig}
\usepackage{epstopdf}
\usepackage{float} 
\usepackage{tensor}
\usepackage{booktabs}
\usepackage{multirow}
\usepackage{cancel}

\usepackage[table]{xcolor}
\usepackage[hypertexnames=false,hyperfootnotes=false,pdfencoding =auto]{hyperref}
\usepackage[all]{hypcap}       

\usepackage[utf8]{inputenc}        

\makeatletter

    \def\CT@@do@color{%
      \global\let\CT@do@color\relax
            \@tempdima\wd\z@
            \advance\@tempdima\@tempdimb
            \advance\@tempdima\@tempdimc
    \advance\@tempdimb\tabcolsep
    \advance\@tempdimc\tabcolsep
    \advance\@tempdima2\tabcolsep
            \kern-\@tempdimb
            \leaders\vrule
                    \hskip\@tempdima\@plus  1fill
            \kern-\@tempdimc
            \hskip-\wd\z@ \@plus -1fill }
    \makeatother

\makeatletter
\g@addto@macro\bfseries{\boldmath}
\makeatother

\newenvironment{Eqnarray}%
     {\arraycolsep 0.14em\begin{eqnarray}}{\end{eqnarray}}
\newcommand{\be}{\begin{equation}}
\newcommand{\ee}{\end{equation}}
\newcommand{\ba}{\begin{Eqnarray}}
\newcommand{\ea}{\end{Eqnarray}}

\newcommand{\diag}{\text{diag}}
\newcommand\Ttstrut{\rule{0pt}{5ex}}         
\newcommand\Bstrut{\rule[-4.2ex]{0pt}{0pt}}   
\newcommand\TBstrut{\Ttstrut\Bstrut}    
\let\Re\relax
\let\Im\relax
\DeclareMathOperator{\Re}{Re}
\DeclareMathOperator{\Im}{Im}
\DeclareMathOperator{\Tr}{Tr}
\def\eq#1{eq.~(\ref{#1})}
\def\Eq#1{Equation~(\ref{#1})}

\def\eqs#1#2{eqs.~(\ref{#1}) and (\ref{#2})}

\def\eqss#1#2#3{eqs.~(\ref{#1}), (\ref{#2}), and (\ref{#3})}
\def\eqst#1#2{eqs.~(\ref{#1})--(\ref{#2})}
\def\Eqst#1#2{Equation~(\ref{#1})--(\ref{#2})}

\newcommand{\UPi}{\ensuremath{{\rm U}(1)\otimes \Pi_2}}
\newcommand{\ZPi}{\ensuremath{\mathbb{Z}_2\otimes\Pi_2}}
\def\T{{\mathsf T}}
\def\phm{\phantom{-}}
\def\pht{\phantom{i}}
\def\half{\tfrac12}
\def\id{\mathbbm{1}}

\newcommand{\YukZeros}[9]{\begin{pmatrix}
	#1 &\,\,\, #2 & \,\,\,#3 \\
	#4 &\,\,\, #5 &\,\,\, #6 \\
	#7 &\,\,\, #8 &\,\,\,#9 \\
\end{pmatrix}}

\newcommand{\YukZerosTwo}[4]{\begin{pmatrix}
	#1 & \,\,\,#2  \\
	#3 &\,\,\, #4  \\
\end{pmatrix}}

\begin{document}
\vspace*{1cm}

\title{Extending the symmetries of the generalized CP-symmetric 2HDM scalar potential to the Yukawa sector}

\author{Sergio Carrolo}%
\email{scarrolo@mpp.mpg.de}
\affiliation{Max-Planck-Institut f\"{u}r Physik,
Werner-Heisenberg-Institut,
D-80805 M\"{u}nchen, Germany}
\author{Howard E.~Haber}
\email{haber@scipp.ucsc.edu}
\affiliation{Santa Cruz Institute for Particle Physics,
University of California,\\
1156 High Street, Santa Cruz, California 95064, USA}
\author{Luis Lourenco}%
\email{luis.marques-lourenco@uni-wuerzburg.de}
\affiliation{Institut f\"{u}r Theoretische Physik und Astrophysik, \\  \small\em
         Universit\"{a}t W\"{u}rzburg, Emil-Hilb-Weg 22, 97074 W\"{u}rzburg, Germany}
\author{João P.\ Silva}%
\email{jpsilva@cftp.tecnico.ulisboa.pt}
\affiliation{Departamento de
	F\'isica and CFTP,  Instituto Superior T\'ecnico,\\  \small\em
	Universidade de Lisboa, Av
	Rovisco Pais, 1, P-1049-001 Lisboa, Portugal}

\begin{abstract}
There are only six independent types of symmetry-constrained (renormalizable) scalar potentials
in the two Higgs doublet model (2HDM).
For example, the scalar sector symmetry known as $\ZPi$,
generated by the simultaneous application of two independent symmetries acting on the
scalar fields, and the generalized CP symmetry known as GCP2 yield equivalent 2HDM scalar
potentials.
A similar situation arises for the scalar sector symmetries known as
$\UPi$ and GCP3, respectively.
In this paper, we show that this ``degeneracy'' remains when the definitions of
the corresponding symmetries are extended to the Yukawa sector with three quark generations.
The proof involves the exploration of all possible extensions of the corresponding symmetries to the Yukawa sector,
consistent with the phenomenological constraints of nonzero quark masses and a nontrivial quark mixing matrix.
Moreover, we find that this result is a peculiarity of a Yukawa sector with three quark generations.
In particular, with two quark generations,
we find that models based on the extension of $\ZPi$ to the Yukawa sector are inequivalent with those based on GCP2.
\end{abstract}

\maketitle

\section{Introduction}
\label{sec:intro}

The discovery at LHC in 2012 of a
neutral scalar ($h_{125}$) with $m_h\simeq 125$~GeV~\cite{ATLAS:2012yve,CMS:2012qbp}
was a major milestone in particle physics.  In particular, the Standard Model
(SM) of fundamental particles and their interactions posits the existence of 
an elementary SU(2)$_L$ doublet of scalar fields.   The spontaneous breaking of the SU(2)$_L\times$U(1)$_Y$
gauge symmetry
generates masses for the gauge bosons, quarks, and charged leptons, while leaving one
physical scalar degree of freedom---the Higgs boson.
In the analysis of LHC data collected over the past thirteen years, the 
observed properties of $h_{125}$ are consistent with the predictions of the SM
within the statistical uncertainties of the measurements~\cite{ATLAS:2022vkf,CMS:2022dwd}.  Nevertheless,
one cannot currently rule out the possibility that additional scalar particles exist with masses
of order the electroweak scale.  Indeed, in light of the nonminimality of the quark
and lepton sectors of the SM, which exhibits three replicas (families/generations)
of each fermion type, one might expect a nonminimal scalar sector as well.
Therefore, it is of fundamental interest to ascertain the number of
elementary scalars in nature.

One of the simplest extensions of the SM scalar sector is the
two-Higgs doublet model (2HDM)~\cite{Branco:2011iw}.
In spite of its apparent simplicity,
the 2HDM has a very rich and vast phenomenology,
having been used, for example, to propose the origin of CP violation as a consequence of a spontaneously broken symmetry
\cite{Lee:1973iz},
to explain the baryon--antibaryon asymmetry \cite{Cohen:1993nk}, and
to provide a plausible dark-matter candidate \cite{Deshpande:1977rw,Barbieri:2006dq}.
The diversity in its phenomenology is due in part to the fact that
the most general 2HDM 
scalar potential initially consists of 14 real parameters,
and the corresponding Higgs--quark Yukawa sector initially consists of 72 real parameters (prior to
identifying the independent physical parameters of the model).
Thus, as in any model of an extended Higgs sector, it is of central importance to
impose additional global (discrete or continuous) symmetries
to increase the predictability of the model by reducing the number of parameters,
to avoid (tree-level) flavor-changing neutral couplings mediated by scalars~\cite{Glashow:1976nt,Paschos:1976ay},
and/or to explain some relations among observables~\cite{Fritzsch:1977za}.

Many theoretical aspects of the 2HDM are now known.
The impact of symmetries in the scalar sector has been studied in detail
by many authors.
Symmetries can be of two types: flavor symmetry transformations, which change a given scalar field into a linear combinations of scalar fields
(but not their complex conjugates),
and generalized CP (GCP) symmetry transformations, which transform a given scalar field into a linear combinations of
complex conjugated scalar fields.
It has been shown that in the 2HDM there are only six inequivalent symmetry-constrained
(renormalizable) scalar potentials~\cite{Ivanov:2007de,Ferreira:2009wh}.
As an example, consider the symmetries
\ba
\mathbb{Z}_2
&:&
\Phi_1 \rightarrow \Phi_1\, , \ \ \ \ \Phi_2 \rightarrow -\Phi_2\, ;
\\
\Pi_2
&:&
\Phi_1 \leftrightarrow \Phi_2\, .
\ea
The scalar potential invariant under a $\ZPi$ (U(1)$\otimes\Pi_2$)
symmetry is the same scalar potential, although in a different scalar field basis,
as the scalar potential invariant under a  generalized CP symmetry which is denoted by GCP2 (GCP3)
\cite{Davidson:2005cw,Ferreira:2009wh,Ferreira:2010yh,Battye:2011jj,Pilaftsis:2011ed,Haber:2021zva}.
These features can be understood by considering basis invariant quantities
\cite{Botella:1994cs,Davidson:2005cw}.  More recently, a more sophisticated
method that employs novel invariant theory
techniques has been shown to yield the same conclusions noted above~\cite{Trautner:2018ipq,Bento:2020jei,Doring:2024kdg}.

The extension of flavor symmetries to the Yukawa sector
was initiated with an examination of abelian symmetries
in Refs.~\cite{Ferreira:2010ir,Ivanov:2011ae,Ivanov:2013bka}.
Extensions of generalized CP transformations into the Yukawa sector of the
2HDM were considered in Ref.~\cite{Ferreira:2010bm},
where it was shown that
there is only one model which is consistent with nonzero quark
masses and a non-diagonal Cabibbo-Kobayashi-Maskawa (CKM) 
matrix.\footnote{Very recently, this study has been generalized to the
GCP-symmetric 3HDM with Yukawa interactions in Ref.~\cite{Bree:2024edl}.}
However, there are subtleties involved in the study of symmetries.
Indeed, when extending symmetries
to the Yukawa sector,
two different scalar symmetries that yield the same scalar potential can result in
different Yukawa couplings \cite{Ferreira:2010ir}.
For example,
the scalar potential of the $\mathbb{Z}_3$-symmetric 2HDM coincides with the scalar potential
of the $\mathbb{Z}_4$-symmetric 2HDM.
Nevertheless,
when extended to the Yukawa sector,
the $\mathbb{Z}_3$-symmetric 2HDM and the $\mathbb{Z}_4$-symmetric 2HDM yield different Yukawa
textures.\footnote{This is analogous to the usual quantum mechanical removal of
degeneracies by the addition of a new term in the Hamiltonian.}

Hence, the following two questions arise, which we propose to address in this paper:
\begin{itemize}
\item Is it possible to impose symmetries in the Yukawa sector in such a way
that the resulting Yukawa matrices are both compatible with a $\ZPi$ ($\UPi$)
symmetry of the scalar sector and consistent with experimental observations?
\item Are the Yukawa sectors of the $\ZPi$ ($\UPi$)-symmetric 2HDM
and the GCP2 (GCP3)-symmetric 2HDM equivalent, or is there
a ``removal of the degeneracy''?
\end{itemize}

In Section~\ref{sec:notation}, we introduce our notation,
examine the consequences of flavor and of GCP symmetries, and 
review the six inequivalent symmetries that can be imposed on the 2HDM scalar potential.
Our primary goal is to determine whether an extension of the $\ZPi$ or $\UPi$ symmetry
to the Yukawa sector can yield a phenomenologically viable model. 
In Section~\ref{sec:ZPi_n=3}, we show that all Yukawa
coupling matrices arising in models with the
 $\ZPi$ symmetry extended to the Yukawa sector with three quark generations must possess at least one massless quark.
For the $\UPi$ model, we show in Section~\ref{sec:U1Pi2_n=3} that only one case exists in 
which the Yukawa coupling matrices yield nonzero quark masses, mixing angles, and a CP-violating
phase that can be compatible with experimental observations.  Moreover, as shown in
Section~\ref{sec:U1PivsGCP3_n=3}, under a
change of scalar field basis, the corresponding Yukawa coupling matrices of the $\UPi$ model coincide with those of the GCP3 model.
Next we consider models with two quark generations that yield
non-vanishing,
non-degenerate quark masses and a non-zero Cabibbo angle.
We show in Section~\ref{sec:Z2PivsGCP2_n=2} that the $\ZPi$ and GCP2 symmetries 
can be extended to the Yukawa sector with two quark generations.
However,
the corresponding Yukawa matrices obtained in $\ZPi$ models with two generations
are \textit{not} compatible with those found for the Yukawa matrices
of the GCP2 model.
This result demonstrates that the two questions posed above 
are indeed relevant.
Although the $\ZPi$ and GCP2 models cannot be extended to the
three-generation Yukawa sector in a way compatible with experiment,
they can both be extended to the two-generation Yukawa sector.  However, the extension
of $\ZPi$ and GCP2 to the Yukawa sector \textit{yields inequivalent models}. 
That is, in this case the degeneracy of the two symmetries of the scalar potential is removed by the extension to the Yukawa sector.
We outline our conclusions in section~\ref{sec:concl}.
For completeness, 
we have included some useful details in four appendices.

\section{Notation}
\label{sec:notation}

Consider the most general (renormalizable) 2HDM scalar potential $V_H$, parametrized by
\begin{align}
	V_H &= m_{11}^2 \Phi_1^{\dagger} \Phi_1+m_{22}^2 \Phi_2^{\dagger}
\Phi_2-\bigl[m_{12}^2 \Phi_1^{\dagger} \Phi_2+\text { h.c. }\bigr]
\nonumber \\
	&+\tfrac{1}{2} \lambda_1\bigl(\Phi_1^{\dagger} \Phi_1\bigr)^2
+\tfrac{1}{2} \lambda_2\bigl(\Phi_2^{\dagger} \Phi_2\bigr)^2
+\lambda_3\bigl(\Phi_1^{\dagger} \Phi_1\bigr)\bigl(\Phi_2^{\dagger} \Phi_2\bigr)
+\lambda_4\bigl(\Phi_1^{\dagger} \Phi_2\bigr)\bigl(\Phi_2^{\dagger} \Phi_1\bigr)
\nonumber \\ 
	&+\bigl[\tfrac{1}{2} \lambda_5\bigl(\Phi_1^{\dagger} \Phi_2\bigr)^2
+\lambda_6\bigl(\Phi_1^{\dagger} \Phi_1\bigr)\bigl(\Phi_1^{\dagger} \Phi_2\bigr)
+\lambda_7\bigl(\Phi_2^{\dagger} \Phi_2\bigr)\bigl(\Phi_1^{\dagger} \Phi_2\bigr)
+\text { h.c. }\bigr] \, ,
\label{VH1}
\end{align}
where \text{h.c.} stands for hermitian conjugate,
$m_{12}^2$, $\lambda_5$, $\lambda_6$, $\lambda_7$
are potentially complex parameters, and all other scalar potential parameters are real. 

The scalar potential may also be written in a more compact notation as:
\be
V_H = Y_{ij}\, (\Phi_i^{\dagger} \Phi_j)
+ Z_{ij,k\ell}\, (\Phi_i^{\dagger} \Phi_j)\, (\Phi_k^{\dagger} \Phi_\ell)\, ,
\label{VH2}
\ee
where $i,j,k,\ell\in\{1,2\}$ with an implicit sum over repeated indices, and hermiticity implies
\be
Y_{ij} = Y_{ji}^*\,,\qquad\quad
Z_{ij,k\ell} \equiv Z_{k\ell,ij} = Z_{ji,\ell k}^*\,.
\ee
We assume that 
the parameters of the scalar potential have been chosen such that 
the minimization of
$V_H$ yields charge preserving scalar field vacuum expectation values (vevs)
$\langle \Phi_i \rangle = (0, v_i)^{\T}$.
In the spirit of Refs.~\cite{Ferreira:2010ir,Ferreira:2010bm,Boto:2020wyf,Bree:2024edl},
we allow the vevs $v_i$ to take any complex value,
consistent with possible soft-symmetry breaking terms that one might
wish to add to the potential for phenomenological reasons.  Note that
\be \label{vdef}
v^2\equiv |v_1|^2+|v_2|^2\,,
\ee
where $v\equiv (2\sqrt{2}\,G_F)^{-1/2}\simeq 174$~GeV is fixed by the value of the Fermi constant.

As for the quark Yukawa sector,
it involves the $n$ generations of
left-handed quark doublets ($q_L$),
right-handed down-type quarks ($n_R$),
and
right-handed up-type quarks ($p_R$).
The Yukawa Lagrangian may be written in the interaction-eigenstate basis as (e.g., see Ref.~\cite{Branco:1999fs}):
\be
\label{yuk}
- \mathscr{L}_\mathrm{Y} =
\bar q_L \left[
\left( \Gamma_1 \Phi_1 + \Gamma_2 \Phi_2 \right) n_R
+
\left( \Delta_1 \widetilde \Phi_1 + \Delta_2 \widetilde \Phi_2 \right) p_R
\right]
+ \mathrm{h.c.},
\ee
where
$\widetilde \Phi \equiv i \tau_2 \Phi^\ast$ [with $i\tau_2\equiv\left(\begin{smallmatrix} \phm 0 &  1 \\ -1 & 0\end{smallmatrix}\right)$],
and $q_L$,
$n_R$,
and $p_R$ are $n$-component vectors
in flavor space.\footnote{In Sections~\ref{subsec:ExtSym}--\ref{sec:U1PivsGCP3_n=3}, we will take the number of
quark generations to be $n=3$.  However, in Section~\ref{sec:Z2PivsGCP2_n=2} and in Appendices~\ref{app:details} and \ref{app:equiv}, we will discuss some toy models
with $n=2$.}
The $n \times n$ matrices
$\Gamma_i$,
$\Delta_i$,
contain the complex Yukawa couplings
to the right-handed down-type quarks and up-type quarks,
respectively.
After spontaneous symmetry breaking,
the quark mass terms appear as
\be
\label{mass}
- \mathscr{L}_\mathrm{Y} \supset
\bar n_L\, M_n\, n_R + \bar p_L\, M_p\, p_R
+ \mathrm{h.c.},
\ee
where
\ba
M_n &=& \Gamma_1 v_1 + \Gamma_2 v_2\, ,
\label{emen}\\
M_p &=& \Delta_1 v_1^* + \Delta_2 v_2^*\, .\label{empee}
\ea
In general, the matrices $M_n$ and $M_p$ are not diagonal,
corresponding to the fact that the  interaction-eigenstate fermion fields $n_L$, $n_R$,
$p_L$, and $p_R$ are not mass-eigenstate fields.
To obtain the physical fermion fields,  we perform the transformations
 \begin{align}
\label{eq:SM_ChangeBasisQuarks}
&
\bar n_L =  \bar d_L V_{dL}^{\dagger} \, ,
\qquad\qquad  \bar p_L = \bar u_L V_{uL}^{\dagger} \, ,
 \\
&n_R = V_{dR}d_R \, , \qquad\qquad p_R = V_{uR} u_R\, , 
\end{align}
where the unitary matrices $V_{dL}$, $V_{dR}$, $V_{uL}$, and $V_{uR}$ are chosen 
such that 
\ba
\text{diag}(m_d,m_s,m_b)
& = D_d = &
V_{dL}^{\dagger} M_n V_{dR}\, ,
\\
\text{diag}(m_u,m_c,m_t)
& = D_u = &
V_{uL}^{\dagger} M_p V_{uR}\,,
\ea
where the diagonal entries of $D_d$ and $D_u$ are real and nonnegative (corresponding to the singular value decomposition of the mass matrices $M_d$ and $M_u$, respectively),
and $d_L$, $d_R$, $u_L$, and $u_R$ are mass-eigenstate fermion fields.
The basis change from interaction eigenstates to mass eigenstates in the left-handed quark sector
yields the couplings of quarks of different generations to the $W$ bosons that are governed by the unitary Cabibbo–Kobayashi–Maskawa (CKM) matrix
\begin{equation}
\label{eq:CKM}
	V_{\rm CKM} \equiv  V_{uL}^{\dagger} V_{dL} \, .
\end{equation} 

We now define the Hermitian matrices
\be
H_d = M_n M_n^\dagger\, ,
\qquad\qquad
H_u = M_p M_p^\dagger\, .
\label{HdHu}
\ee
Notice that these matrices are bilinear in left-handed spaces
($d_L$ and $u_L$, respectively);
effectively, the right-handed spaces have been traced over.
It immediately follows that
\be
D_d^2 =  V_{dL}^{\dagger} H_d V_{dL}\, ,
\qquad\qquad
D_u^2 =  V_{uL}^{\dagger} H_u V_{uL}\, .
\ee
For definiteness, let us take the usual generation number $n=3$.   Then, several consequences follow.   
First,
the quark masses (say, those of the down-type quarks)
may be accessed by  looking at the three invariants that are obtained from $H_d$,
which can be taken to be either:
(i) the three eigenvalues of $H_d$;
(ii) $\Tr(H_d)$,  $\textrm{det}(H_d)$,
and the third coefficient of the characteristic equation;
or (iii) the traces of $H_d$, $H_d^2$, and $H_d^3$. 
Second, the elements of the CKM matrix can be accessed by beating $H_d$ against $H_u$ and
taking traces.
For example,
one can find the four independent magnitudes of CKM matrix elements
by calculating  $\Tr(H_d H_u)$, $\Tr(H_d H_u^2)$, 
$\Tr(H_d^2 H_u)$,  and $\Tr(H_d^2 H_u^2)$
\cite{Branco:1987mj,Bree:2023ojl}.
This does not define the sign of the CP violating CKM phase,
which can be inferred through the Jarlskog invariant $J_{\rm CP}$~\cite{Jarlskog:1985ht,Jarlskog:1985cw,Dunietz:1985uy}.
Following Refs.~\cite{Botella:1994cs,Silva:2004gz},
\ba
J_{\rm CP}
&=&
\textrm{Im} \left\{
\Tr\left(
H_u H_d H_u^2 H_d^2
\right)
\right\}=
(m_t^2 - m_c^2)(m_t^2-m_u^2)(m_c^2-m_u^2)
(m_b^2 - m_s^2)(m_b^2-m_d^2)(m_s^2-m_d^2)\ 
J_\textrm{CKM}\, ,
\label{J_first}
\ea
where
\be
J_\textrm{CKM} =
\left|
\textrm{Im}
\left(
V_{\alpha a} V_{\beta b}
V_{\alpha b}^* V_{\beta a}^*
\right)
\right|\, ,
\ee
for any choice of up-type and down-type quark flavor indices, $\alpha \neq \beta$ and $a\neq b$, respectively.
Alternatively, one can note that
\cite{Jarlskog:1985ht,Dunietz:1985uy,Bernabeu:1986fc}:
\be
\Tr\left[ H_u, H_d \right]^3
= 3\, \textrm{det}\left[ H_u, H_d\right]
= 6 i\, \textrm{Im} \left\{
\Tr\left(
H_u H_d H_u^2 H_d^2
\right)
\right\}\, .
\label{J_second}
\ee
Clearly, through the squared-mass prefactors,
$J_{\rm CP}$ vanishes whenever two same-charge quark masses are degenerate.
Moreover, $J_{\rm CP}$ vanishes (through $J_{\rm CKM}$) if the CP-violating
phase in the CKM matrix vanishes.
Finally, $J_{\rm CP}$ vanishes (again through $J_{\rm CKM}$) whenever
the CKM matrix is block diagonal.

\subsection{\label{subsec:flavor} Basis transformations and flavor symmetries}

Physical observables are independent of the choice of the basis for the scalar fields and fermion fields employed in
the Higgs Lagrangian.  Since the Lagrangian parameters are basis-dependent quantities, some care is needed in
identifying the physical parameters of the theory.
Moreover, the presence of a symmetry can impart physical significance to the parameters
of a particular basis choice.

The Higgs Lagrangian specified in Section~\ref{sec:notation} was written in terms of fields
$\Phi_i$, $q_L$, $n_R$ and $p_R$.  The most general \textit{basis transformation} that preserves the form of the
gauge covariant kinetic energy terms yields new scalar and fermion fields,
\begin{align}
\Phi_i
&\rightarrow \Phi_i^{\prime}
= U_{ij} \Phi_j \, , \qquad \quad \ \ q_L \rightarrow q_L^\prime
= U_L q_L \,,
\label{basis_transformation1} \\
n_R
&\rightarrow n_R^\prime = U_{n_R} n_R \, , \qquad
\quad p_R \rightarrow p_R^\prime
= U_{p_R} p_R \,,
\label{basis_transformation2}
\end{align}
where $U$ is an arbitrary $2 \times 2$ unitary matrix 
and $U_L, U_{n_R}, U_{p_R}$ are arbitrary $n \times n$ unitary matrices (where $n$ is the number of quark generations).
With respect to the transformed basis of scalar and quark fields, the scalar potential parameters and vevs
likewise transform as
\begin{align}
Y_{ij}
& \rightarrow Y_{ij}^{\prime}= U_{ik}Y_{k\ell}U_{j\ell}^*,
\label{eq:newbasis_2}
\\
Z_{ij,k\ell}
& \rightarrow Z_{ij,k\ell}^{\prime}=U_{im}U_{ko}Z_{mn,op}U_{jn}^*U_{\ell p}^*,
\label{eq:newbasis_4}
\\
v_i
& \rightarrow v^{\prime}_i
= U_{ij} v_j \, , \label{vtrans}
\end{align}
whereas the Yukawa matrices transform as
\begin{align}
&\Gamma_i \rightarrow \Gamma_i^{\prime}
=U_L \Gamma_j U_{n_R}^{\dagger}\left(U^{\dagger}\right)_{ji} \, , \label{Gamtransform}
\\
&\Delta_i \rightarrow \Delta_i^{\prime}
=U_L \Delta_j U_{p_R}^{\dagger}\left(U^{\T}\right)_{ji}\, . \label{Deltransform}
\end{align}
Since physical observables do not depend on the choice of basis, 
only basis invariant combinations of the Higgs Lagrangian parameters are physical~\cite{Botella:1994cs,Davidson:2005cw} (which can be
experimentally measured). 

In contrast to basis transformations, consider the implications of a \textit{flavor symmetry} transformation of fields.
The  flavor symmetry transformation groups are subgroups of the corresponding groups of basis transformations of scalar and fermion fields that leave the Higgs Lagrangian invariant.
We shall denote the corresponding  Higgs and fermion flavor symmetry transformations by
\begin{align}
\Phi_i
&\rightarrow \Phi_i^{S}= S_{ij} \Phi_j \, ,
\qquad\quad \  q_L \rightarrow q_L^S = S_L q_L \,,
\label{S_symmetry1}
\\
n_R
&\rightarrow n_R^S = S_{n_R} n_R \, ,
\qquad\quad p_R \rightarrow p_R^S = S_{p_R} p_R \, ,
\label{S_symmetry2}
\end{align}
where $S$ is a $2 \times 2$ unitary matrix and
$S_L, S_{n_R}, S_{p_R}$ are unitary $n \times n$ matrices such that
the corresponding Higgs Lagrangian parameters are left invariant.
The flavor symmetry corresponding to the transformation of scalar fields is sometimes called a Higgs family (HF) symmetry.
If the Higgs Lagrangian is invariant under a HF symmetry transformation, then the scalar potential parameters defined in \eq{VH2} satisfy
\begin{align}
	Y_{ij} &= Y_{ij}^{S} = S_{ik}Y_{k\ell}S_{j\ell}^*, \label{eq:S_on_Y} \\
	Z_{ij,k\ell} &= Z_{ij,k\ell}^{S}=S_{im}S_{kp}Z_{mn,pr}S_{jn}^*S_{\ell r}^* \,.
	\label{eq:S_on_Z}
\end{align}
Likewise, if the Yukawa Lagrangian is invariant under the HF and quark flavor symmetry transformations, then
the Yukawa matrices defined in \eq{yuk} satisfy
\begin{align}
&\Gamma_i = S_L \Gamma_j S_{n_R}^{\dagger}\left(S^{\dagger}\right)_{ji}
\, ,
\label{eq:S_on_Gamma}
\\
&\Delta_i = S_L \Delta_j S_{p_R}^{\dagger}\left(S^{T}\right)_{ji}\, .
\label{eq:S_on_Delta}
\end{align}
The HF symmetry is unbroken if $v_i=v_i^S$, and it is spontaneously broken if $v_i\neq v_i^S$, where
\begin{equation}
 v_i^S\equiv S_{ij} v_j \, .
\label{eq:Sv}
\end{equation}

Note that the symmetry transformation matrices introduced in \eqs{S_symmetry1}{S_symmetry2} are defined with respect to a particular basis choice for the scalar and fermions fields.   
One can perform a basis transformation specified in \eqs{basis_transformation1}{basis_transformation2} to express
the corresponding symmetry transformation matrices with respect to the new basis of scalar and fermion fields~\cite{Ferreira:2010ir},
\ba
	S^\prime&=&U S U^\dagger\,, \qquad\qquad\qquad
	S_L^\prime = U_L S_L U_L^{\dagger}\,,
\label{Sprime1}\\
	S_{n_R}^\prime &=& U_{n_R} S_{n_R} U_{n_R}^{\dagger}\,,\qquad\quad
	S_{p_R}^\prime = U_{p_R} S_{p_R} U_{p_R}^{\dagger}\,.\label{Sprime2}
\ea
When expressed in terms of the new basis fields, one can check that the corresponding basis-transformed parameters
are invariant with respect to the symmetry transformations exhibited in \eqs{Sprime1}{Sprime2}.
In studies of all possible inequivalent symmetries, it is often
useful to employ a basis corresponding to the choice of unitary matrices $U$, $U_L$, $U_{n_R}$, and $U_{p_R}$
such that $S'$, $S'_L$, $S'_{n_R}$, and $S'_{p_R}$ are diagonal matrices.

\subsection{\label{subsec:GCP} GCP Symmetries}

The flavor symmetries discussed in the previous section are not the only type of symmetries that leave the gauge covariant kinetic
terms invariant.  In addition, one can also consider GCP symmetries
that transform the fields into linear combinations of the corresponding CP conjugate fields.
These symmetries act on the scalar and fermion fields as 
\begin{align}
	\Phi_i(t,\boldsymbol{\vec{x}}) &\rightarrow \Phi_i^{\text{GCP}} (t,\boldsymbol{\vec{x}})
= X_{ij}\Phi_{j}^*(t,-\boldsymbol{\vec{x}}) \, , \label{eq:GCPSymmetry1}
\\
	Q_L(t,\boldsymbol{\vec{x}}) &\rightarrow Q_L^{\text{GCP}}(t,\boldsymbol{\vec{x}})
= X_L \gamma^0 C \overline{Q}_L^{\T}(t,-\boldsymbol{\vec{x}}) \, ,\label{eq:GCPSymmetry2}
\\ 
	n_R(t,\boldsymbol{\vec{x}}) &\rightarrow n_R^{\text{GCP}}(t,\boldsymbol{\vec{x}})
= X_{n_R} \gamma^0 C \overline{n}_R^{\T}(t,-\boldsymbol{\vec{x}}) \, ,\label{eq:GCPSymmetry3}
\\ 
	p_R(t,\boldsymbol{\vec{x}}) &\rightarrow p_R^{\text{GCP}}(t,\boldsymbol{\vec{x}})
= X_{p_R} \gamma^0 C \overline{p}_R^{\T}(t,-\boldsymbol{\vec{x}}) \, ,
\label{eq:GCPSymmetry4}
\end{align}
where $X$, $X_L$, $X_{n_R}$, and $X_{p_R}$ are generic unitary
matrices in the respective flavor spaces, $\gamma^0$ is a Dirac matrix,
and $C$ is the charge conjugation matrix.
Henceforth, we will suppress the reference to the space-time coordinates.

Invariance of the Higgs Lagrangian under the GCP transformations exhibited in
\eqst{eq:GCPSymmetry1}{eq:GCPSymmetry4} implies that
\ba
Y_{ij}^*
&=&
X_{k i}^* Y_{k\ell} X_{\ell j} = ( X^\dagger\, Y\, X )_{ij}\, ,
\label{GCP_invariant1}
\\
Z_{ij,k\ell}^*
&=&
X_{mi}^* X_{pk}^*
Z_{mn,pr} X_{nj} X_{r\ell}\, ,
\label{GCP_invariant2}
\\
\Gamma_j^* 
&=&
X_L^\dagger
X_{ij} \Gamma_i X_{n_R}\, ,
\label{GCP_invariant3}
\\
\Delta_j^*
&=&
X_L^\dagger
X_{ij}^*\Delta_i  X_{p_R} \, .
\label{GCP_invariant4}
\ea
One can again perform a basis transformation specified in \eqs{basis_transformation1}{basis_transformation2} to express
the corresponding GCP symmetry transformation matrices with respect to the new basis of scalar and fermion fields,
\ba
	X^\prime &=& U X U^{\T} \,,\qquad\qquad\qquad\,
	X_L^\prime = U_L^{} X_{L}^{} U_L^{\T} \, ,
\label{X_basis_change1} \\
	X_{n_R}^\prime &=& U_{nR}^{} X_{n_R}^{} U_{nR}^{\T}  \,,\qquad\quad\,
	X_{p_R}^\prime = U_{pR}^{} X_{p_R}^{} U_{pR}^{\T} \, . 
\label{X_basis_change2}
\ea
In the special cases where the unitary $X$ matrices are also symmetric, then one can prove that a unitary matrix $V$ exists such that $X=V^{\T}V$ (e.g., see Appendix B of Ref.~\cite{Gunion:2005ja}).  Then, the basis choice of $U=V$ yields
$X'=\mathds{1}$ (the identity matrix), with a similar result for $X'_L$, $X'_{n_R}$, and $X'_{p_R}$.  In these special cases, the GCP transformations reduce to ordinary CP transformations.
More generally, the unitary $X$ matrices are not symmetric, in which case
one cannot employ a basis (corresponding to a choice of unitary matrices $U$, $U_L$, $U_{n_R}$, and $U_{p_R}$)
in which the transformed $X$ matrices are diagonal. 
However, it is still possible to reduce the general form of
the $X$ matrices by using a theorem proved in
Ref.~\cite{Ecker:1987qp},
which states that for any unitary matrix $X$, there exists a unitary matrix $U$
such that the transformed $X$ matrices above can be reduced to the forms
\begin{align}
	X' = 
	\begin{pmatrix}
		\phm c_\theta & \,\,\,s_\theta \\ 
		-s_\theta & \,\,\, c_\theta	
	\end{pmatrix} \, , \qquad \quad  X'_{\sigma} = 
	\begin{pmatrix}
		\phm c_{\theta_\sigma} &  \,\,\,s_{\theta_\sigma}  & \,\,\, 0\\ 
		-s_{\theta_\sigma} & \,\,\,  c_{\theta_\sigma} & \,\,\, 0\\
		\phm 0 &\,\,\,  0 & \,\,\, 1
	\end{pmatrix} \, , \quad \text{ with $\sigma \in \{L,n_R,p_R\}$} \, ,
\label{eq:GCPReduced}
\end{align}
where $c_\theta \equiv \cos\theta$, $s_\theta \equiv\sin \theta$,
and all angles lie in the closed interval $[0,\pi/2]$.
This result is very useful in classifying the GCP symmetries
and in studying their effects in the Yukawa sector.

Alternatively,
one can start from \eq{eq:GCPReduced} and make a further basis transformation using
 \eqs{basis_transformation1}{basis_transformation2}
with
\begin{align}
	U = \frac{ e^{3 i \pi/4}}{\sqrt{2}}
	\begin{pmatrix}
		1 &\,\,\,  i \\ 
		i &\,\,\, 1	
	\end{pmatrix} \,,\qquad\quad 
\quad U_{L} = U_{n_R} = U_{p_R} = \frac{ e^{3 i \pi/4}}{\sqrt{2}}
	\begin{pmatrix}
		1 & \quad i  & 0\\ 
		i & \quad 1 & 0\\
		0 & \quad 0 & \sqrt{2}\, e^{-3 i \pi/4}
	\end{pmatrix} \, ,
\label{U_for_sin/cos_to_exp}
\end{align}
to obtain the corresponding GCP symmetry transformation matrices 
\begin{align}
	X^{\prime\prime} = 
	\begin{pmatrix}
		0 & e^{-i\theta} \\ 
		 e^{i\theta} & 0	
	\end{pmatrix} \, ,\qquad  \quad X_{\sigma}^{\prime\prime} = 
	\begin{pmatrix}
		0 & e^{-i\theta_\sigma}  & 0\\ 
		e^{i\theta_\sigma} & 0 & 0\\
		0 & 0 & 1
	\end{pmatrix} \, ,\quad  \text{ with $\sigma \in \{L,n_R,p_R\}$} \, .
\label{eq:GCPReduced_exp}
\end{align}
Note that the basis choices introduced above can be made independently on each of the four spaces
(i.e., the scalar, $q_L$, $n_R$, and $p_R$ spaces). 

\subsection{ \label{subsec:ClassSymmetries}The inequivalent symmetries of the 2HDM potential}

Consider the most general renormalizable 2HDM scalar potential $V_H$ given in \eq{VH1}.
In order to reduce the number of scalar potential parameters,
one can impose symmetries on the Higgs Lagrangian that leave the gauge covariant scalar and fermion kinetic terms unchanged.  
These symmetries 
can be classified according to two different types:
flavor symmetries 
exhibited in \eqs{S_symmetry1}{S_symmetry2},
or GCP symmetries exhibited in \eqst{eq:GCPSymmetry1}{eq:GCPSymmetry4}.
Moreover, contrary to what one might expect,
imposing two different symmetries on the scalar fields
does not necessarily give rise to two different scalar potentials.
As shown in Refs.~\cite{Ivanov:2005hg,Ivanov:2006yq,Ferreira:2009wh},
there are only six inequivalent symmetries that can be imposed on the scalar potential which leave invariant the gauge covariant kinetic
energy terms of the scalar fields.  Of these six inequivalent symmetries, three are HF symmetries and three are GCP symmetries.
The three HF symmetries are $\mathbb{Z}_2$,
the Peccei-Quinn symmetry U(1)~\cite{Peccei:1977ur},
and the maximal Higgs flavor symmetry $U(2)/U(1)_Y$.
With respect to a certain basis, the HF symmetries
act on the scalar fields as\footnote{In light of electroweak gauge invariance, the scalar potential is invariant under a hypercharge U(1)$_Y$ transformation $\Phi_i\to e^{i\theta}\Phi_i$ (for $i=1,2$)
for \textit{arbitrary} choice of scalar potential parameters.   Thus, we remove this symmetry from the definition of the maximal U(2) flavor symmetry in \eq{eq:HF3}.}
\begin{align}
\mathbb{Z}_2:\ 
&\Phi_1 \rightarrow \Phi_1 \, , \quad \Phi_2 \rightarrow -\Phi_2 \, ,
\label{eq:HF1}
\\
U(1):\ 
&\Phi_1 \rightarrow \Phi_1 \, , \quad \Phi_2 \rightarrow e^{i\theta} \Phi_2 \, , \quad 0<\theta<2\pi\,,
\label{eq:HF2}
\\
U(2)/U(1)_Y:\
&\Phi_a \rightarrow \Phi^S_a = S_{ab} \Phi_b \, ,
\quad \text{with } S \in U(2)/U(1)_Y \, .
\label{eq:HF3}
\end{align}
The three GCP symmetries are the standard
CP transformation (sometimes called GCP1), GCP2, and GCP3.
Their action on the scalar fields is given by
\ba
\textrm{Standard CP}&:&
\,\,\Phi_1 \rightarrow \Phi_1^* \, ,
\quad \Phi_2 \rightarrow \Phi_2^* \, ,
\label{eq:CP1}
\\[4pt]
\textrm{GCP2}&:&
\,\,\Phi_1 \rightarrow \Phi_2^* \, ,
\quad \Phi_2 \rightarrow -\Phi_1^* \, ,
\label{eq:CP2}
\\[4pt]
\textrm{GCP3}&:& 
\begin{cases}
		\Phi_1 \rightarrow c_\theta \Phi_1^* + s_\theta \Phi_2^*  \\
		\Phi_2 \rightarrow c_\theta \Phi_2^* - s_\theta \Phi_1^*
	\end{cases}\, , 
\quad  0<\theta<\half\pi \, . \label{eq:CP3}
\ea
In the case of GCP3, any choice of  $0<\theta<\half\pi$ imposes the same conditions on the scalar potential parameters.

In principle, any subgroup of U(2) provides a possible HF symmetry that can be imposed on the 2HDM scalar potential.   But, any such symmetry must be equivalent to
$\mathbb{Z}_2$, U(1), or U(2)/U(1)$_Y$.  For example, suppose one
imposes a discrete symmetry $\mathbb{Z}_n$ (with integer $n \geq 3$) on the scalar
potential.  Then, one obtains a scalar potential that is in fact (accidentally) invariant
under the full continuous U(1) symmetry.

In this work, we will consider yet another HF symmetry, $\Pi_2$, defined as
\begin{equation}
	\Pi_2 \text{ : } \Phi_1 \leftrightarrow \Phi_2 \, . \label{eq:Pi2}
\end{equation}
However, the $\Pi_2$ symmetry is equivalent to the $\mathbb{Z}_2$ symmetry specified in \eq{eq:HF1} after performing a change of scalar field basis~\cite{Davidson:2005cw}. In particular, with $U$ given by 
\begin{equation}
	U =
	\frac{1}{\sqrt{2}} 
	\begin{pmatrix}
		1 & \phm 1 \\
		1 & -1
	\end{pmatrix} \, ,
\label{ULuis}
\end{equation}
\eq{Sprime1} implies that the $\Pi_2$ symmetry transformation matrix changes to
\begin{equation}
	\frac{1}{\sqrt{2}}
	\begin{pmatrix}
	1 & \phm 1 \\ 
	1 & -1 	
	\end{pmatrix}
	\begin{pmatrix}
		0 & \phm 1 \\ 
		1 & \phm 0 
	\end{pmatrix}
	\frac{1}{\sqrt{2}}
	\begin{pmatrix}
		1 & \phm 1 \\ 
		1 & -1 
	\end{pmatrix} = 
	\begin{pmatrix}
		1 & \phm 0 \\
		0 & -1
	\end{pmatrix}\, ,
\label{UZUPi}
\end{equation}
which we recognize as the $\mathbb{Z}_2$ symmetry transformation matrix.

If we now impose the symmetries discussed above in the scalar field
basis $\{\Phi_1, \Phi_2\}$ where they are written as
\eqst{eq:HF1}{eq:Pi2},
we obtain the constraints on the parameters of the scalar potential listed in the Table~\ref{tab:cases}.
\begin{table}[ht!]
\begin{tabular}{|c|ccccccccccc|}
\hline
\pht symmetry &  $\phantom{m_{11}^2}$ & $m_{22}^2$ &\quad $m_{12}^2$ \quad & $\phantom{\lambda_1}$ &
 $\lambda_2$ & $\phantom{\lambda_3}$ & $\lambda_4$ &
$\Re\lambda_5$ &\pht  $\Im\lambda_5$  \pht & $\lambda_6$\quad  & \quad $\lambda_7$ \qquad \\
\hline
$\mathbb{Z}_2$ & &   & $0$
   &  &  &  & & &
   & $0$ & $\phm 0$ \\
U(1) & &  & $0$ 
 &  & &  & &
$0$ & $0$ & $0$ &  $\phm 0$ \\
U(2)/U(1)$_Y$  && $ m_{11}^2$ & $0$
   && $\lambda_1$ &  & \pht $\lambda_1 - \lambda_3$\pht  &
$0$ & $0$ & $0$ & $\phm 0$ \\
\hline
CP   & & & real
 & &  &  &&
& $0$ &\pht  real \pht & \pht real \pht \\
GCP2   && $m_{11}^2$ & $0$
  && $\lambda_1$ &  &  &
&   &  & $- \lambda_6\pht$ \\
GCP3   && $m_{11}^2$ & $0$
   && $\lambda_1$ &  &  &
\pht  $\lambda_1 - \lambda_3 - \lambda_4$\pht  & $0$ & $0$ & $\phm 0$ \\
\hline\hline
$\Pi_2$  &  &$ m_{11}^2$ &\pht  real \pht &&
    $ \lambda_1$ & &  &  
 & $0$ &  & $\phm\lambda_6^\ast$
\\
$\mathbb{Z}_2\otimes\Pi_2$ & & $m_{11}^2$ & $0$ && $\lambda_1$ &&&  & $0$ &  $0$ & $\phm 0$
\\
U(1)$\otimes\Pi_2$  & & $m_{11}^2$ & $0$ && $\lambda_1$ &&&  $0$ & $0$ & $0$ & $\phm 0$
\\
\hline
\end{tabular}
\caption{Classification of 2HDM scalar potential symmetries~\cite{Ivanov:2007de,Ferreira:2009wh} defined in \eqst{eq:HF1}{eq:CP3} and their impact on the parameters of the scalar potential with respect to the basis $\{\Phi_1,\Phi_2\}$ defined in \eq{VH1}.
Empty entries correspond to a lack of constraints on the corresponding parameters. Note that $\Pi_2$ [defined in \eq{eq:Pi2}], $\mathbb{Z}_2\otimes\Pi_2$ and
U(1)$\otimes\Pi_2$ are
not independent from other symmetry conditions, since a change of scalar field basis can be performed in each case to yield a new basis in which the $\mathbb{Z}_2$, GCP2 and GCP3 symmetries, respectively, are manifestly realized. 
\label{tab:cases}}
\end{table}

So far, we have only considered the case where we impose symmetries on the scalar sector Lagrangian
with one symmetry generator, dubbed simple symmetries in Ref.~\cite{Ferreira:2008zy}.
However,
one could also require the scalar potential to be invariant under multiple
symmetries in the $(\Phi_1, \Phi_2)$ basis.
For example, consider a potential invariant under
$\mathbb{Z}_2$ or U(1).
One can, in addition, impose \textit{in the same basis} that the scalar potential
is also symmetric under the action of another simple symmetry such as
$\Pi_2$.
In such cases, we say that the potential is invariant
under $\ZPi$ or U(1)$\otimes \Pi_2$, respectively.\footnote{Because we are requiring the symmetries to be imposed on the same basis,
there is no unitary change of basis one can make that simultaneously
diagonalizes both the generator of $\mathbb{Z}_2$ [or U(1)]
and that of $\Pi_2$.}
Furthermore,
it follows from the analysis in Ref.~\cite{Ferreira:2009wh},
that both of these must be equivalent to one of the six symmetries presented above.
In particular,
concerning the impact on the scalar potential,
$\ZPi$ was shown to be equivalent to GCP2
and U(1)$\otimes \Pi_2$ was shown to be equivalent to GCP3.
To be more precise, there is a change of basis that can take
the potential invariant under $\ZPi$ to the expression of the potential
invariant under GCP2,
and similarly for U(1)$\otimes \Pi_2$ and GCP3.
The specific basis choices that relate these symmetry relations can be found in Ref.~\cite{Haber:2021zva}.

 From a phenomenological point of view, one must be careful in
imposing a U(1), U(2)/U(1)$_Y$, or GCP3 symmetry on the scalar potential.  In particular, if any of these symmetries are spontaneously broken, then the scalar spectrum will contain
an unwanted massless scalar~\cite{Ferreira:2008zy}.  In such cases, one will need to softly break the corresponding symmetry with dimension-two squared-mass terms to give a phenomenologically acceptable mass to the would-be Goldstone boson.

\section{\label{subsec:ExtSym}Extensions of symmetries to the Yukawa sector}

So far, we have only considered and classified
the effect of symmetries in the scalar sector.
Now, we will be interested in determining how these symmetries can be extended to the Yukawa sector.

We will start by considering flavor symmetries.
Recall from Section~\ref{subsec:flavor} that requiring invariance under this type
of symmetries implies that the Yukawa matrices must satisfy
\eqs{eq:S_on_Gamma}{eq:S_on_Delta}.
On the other hand, if we consider GCP symmetries, we require that the Lagrangian
is invariant under the transformations in \eqst{eq:GCPSymmetry1}{eq:GCPSymmetry4},
which in turn means that the Yukawa matrices must satisfy \eqst{GCP_invariant1}{GCP_invariant4}.
What we then mean by extending the symmetry $S$ ($X$) of the scalar sector
to the Yukawa sector is to find a set
\{$S_L$,$S_{nR}$,$S_{pR}$\}
(\{$X_L$,$X_{n_R}$,$X_{p_R}$\})
such that the Yukawa couplings are compatible
with experimental observations;
i.e., 
(i) non-vanishing quark masses
and
(ii) a non zero Jarlskog invariant $J_{\rm CP}$.
Recall from \eqst{J_first}{J_second} that the latter means 
that there exists a nonzero CP-violating CKM phase \textit{and} the CKM matrix is not block diagonal.
The first condition is equivalent to requiring
\begin{equation}
	\text{det} H_d \neq 0 \, , \qquad \text{det} H_u \neq 0\, .
\label{eq:CompCond1}
\end{equation}
The second condition can be directly extracted from 
\begin{equation}
	\text{det}\bigl\{[H_u,H_d]\bigr\} \neq 0 \, .
\end{equation}

In 2010, Ferreira and Silva \cite{Ferreira:2010bm} performed a complete study of
the extensions of GCP symmetries to the Yukawa sector and have found two interesting results.
First, they proved that GCP2 cannot be extended to the Yukawa sector
in a way compatible with the two criteria discussed above.
Second, they proved that there was only one
possible extension of the GCP3 symmetry to the fermions,
with the angles in the reduced form of \eq{eq:GCPReduced} given by
$\theta = \theta_L = \theta_{n_R}= \pi/3$.
In this model, the corresponding down-type quark Yukawa matrices were given by 
\begin{equation}
	\Gamma_1 = 
	\begin{pmatrix}
		i a_{11} &\phm  i a_{12} &\phm  a_{13} \\
		i a_{12} & -i a_{11} &\phm  a_{23} \\
		a_{31} & \phm a_{32} &\phm  0
	\end{pmatrix} \, , \qquad\quad \Gamma_2 = 
	\begin{pmatrix}
		\phm i a_{12} & -i a_{11} & -a_{23} \\
		-i a_{11} & -i a_{12} &\phm  a_{13} \\
		-a_{32} & \phm  a_{31} & \phm  0
	\end{pmatrix} \, ,
	\label{eq:CP3Yuk}
\end{equation}
where the $a_{ij}$ are real parameters. 

This result prompts us to ask the following question:
Can the symmetries $\ZPi$  and U(1)$\otimes \Pi_2$ be extended to the quark
sector in a way compatible with the two conditions discussed above?
And, if so, how do they relate to the extensions
of GCP2 and GCP3?
This is what we propose to address in the following sections.

\section{$\ZPi$ for three generations}
\label{sec:ZPi_n=3}

We begin our analysis by considering how the symmetry $\ZPi$ should act on the quark fields such that
the Yukawa Lagrangian is invariant with respect to $\ZPi$ transformations of the scalars and quarks.
In their study of the extension of abelian symmetries to the Yukawa sector, the authors of
Ref.~\cite{Ferreira:2010ir} found all possible extensions of the $\mathbb{Z}_2$ symmetry
to the fermions.  For the convenience of the reader, 
the list given in Ref.~\cite{Ferreira:2010ir} of all possible extensions of the $\mathbb{Z}_2$ symmetry to the down-type quark Yukawa couplings is provided 
in Table \ref{tab:Z2Models} of Appendix~\ref{appEuler:derivation}.   The various models listed there correspond to choosing the $\mathbb{Z}_2$ symmetry
matrices $S^{(\mathbb{Z}_2)}={\rm diag}\{1\,,\,-1\}$ while surveying over possible choices for the $\mathbb{Z}_2$ symmetry
matrices $S_L^{(\mathbb{Z}_2)}$ and  $S_{n_R}^{(\mathbb{Z}_2)}$ such that 
the $\mathbb{Z}_2$ symmetry equations  given in \eqs{eq:S_on_Y}{eq:S_on_Gamma} are satisfied.   For example,
the models 67, 71, and 73 of Ref.~\cite{Ferreira:2010ir}\footnote{The model numbers refer to the corresponding equation numbers appearing in Ref.~\cite{Ferreira:2010ir}, which are replicated in Appendix~\ref{appEuler:derivation}.}
can be obtained by choosing\footnote{There is some phase freedom in defining the symmetry matrices as
the matrices employed in \eqst{SSLSnI}{SSLSnIII} are not the unique choices that yield the Yukawa matrices exhibited in \eqst{eq:CaseIZ2}{eq:CaseIIIZ2}.  For example, in \eq{SSLSnI}, one would also obtain \eq{eq:CaseIZ2} by choosing $S={\rm diag}\{1,e^{i\theta}\}$, $S_L=e^{i\eta}\id$, and $S_{n_R}=e^{i\eta}{\rm diag}\{1,1, e^{-i\theta}\}$ for $0<\theta<2\pi$ and $0\leq\eta<2\pi$.  Strictly speaking, the corresponding symmetry group is $\mathbb{Z}_2$ only in the case of $\theta=\half\pi$.  Nevertheless, the resulting constraint on the down-type quark Yukawa matrices yields \eq{eq:CaseIZ2} for \textit{all} allowed values of $\theta$ and $\eta$.\label{fn:phases}}
\ba
\text{Case I: } \quad S^{(\mathbb{Z}_2)} &=& \textrm{diag}\{ 1, -1\}\, ,
\qquad
S^{(\mathbb{Z}_2)}_L = \id\, ,
\qquad
S^{(\mathbb{Z}_2)}_{n_R} = \textrm{diag}\{1,1, -1\}\,, \label{SSLSnI} \\
\text{Case II: } \quad S^{(\mathbb{Z}_2)} &=&\textrm{diag}\{ 1, -1\}\, ,
\qquad
S^{(\mathbb{Z}_2)}_L = \textrm{diag}\{1,1, -1\}\,, \qquad S^{(\mathbb{Z}_2)}_{n_R} =\id\,, \label{SSLSnII}  \\
\text{Case III: } \quad 
S^{(\mathbb{Z}_2)} &=&\textrm{diag}\{ 1, -1\}\, ,
\qquad
S^{(\mathbb{Z}_2)}_L =S^{(\mathbb{Z}_2)}_{n_R} =\textrm{diag}\{1,1, -1\}\,, \label{SSLSnIII} 
\ea
where $\id$ is the $3\times 3$ identity matrix.  These choices yield the following forms for the down-type quark Yukawa coupling matrices,
\begin{align}
	\text{Case I: } \quad  \Gamma_1
&= \YukZeros{\text{x}}{\text{x}}{0}{\text{x}}{\text{x}}{0}{\text{x}}{\text{x}}{0} \, ,
\qquad\quad \Gamma_2 = \YukZeros{0}{0}{\text{x}}{0}{0}{\text{x}}{0}{0}{\text{x}} \, ,
\label{eq:CaseIZ2}\\[4pt] 
	\text{Case II: } \quad  \Gamma_1
&= \YukZeros{\text{x}}{\text{x}}{\text{x}}{\text{x}}{\text{x}}{\text{x}}{0}{0}{0} \, ,
\qquad\quad \Gamma_2 = \YukZeros{0}{0}{0}{0}{0}{0}{\text{x}}{\text{x}}{\text{x}} \, ,
\label{eq:CaseIIZ2} \\[4pt] 
	\text{Case III: } \quad  \Gamma_1
&= \YukZeros{\text{x}}{\text{x}}{0}{\text{x}}{\text{x}}{0}{0}{0}{\text{x}} \, ,
\qquad\quad \Gamma_2 = \YukZeros{0}{0}{\text{x}}{0}{0}{\text{x}}{\text{x}}{\text{x}}{0} \, ,
\label{eq:CaseIIIZ2}
\end{align}
where 
x stands for an arbitrary complex number.
These three cases will be of particular interest in the analysis that follows.

With the list all possible extensions of the $\mathbb{Z}_2$ symmetry
to the Yukawa sector in hand, the problem of finding the extensions of $\ZPi$
to the Yukawa sector is equivalent to imposing the $\Pi_2$ symmetry equations on the Yukawa coupling matrices listed in Appendix~\ref{appEuler:derivation},
\ba
	\Gamma_1 &=& S_L^{} \Gamma_2 S_{nR}^\dagger \, , \qquad\qquad
	\,\,\Gamma_2 = S_L^{} \Gamma_1 S_{nR}^\dagger \,, \label{eq:Pi2_On_Gamma}\\
	\Delta_1 &= &S_L^{} \Delta_2 S_{pR}^\dagger \, , \qquad\qquad 
	\Delta_2 = S_L^{} \Delta_1 S_{pR}^\dagger \,,
\label{eq:Pi2_On_Delta}
\ea
after employing \eqs{eq:S_on_Gamma}{eq:S_on_Delta} with $S=\left(\begin{smallmatrix} 0 & 1 \\ 1 & 0\end{smallmatrix}\right)$, 
where $S_L$, $S_{nR}$, and $S_{pR}$ are arbitrary $3\times 3$ unitary matrices.\footnote{Note that when applying the $\Pi_2$ symmetry conditions, the matrices $S_L$, $S_{nR}$, and $S_{pR}$
 are taken to be arbitrary because the freedom of choosing a basis for the quarks is fixed,
up to permutations, once the 
$\mathbb{Z}_2$ symmetry matrix $S^{(\mathbb{Z}_2)}$
is chosen to be diagonal.  See Section (II.C) of Ref.~\cite{Ferreira:2010ir} for further details.}
We can now immediately exclude
models 66 and 69 of Appendix~\ref{appEuler:derivation},
since for these two models one obtains $\Gamma_1=\Gamma_2=0$ after imposing \eq{eq:Pi2_On_Gamma}, which would imply vanishing quark masses.
One can further reduce the number
of inequivalent models by noting that the $\Pi_2$ symmetry that interchanges $\Phi_1$ and $\Phi_2$ retains the same form under the change of basis
\begin{equation}
	\begin{pmatrix}
		\Phi_1' \\
		\Phi_2' 
	\end{pmatrix} = \begin{pmatrix}
		0 & 1 \\ 
		1 & 0 
	\end{pmatrix}
	\begin{pmatrix}
		\Phi_1 \\
		\Phi_2 
	\end{pmatrix} \,.
	\label{eq:Pi2_Basis_change}
\end{equation} 
Hence, when imposing $\Pi_2$, it is equivalent to work in the original $\{\Phi_1,\Phi_2\}$-basis or in the transformed $\{\Phi'_1,\Phi'_2\}$-basis.
In particular, any two models that are related by
the basis change specified in \eq{eq:Pi2_Basis_change} will be affected by the $\Pi_2$ symmetry in the same way.
Consequently, the elements of the model pairs
$\{67,68\}$, $\{71,79\}$, and $\{73,75\}$ 
can be viewed as equivalent models.
We are therefore left with three models, $\{67,71,73\}$, that cannot be related by any 
family permutations, with corresponding down-type quark Yukawa matrices given by \eqst{eq:CaseIZ2}{eq:CaseIIIZ2} prior to imposing the $\Pi_2$ symmetry.
We will treat the models corresponding to these three cases independently in the following.

Starting from the basis where $\Gamma_1$ and $\Gamma_2$ take the forms shown in \eqst{eq:CaseIZ2}{eq:CaseIIIZ2}, 
we now impose the $\Pi_2$ symmetry, which adds the additional constraints exhibited in \eq{eq:Pi2_On_Gamma}.  The impact of the additional constraints can be determined
by employing the following analysis.  Consider the quantity,
\be \label{H}
H(c_1,c_2)\equiv (c_1\Gamma_1+c_2\Gamma_2)(c_1\Gamma_1+c_2\Gamma_2)^\dagger=|c_1|^2\Gamma_1\Gamma_1^\dagger+c_1 c_2^*\,\Gamma_1\Gamma_2^\dagger  + c_1^*c_2\Gamma_2\Gamma_1^\dagger+ |c_2|^2\Gamma_2\Gamma_2^\dagger\,,
\ee
where $c$ is an arbitrary complex number.  In light of \eq{eq:Pi2_On_Gamma},
\ba
	\Gamma_1 \Gamma_1^\dagger &=& S_L \left(\Gamma_2 \Gamma_2^\dagger\right) S_L^\dagger \,,\qquad\qquad
	\Gamma_1 \Gamma_2^\dagger = S_L \left(\Gamma_2 \Gamma_1^\dagger\right) S_L^\dagger \,,\\
	\Gamma_2 \Gamma_1^\dagger &=& S_L \left(\Gamma_1 \Gamma_2^\dagger\right) S_L^\dagger \,,\qquad\qquad
	\Gamma_2 \Gamma_2^\dagger = S_L \left(\Gamma_1 \Gamma_1^\dagger\right) S_L^\dagger \,.
\ea
Inserting these results back into \eq{H}, we see that under the $\Pi_2$ symmetry $H(c_1,c_2)$ is transformed into
\be \label{Hp}
H^\prime(c_1,c_2)\equiv S_L H(c_1,c_2) S_L^\dagger = |c_2|^2\Gamma_1\Gamma_1^\dagger+c_1^*c_2\,\Gamma_1\Gamma_2^\dagger  + c_1 c_2^*\,\Gamma_2\Gamma_1^\dagger+ |c_1|^2\Gamma_2\Gamma_2^\dagger=H(c_2,c_1)\,.
\ee
Note that $\det H^\prime(c_1,c_2)=\det H(c_1,c_2)$ since $S_L$ is a unitary matrix.

In both Case I [\eq{eq:CaseIZ2}] 
and Case II [\eq{eq:CaseIIZ2}], an explicit computation of the determinants of the right-hand sides of \eqs{H}{Hp} yields: 
\ba
	\det H(c_1,c_2) &=& |c_1|^4 |c_2|^2 \left| \det\left( \Gamma_1+\Gamma_2 \right)\right|^2  \label{H2} \,,\\
	\det H'(c_1,c_2) &=& |c_1|^2 |c_2|^4\left| \det\left( \Gamma_1+\Gamma_2 \right)\right|^2 \label{H3}\,,
\ea
Consequently,
\be \label{H4}
\det H(c_1,c_2)-\det H^\prime(c_1,c_2)= |c_1|^2 |c_2|^2\bigl(|c_1|^2-|c_2|^2\bigr) \left|\det\left( \Gamma_1+\Gamma_2 \right)\right|^2=0\,.
\ee
Noting that a multinomial (in $|c_1|$ and $|c_2|$) is zero if and only if all of its coefficients are zero, it follows that 
$\det( \Gamma_1+\Gamma_2)=0$.  That is, the impact of the $\Pi_2$ symmetry is to impose the condition $\det( \Gamma_1+\Gamma_2)=0$ on the matrices specified in \eqs{eq:CaseIZ2}{eq:CaseIIZ2}.  Inserting this result back into \eq{H2}, we obtain
$\det H(c_1,c_2)=0$.
Finally, using \eqs{emen}{HdHu}, it follows that
\be \label{hdzero}
\det H_d =\det\bigl[ (\Gamma_1 v_1 + \Gamma_2 v_2) (\Gamma_1 v_1 + \Gamma_2 v_2)^\dagger \bigr]= \det H(v_1,v_2)=0\,,
\ee
which implies that at least one of the down-type quarks is massless.

In Case III [\eq{eq:CaseIIIZ2}], an
explicit calculation yields 
\ba
\det H(c_1,c_2)&=& |c_1|^2\Bigl|c^2_1\det\Gamma_1+c_2^2\det(\widetilde{\Gamma}_1+\Gamma_2)\Bigr|^2\,, \label{explicit}\\
\det H^\prime(c_1,c_2)&=& |c_2|^2\Bigl|c^2_2\det\Gamma_1+c_1^2\det(\widetilde{\Gamma}_1+\Gamma_2)\Bigr|^2\,,
\ea
where $\widetilde\Gamma_1$ is obtained from $\Gamma_1$ by setting $(\Gamma_1)_{33}=0$.  That is, 
\be \label{gamtil}
\widetilde\Gamma_1= \YukZeros{\text{x}}{\text{x}}{0}{\text{x}}{\text{x}}{0}{0}{0}{\text{0}}\,.
\ee
Consequently,
\ba \label{Hcpoly}
\det H(c_1,c_2)-\det H^\prime(c_1,c_2)&=&\bigl(|c_1|^6-|c_2|^6\bigr)|\det\Gamma_1|^2-|c_1|^2|c_2|^2\bigl(|v_1|^2-|v_2|^2\bigr)|\det(\widetilde\Gamma_1+\Gamma_2)|^2\nonumber \\
&&\qquad\qquad +\bigl[c_1 c_2\bigl(c_1^{*\,3} c_2-c_2^{*\,3} c_1\bigr)\det\Gamma_1^*\det(\widetilde\Gamma_1+\Gamma_2)+{\rm c.c.}\bigr]=0\,,
\ea
where c.c. stands for complex conjugate of the preceding term.
Since \eq{Hcpoly} is a multinomial in the variables $c_{1}$, $c_1^*$, $c_2$, and $c_2^*$, it is equal to zero if and only if all of its coefficients are zero.  
It follows that\footnote{Note that $\det\Gamma_1=0$ implies that either  
$(\Gamma_1)_{33}=0$ or $(\Gamma_1)_{11}(\Gamma_1)_{22}=(\Gamma_1)_{12}(\Gamma_1)_{21}$.  In either case, $\det(\Gamma_1+\Gamma_2)$ is independent of the value of 
$(\Gamma_1)_{33}=0$ due to the form of $\Gamma_2$ specified in \eq{eq:CaseIIIZ2}.  It then follows that $\det(\Gamma_1+\Gamma_2)=\det(\widetilde\Gamma_1+\Gamma_2)=0$.
In particular, the impact of the $\Pi_2$ symmetry is to impose the conditions $\det\Gamma_1=\det( \Gamma_1+\Gamma_2)=0$ on the matrices specified in \eq{eq:CaseIIIZ2}.}
\be \label{zeros}
\det\Gamma_1=\det(\widetilde{\Gamma}_1+\Gamma_2)=0\,.
\ee
In light of \eqss{hdzero}{explicit}{zeros} we again obtain $\det H_d=0$.
As in Case I and Case II, it follows that at least one of the down-type quarks is massless.

It is noteworthy that the analysis presented above does not require one to determine the exact
form of the Yukawa matrices after the application of both the $\mathbb{Z}_2$ and $\Pi_2$ symmetries.
We were able to exclude all cases based exclusively on basis invariant considerations.
Since all the possible extensions of $\ZPi$ to the Yukawa sector lead to
at least one massless down-type quark, we conclude that \textit{it is impossible to extend $\ZPi$
to the fermions in a way compatible with experimental results}.
Thus, in the 2HDM with three fermion generations, there is no difference between imposing GCP2 (as shown in Ref.~\cite{Ferreira:2010bm})
and $\ZPi$ on the Yukawa Lagrangian.  In both cases, the 
corresponding models are phenomenologically unacceptable due to the presence of at least one massless down-type quark.

One can now examine the consequences of \eq{eq:Pi2_On_Delta}, which governs the up-type quark Yukawa couplings.
The first step would be to determine the possible structures of $\Delta_1$ and $\Delta_2$ after imposing the $\mathbb{Z}_2$ symmetry
constraints.  Using \eq{eq:S_on_Delta}, we would make use of the choices of $S^{(\mathbb{Z}_2)}$ and $S^{(\mathbb{Z}_2)}_L$ employed in \eqst{SSLSnI}{SSLSnIII}, while
surveying all possible choices for $S^{(\mathbb{Z}_2)}_{pR}$.  Of course, since we have already concluded that none of the models governed by
Cases I--III are phenomenologically viable, there is no need to study further the possible extensions of the $\mathbb{Z}_2$ symmetry to the up-type quark sector.

\section{Extending U(1)$\otimes \Pi_2$ for three generations}
\label{sec:U1Pi2_n=3}

In this section, we consider how the symmetry U(1)$\otimes \Pi_2$ should act on the quark fields such that
the Yukawa Lagrangian is invariant with respect to U(1)$\otimes \Pi_2$ transformations of the scalars and quarks.
Following a similar strategy to the one employed in Section~\ref{sec:ZPi_n=3},
we begin with a list all possible extensions of U(1) to the fermions obtained in 
Ref.~\cite{Ferreira:2010ir}, which we have organized in Tables \ref{tab:1U1}--\ref{tab:4U1}.

Examining the possible models, one can check that most of them are actually
equivalent to or subcases of 
the extensions of $\mathbb{Z}_2$ analyzed before (up to permutations of the quark doublets 
and/or scalar doublets).
This should not come as a surprise, as U(1) contains a $\mathbb{Z}_2$.  
However, one can also consider a discrete $\mathbb{Z}_3$ symmetry, which when applied to the scalar sector results in a U(1)-symmetric scalar potential.
Extensions of $\mathbb{Z}_3$ to the fermions do not necessarily yield a Yukawa Lagrangian that is invariant 
under the $\mathbb{Z}_2$ symmetry previously analyzed (where the Yukawa matrices 
$\Gamma_1$ and $\Gamma_2$ are specified by one of the models of Table \ref{tab:Z2Models}).
That is, not all models with a U(1)-symmetric scalar potential, when extended to the Yukawa sector, are equivalent to or subcases of 
the extensions of the $\mathbb{Z}_2$ symmetry analyzed previously.   These are the models that we focus on
in this section.
Once again, we discard models that after imposing \eq{eq:Pi2_On_Gamma} yield $\Gamma_1=\Gamma_2=0$.
Then, there are only two inequivalent classes of models, $\{57, 92\}$, that cannot be related by any family
permutations,
\begin{align}
	\text{Case IV: } \quad  \Gamma_1
&= \YukZeros{0}{0}{0}{0}{0}{\text{x}}{\text{x}}{\text{x}}{0} \, , \qquad\qquad
\Gamma_2 = \YukZeros{\text{x}}{\text{x}}{0}{0}{0}{0}{0}{0}{\text{x}} \, ,
\label{eq:CaseIU1} \\[8pt] 
	\text{Case V: } \quad  \Gamma_1
&= \YukZeros{\text{x}}{0}{0}{0}{\text{x}}{0}{0}{0}{\text{x}} \, , \qquad\qquad
\Gamma_2 = \YukZeros{0}{\text{x}}{0}{0}{0}{\text{x}}{\text{x}}{0}{0} \, ,
\label{eq:CaseIIU1}
\end{align}
where x stands for an arbitrary complex number.
Note that we have obtained \eqs{eq:CaseIU1}{eq:CaseIIU1} 
by requiring that the Higgs Lagrangian is invariant under a $\mathbb{Z}_3$ 
Higgs and quark flavor symmetry transformation
(the former resulting in a U(1)-symmetric scalar potential).
In particular, 
the $\mathbb{Z}_3$ symmetry equations [\eqs{eq:S_on_Gamma}{eq:S_on_Delta}] have been applied 
with\footnote{As noted in footnote~\ref{fn:phases}, there is some additional phase freedom in defining the symmetry matrices exhibited in \eqs{zthreeIV}{zthreeV}.
Without loss of generality, we have simplified the form of the symmetry matrices by setting such phases to zero.}
\ba
\text{Case IV: } \quad S^{(\mathbb{Z}_3)} &=& \textrm{diag}\{ \omega, 1\}\, ,
\qquad
S^{(\mathbb{Z}_3)}_L = \textrm{diag}\{1, \omega^2, \omega\}\, ,
\qquad
S^{(\mathbb{Z}_3)}_{n_R} = \textrm{diag}\{1,1, \omega\}\, , \label{zthreeIV}
\\
\text{Case V: } \quad S^{(\mathbb{Z}_3)} &=& \textrm{diag}\{1, \omega\}\, ,
\qquad
S^{(\mathbb{Z}_3)}_L = S^{(\mathbb{Z}_3)}_{n_R} =\textrm{diag}\{1, \omega^2, \omega\}\,, \label{zthreeV}
\ea
where $1$, $\omega$, and $\omega^2$ 
[with $\omega \equiv \exp{(2i\pi/3)}$] are the three cube roots of unity. 
We are now ready to find the corresponding extension of $\UPi$ to the Yukawa interactions of the down-type quarks
by imposing the $\Pi_2$ symmetry equations given in 
\eq{eq:Pi2_On_Gamma}.
We will treat the two models corresponding to Cases IV and V independently in the following.

\subsection{Check for massless quarks}

We begin by applying to these cases the same test of looking at the effect of the $\Pi_2$ symmetry equations [\eq{eq:Pi2_On_Gamma}] on $\det(H_d)$.
We again introduce $H(c_1,c_2)$ as in \eq{H}.  Then, in  Case IV we find that \eqst{H2}{H4} are satisfied, which implies that $\det(\Gamma_1+\Gamma_2)=0$.
Hence, $\det H_d=H(v_1,v_2)=0$, which implies the existence of at least one massless down-type quark.

In Case V [\eq{eq:CaseIIU1}], an
explicit calculation yields 
\ba
\det H(c_1,c_2)&=& \Bigl|c_1^3\det\Gamma_1+c_2^3\det\Gamma_2\Bigr|^2\,, \\
\det H^\prime(c_1,c_2)&=&\Bigl|c_2^3\det\Gamma_1+c_1^3\det\Gamma_2\Bigr|^2\,.
\ea
Using
\be
\det H(c_1,c_2)-\det H^\prime(c_1,c_2)=\bigl(|c_1|^6-|c_2|^6\bigr)\bigl(|\det\Gamma_1|^2-|\det\Gamma_2|^2\bigr)-4\Im(c_1^3 c_2^{*\,3})\Im(\det\Gamma_1\det\Gamma_2^*)=0\,,
\ee
it follows that
\be
|\det\Gamma_1|=|\det\Gamma_2|\,,\qquad\qquad \Im(\det\Gamma_1\det\Gamma_2^*)=0\,,
\ee
which implies that $\det\Gamma_2=\pm\det\Gamma_1$.  In particular,
\be \label{HdCaseV}
\det H_d(v_1,v_2)=|v_1^3\pm v_2^3|^2|\det\Gamma_1|^2\,,
\ee
which is nonzero in general.  Hence,  Case V is compatible with the existence of non-zero down-type quark masses.

It is noteworthy that \eq{HdCaseV} has been derived under the assumption that a specific basis for the scalar fields and quark fields has been chosen.   
In particular, under a change basis of the scalar fields and quark fields,
$H_d\to U_L H_d U_L^\dagger$, after making use of eqs.~(\ref{emen}), (\ref{HdHu}), (\ref{vtrans}), and (\ref{Gamtransform}).  Consequently, the left-hand side of 
\eq{HdCaseV} is a manifestly basis independent quantity, whereas the right-hand side has been obtained in a specific basis.  It is straightforward to show that one cannot flip the sign in \eq{HdCaseV} by the transformation $\Phi_2\to -\Phi_2$, since such a transformation also changes $\Gamma_2\to -\Gamma_2$ [cf.~\eq{Gamtransform}], in which case $M_n=\Gamma_1 v_1+\Gamma_2 v_2$ (and likewise, $H_d=M_n M_n^\dagger$) are unchanged.

\subsection{Case V}
\label{vee}

We now take a closer look at the down-type quark mass matrix in Case V and the corresponding constraints imposed by the $\Pi_2$ symmetry.  We parametrize the down-type
Yukawa matrices as
\begin{equation} \label{exes}
	\Gamma_1 = 
	\begin{pmatrix}
	x_{11} & 0 & 0 \\
	0 & x_{22} & 0 \\
	0 & 0 & x_{33}	
	\end{pmatrix} \, , \qquad \qquad
	\Gamma_2 = 
	\begin{pmatrix}
		0 & x_{12} & 0 \\
		0 & 0 & x_{23} \\
		x_{31} &0 &0 
	\end{pmatrix}
	\, ,
\end{equation}
where the $x_{ij}$ are complex numbers.
Using the first relation given in \eq{eq:Pi2_On_Gamma}, 
we obtain
\begin{equation}
	\Gamma_1 = \text{diag}(x_{11},x_{22},x_{33}) = S_L \Gamma_2 S_{nR}^\dagger \, .
	\label{eq:BiDiagU1}
\end{equation}
It follows that
\ba
\diag\bigl(|x_{11}|^2,|x_{22}|^2,|x_{33}|^2\bigr)&=&\Gamma_1\Gamma_1^\dagger =S_L(\Gamma_2\Gamma_2^\dagger)S_L^\dagger=S_L\, \diag\bigl(|x_{12}|^2,|x_{23}|^2,|x_{31}|^2\bigr)S_L^\dagger\,,\\[4pt]
\diag\bigl(|x_{11}|^2,|x_{22}|^2,|x_{33}|^2\bigr)&=&\Gamma_1^\dagger \Gamma_1= S_{nR}(\Gamma_2^\dagger\Gamma_2)S_{nR}^\dagger=S_{nR}\, \diag\bigl(|x_{31}|^2,|x_{12}|^2,|x_{23}^2|\bigr)S_{nR}^\dagger\,.
\ea
Since $\Gamma_1\Gamma_1^\dagger$ and $\Gamma_2\Gamma_2^\dagger$ are related by a similarity transformation, they possess the same eigenvalues (and likewise for
$\Gamma_1^\dagger\Gamma_1$ and $\Gamma_2^\dagger\Gamma_2$).  Hence, 
$S_L$ and $S_{nR}$ are permutation matrices multiplied by a diagonal matrix of phases.

In general, a $n\times n$ permutation matrix $P$ is a real orthogonal matrix that consists of matrix elements such that~1 appears exactly once in each row and column while the remaining $n^2-n$
entries are 0.   For an arbitrary $n\times n$ matrix $A$, the matrix $AP$ permutes the columns of $A$ and the matrix $PA$ permutes the rows of $A$.
For the $n!$ possible permutations corresponding to the symmetric group $S_n$, we shall employ the cycle notation~\cite{Artin}.  In particular, $S_3$ is the group of 
permutation of three numbers $\{1,2,3\}$, whose elements are denoted by $\{Id,(12),(13),(23),(123),(132)\}$, where $Id$ is the identity element, $(12)$ corresponds to $1\leftrightarrow 2$, $3\to 3$; $(123)$ corresponds to $1\to 2$, $2\to 3$, $3\to 1$; etc.  The corresponding permutation matrices will be denoted by $P_g$ where $g\in S_3$.  For example~\cite{Artin},
\be
P_{(123)}=\begin{pmatrix} 0 & 0 & 1 \\ 1 & 0 & 0 \\ 0 & 1 & 0\end{pmatrix}\,.
\ee
One can check that $P_{(123)} (\Gamma_2 \Gamma_2^\dagger) P_{(123)}^{\T} = \Gamma_2^\dagger \Gamma_2$.  It then follows that
\begin{equation}\label{eq:relation_SL_SnR}
	KS_L = S_{nR} P_{(123)} \,,
\end{equation}
where $K$ is a diagonal matrix of phases.  Without loss of generality, we may choose $S_{nR}$ to be a permutation matrix by absorbing any additional phases of $S_{nR}$ into $K$.

Using the second equation of \eq{eq:Pi2_On_Gamma}, the roles of $\Gamma_1$ and $\Gamma_2$ are interchanged.   
Thus, instead of \eq{eq:BiDiagU1},
we now have $\Gamma_2=S_L\Gamma_1 S^\dagger_{nR}$.   Inserting this result back into \eq{eq:BiDiagU1} yields
\be \label{Kconstraint}
\Gamma_1 S_{nR}^2=S_L^2\Gamma_1\,. 
\ee
We will see shortly that \eqs{eq:relation_SL_SnR}{Kconstraint} constrain the diagonal matrix of phases, $K$.
Moreover, 
one can use the second equation of \eq{eq:Pi2_On_Gamma} to derive a relation analogous to 
\eq{eq:relation_SL_SnR}, $(KS_L)^{\dagger} = S_{nR}^{\dagger} P_{(123)}$, which is equivalent to
\begin{equation} \label{eq:relation_SL_SnR2}
	KS_L = P_{(123)}^{\T} S_{nR} \,.
\end{equation}
Combining \eqs{eq:relation_SL_SnR}{eq:relation_SL_SnR2} yields
\be \label{PSP}
P_{(123)} S_{nR} P_{(123)} = S_{nR}\,.
\ee
Having defined $S_{nR}$ to be a permutation matrix, one can easily check (using the $S_3$ group multiplication table) that only the odd permutation matrices,\footnote{The odd permutations are permutations that can be expressed as products of an odd number of 2-cycles.  Thus, $\{Id,(123),(132)\}$ are even permutations whereas $\{(12),(23),(13)\}$ are odd permutations.}
 $P_{(12)}$, $P_{(23)}$, and $P_{(13)}$, satisfy \eq{PSP}. 
Moreover, when $S_{nR}$ is an odd permutation,
$KS_L$ given by \eq{eq:relation_SL_SnR} is also an odd permutation.
Hence there are three possible cases:
\begin{enumerate}
	\item $S_{nR}=P_{(12)}\,$, and $KS_L=P_{(23)}$;
	\item $S_{nR}=P_{(23)}\,$, and $KS_L=P_{(13)}$;
	\item $S_{nR}=P_{(13)}\,$, and $KS_L=P_{(12)}$.
\end{enumerate}
Note that these cases are related by a
basis change of the quark fields where two of the generations
are interchanged.  Thus, without loss of generality, it suffices to consider one of the three models listed above.

Focusing on the third case above,
\begin{equation}
	KS_L = 
	\begin{pmatrix}
		0 & 1 & 0 \\
		1 & 0 & 0 \\
		0 & 0 & 1
	\end{pmatrix} \, , \qquad\qquad 
	S_{nR}=S_{nR}^\dagger = 
	\begin{pmatrix}
		0 & 0 & 1 \\
		0 & 1 & 0 \\
		1 & 0 & 0
	\end{pmatrix} \,.
	\label{eq:SymsU1Pi2}
\end{equation}
In light of \eq{Kconstraint}, it follows that 
\be
K={\rm diag}\bigl(e^{i\theta}\,,\,e^{-i\theta}\,,\, \pm 1\bigr)\,,
\ee
where $\theta$ is arbitrary (mod $2\pi$) and either choice of sign is allowed.
Then, the right-hand side of \eq{eq:BiDiagU1} yields:
\begin{equation}
	\text{diag}(x_{11},x_{22},x_{33}) = \text{diag}(e^{-i\theta}x_{23}\,,\,e^{i\theta}x_{12}\,,\,\pm x_{31}) \, .
	\label{eq:G1G2Diags}
\end{equation}
Hence, \eqs{exes}{eq:G1G2Diags} yield
\be \label{exes2}
\Gamma_2=\begin{pmatrix}
		0 &e^{-i\theta} x_{22} & 0 \\
		0 & 0 & e^{i\theta}x_{11} \\
		\pm x_{33} &0 &0 
	\end{pmatrix} .
\ee

Using \eqss{emen}{exes}{exes2}, it follows that
\be
	M_n=v_1\Gamma_1+v_2\Gamma_2 =
	\begin{pmatrix}
		\phm v_1 x_{11} &\quad  v_2 x_{22} e^{-i\theta} & \quad0 \\
		0 & \quad v_{1}x_{22} &\quad  v_2 x_{11} e^{i\theta} \\
		\pm v_{2}x_{33} & \quad 0 & \quad v_{1}x_{33}
	\end{pmatrix} \,,
	\label{eq:GammaU1Pi2}
\ee
and
\be \label{Hdee}
\det H_d\equiv \det (M_n M_n^\dagger)=|v_1^3 \pm v_2^3|^2|\det\Gamma_1|^2\,,
\ee
independently of the value of $\theta$,
in agreement with the result obtained in \eq{HdCaseV}.   As noted in the previous subsection, \eq{Hdee} generically implies that all down-type quark masses are nonzero,\footnote{In the special cases where $v^3_1=\mp v_2^3$, the model will contain a massless down-type quark and thus must be excluded.}  
as required for a viable candidate for a model that is invariant under a U(1)$\otimes \Pi_2$ symmetry transformation.

One can now examine the consequences of \eq{eq:Pi2_On_Delta}, which governs the up-type quark Yukawa couplings.
The first step would be to determine the possible structures of $\Delta_1$ and $\Delta_2$ after imposing the $\mathbb{Z}_3$ symmetry
constraints.  Using \eq{eq:S_on_Delta}, we would make use of the choices of $S^{(\mathbb{Z}_3)}$ and $S^{(\mathbb{Z}_3)}_L$ employed in \eqs{zthreeIV}{zthreeV}, while
surveying all possible choices for $S^{(\mathbb{Z}_3)}_{pR}$.   If we now impose a $\Pi_2$ symmetry with
$S_{pR}$ given by $S_{nR}$ in \eq{eq:SymsU1Pi2},
we obtain $\Delta_1$ and $\Delta_2$ with the same textures as $\Gamma_1$ and $\Gamma_2$, respectively.
This provides us with enough freedom to produce a model with
non-zero CKM mixing angles and a non-vanishing $J_{\rm CP}$.
We conclude that \textit{it is possible to extend U(1)$\otimes \Pi_2$
to the fermions in a way compatible with phenomenological constraints}.

\section{Equivalence of U(1)$\otimes \Pi_2$ and GCP3 models with three quark generations}
\label{sec:U1PivsGCP3_n=3}

In Section~\ref{subsec:ClassSymmetries},
we recalled that the GCP3-symmetric and the U(1)$\otimes \Pi_2$-symmetric scalar potentials were related by
a change of scalar field basis~\cite{Ferreira:2009wh}.
Moreover, as noted 
in Section~\ref{subsec:ExtSym}, it was shown in Ref.~\cite{Ferreira:2010bm} that there is
only one possible extension of GCP3 to the fermions in a model with three generations.
We also showed in Section~\ref{sec:U1Pi2_n=3} that it is possible to extend
the U(1)$\otimes \Pi_2$ symmetry of the scalar potential to the Yukawa sector.
Hence, one can now pose the following question: 
do the Yukawa extended GCP3 model and the Yukawa extended
U(1)$\otimes \Pi_2$ model coincide?

In Ref.~\cite{Ferreira:2009wh}, it was shown that given a GCP3-invariant scalar potential $V(\Phi^\prime)$, expressed in the scalar field
basis $\{\Phi_1^\prime,\Phi_2^\prime\}$, one can find a basis transformation,
\begin{equation} \label{basischange}
	\begin{pmatrix}
		\Phi_1 \\
		\Phi_2
	\end{pmatrix} = 
	\frac{1}{\sqrt{2	}}
	\begin{pmatrix}
		1 & 0 \\
		0 & e^{i\lambda} 
	\end{pmatrix}
	\begin{pmatrix}
		\phm 1 & -i \\
		-i & \phm 1 
	\end{pmatrix}
	\begin{pmatrix}
		\Phi_1^\prime \\
		\Phi_2^\prime
	\end{pmatrix} 
	= \frac{1}{\sqrt{2	}}
	\begin{pmatrix}
		\phm 1 & -i \\
		-i e^{i\lambda}  &\phm  e^{i\lambda} 
	\end{pmatrix}
	\begin{pmatrix}
		\Phi_1^\prime \\
		\Phi_2^\prime
	\end{pmatrix} 
	\,,
\end{equation}
such that the scalar potential $V(\Phi)$, expressed in the scalar field basis $\{\Phi_1,\Phi_2\}$,  is invariant with respect to U(1)$\otimes\Pi_2$ transformations.   In \eq{basischange},
we have introduced an arbitrary phase factor $e^{i\lambda}$ that will prove useful in the following.   One can check
that if the parameters of $V(\Phi^\prime)$ satisfy the GCP3 parameter relations specified in Table~\ref{tab:cases}, then the parameters of $V(\Phi)$ satisfy the
U(1)$\otimes\Pi_2$ parameter relations given in the same Table.

Suppose that the down-type quark Yukawa matrices satisfy the GCP3 conditions specified in \eq{eq:CP3Yuk}, which we shall henceforth denote by $\Gamma_1^\prime$ and $\Gamma_2^\prime$.  
Using the inverse of \eq{Gamtransform}, we can perform basis transformations both in 
the scalar and Yukawa sectors to obtain the corresponding Yukawa matrices in the new basis (denoted by $\Gamma_1$ and $\Gamma_2$, respectively), which we will then compare with the results obtained in Section~\ref{sec:U1Pi2_n=3}.  It then follows that
\begin{equation} \label{GammaUU}
	\Gamma_1 = U_L^\dagger \frac{1}{\sqrt{2}}(\Gamma_1^\prime  + i \Gamma_2^\prime)U_{nR} \, ,
\qquad\quad \Gamma_2 = U_L^\dagger \frac{e^{-i\lambda}}{\sqrt{2}}(i\Gamma_1^\prime  + \Gamma_2^\prime)U_{nR} \, .
\end{equation}  
In light of \eq{exes}, we perform a singular value decomposition by choosing the matrices $U_L$ and $U_{nR}$ such that $\Gamma_1$ is diagonal:
\begin{equation}
	U_L^\dagger = \frac{1}{\sqrt{2}}	
	\begin{pmatrix}
		-i & \phm 1 & \phm 0 \\
		\phm1 & -i &\phm  0 \\
		\phm 0 &\phm  0 &\phm  \sqrt{2}
	\end{pmatrix} \, , \qquad \quad
	U_{nR} = \frac{1}{\sqrt{2}}	
	\begin{pmatrix}
		-i & \phm 0 & \phm 1 \\ 
		\phm 1 &\phm 0 & -i \\
		\phm 0 &\phm \sqrt{2} &\phm 0
	\end{pmatrix} \,.
\end{equation}
\Eq{GammaUU} then yields,
\be \label{svdGam1}
	\Gamma_1 = {\rm diag}\Bigl(
	 	\sqrt{2} (a_{12}-i a_{11}) \,,\,a_{13}-i a_{23} \,,\, a_{31}-i a_{32} \Bigr)={\rm diag}\Bigl(x_{11}\,,\,x_{22}\,,\,x_{33}\Bigr)\,,
\ee
where we have introduced $x_{11}$, $x_{22}$, and $x_{33}$ to match the notation of \eq{exes}, and
\be
	\Gamma_2 = e^{-i\lambda}
	\begin{pmatrix}
		0 & a_{13}+i a_{23} & 0 \\
 		0 & 0 & -\sqrt{2} (a_{11}-i a_{12}) \\
		a_{31}+i a_{32} & 0 & 0 \\
	\end{pmatrix} = e^{-i\lambda}
	\begin{pmatrix}
		0 & \phm x^*_{22} &\phm  0 \\
 		0 & \phm 0 &\phm  i\,\!x^*_{11} \\
		x^*_{33} &\phm  0 & \phm 0 \\
	\end{pmatrix} \, .
\ee

Note that $\Gamma_2$ does not quite match the corresponding Yukawa matrix that appears in \eq{exes} after identifying its parameters using \eq{eq:G1G2Diags}.
However, we still have some phase freedom in defining the matrices $U_L$ and $U_R$ of the singular value decomposition.
Indeed, if we make the following replacements,
\be
U^\dagger_L\longrightarrow {\rm diag}\bigl(e^{-i\phi_1}\,,\,e^{-i\phi_2}\,,\,e^{-i\phi_3}\bigr)\,U^\dagger_L\,, \qquad\quad
U_R\longrightarrow U_R\,{\rm diag}\bigl(e^{i\phi_1}\,,\,e^{i\phi_2}\,,\,e^{i\phi_3}\bigr)\,,
\ee
then \eq{svdGam1} is unmodified, whereas $\Gamma_2$ is transformed into 
\be \label{gammatwo}
\Gamma_2 = 
	\begin{pmatrix}
		0 &\,\, e^{-i(\phi_{12}+\alpha_2+\lambda)} |x_{22}| & \,\,0 \\
		0 &\,\, 0 &\,\, i\,\!e^{-i(\phi_{23}+\alpha_1+\lambda)} |x_{11}| \\
		e^{-i(\phi_{31}+\alpha_3+\lambda)} |x_{33}| &\,\,& \,\,0 
	\end{pmatrix} \,,
\ee
where $\phi_{ij}\equiv\phi_i-\phi_j$ and $\alpha_i\equiv \arg x_{ii}$.

In \eq{exes2}, the expression obtained for $\Gamma_2$ contained the parameter $\theta$ and a choice of sign, 
\be \label{gammatwo2}
\Gamma_2=\begin{pmatrix}
		0 &\,\, |x_{22}|e^{i(\alpha_2-\theta)} &\,\, 0 \\
		0 &\,\,0 &\,\, |x_{11}|e^{i(\alpha_1+\theta)} \\
		\pm |x_{33}|e^{i\alpha_3} &\,\, 0 &\,\, 0 
	\end{pmatrix} \,.
\ee
We can therefore equate \eqs{gammatwo}{gammatwo2}
if the following equations are satisfied mod $2\pi$:
\ba
2\alpha_1&=&\half\pi-\theta-\lambda-\phi_{23}\,,\label{e1}\\
2\alpha_2&=&\theta-\lambda-\phi_{12}\,,\label{e2} \\
2\alpha_3+m\pi&=&-\lambda+\phi_{12}+\phi_{23}\,,\label{e3}
\ea
where $m=0$ or 1  and we have used $\phi_{12}+\phi_{23}+\phi_{31}=0$ to eliminate $\phi_{31}$.   We can use \eqst{e1}{e3} to solve for
$\lambda$, $\phi_{12}$ and $\phi_{23}$ in terms of the $\alpha_i$ , $m$, and $\theta$.   Thus, we have demonstrated that the down-type Yukawa coupling matrices in
the GCP3 model and the U(1)$\otimes \Pi_2$ model can be related by an appropriate change of basis of the scalar fields and quark fields.

We conclude that \textit{there is a one-to-one correspondence between the Yukawa extended GCP3 model and the Yukawa extended
{\rm U(1)}$\otimes \Pi_2$ models that simply reflects a different choice of the scalar field and quark field basis}.
The addition of the three-generation Yukawa sector did not ``remove the degeneracy'' of
the GCP3 and U(1)$\otimes \Pi_2$ models that was present when only the symmetries of the scalar potential were considered.

\section{(Non-)Equivalence of $\mathbb{Z}_2\otimes \Pi_2$ and GCP2 models with two quark generations}
\label{sec:Z2PivsGCP2_n=2}

In previous sections, we showed that although there is no distinction between a scalar potential of the 2HDM that respects the GCP2 or $\ZPi$ symmetry, no viable model (i.e., a model that
possesses nonzero quark masses and a nontrivial CKM matrix) exists in which the GCP2 and/or the $\ZPi$ symmetry can be consistently extended to the Yukawa sector with $n=3$ generations of quarks.   Likewise, there is no distinction between
a scalar potential that respects the GCP3 or $\UPi$ symmetry, although both these symmetries can be extended to the three-generation Yukawa sector to produce a viable model.    
However, we also showed that this extension does \textit{not} ``remove the degeneracy'' present in the scalar sector between GCP3 and U(1)$\otimes \Pi_2$, as the two models are related by a particular change of the scalar field and quark field basis.

To see whether
these results are specific to the 2HDM with $n=3$ quark generations, we shall consider a version of the
2HDM with $n$ different from three.   The case of $n=1$ can be discarded,\footnote{If one were to expand the 2HDM Yukawa sector to include, e.g., up-type and down-type vectorlike quarks~\cite{Arhrib:2016rlj},
then it would be possible to extend the $\ZPi$ and $\UPi$ symmetries of the scalar sector to a one-generation Yukawa sector~\cite{Draper:2016cag,Draper:2020tyq}.  Such models are beyond the scope of this work.} 
since one would quickly conclude using the methods employed in 
this section that there are no consistent extensions of these symmetries to the Yukawa sector
unless all Yukawa couplings vanish~\cite{CarroloThesis}.   In this section, we consider the 2HDM with $n=2$
quark generations.  In particular, we shall examine the relation between two-generation models 
with GCP2 and $\ZPi$ symmetries appropriately extended to the Yukawa sector.

\subsection{GCP2 for two generations}
\label{sec:GCPfortwo}

In the case of two generations of fermions,
the Yukawa couplings are $2 \times 2$ matrices that can be parametrized as 
\begin{equation}
	\Gamma_1 = 
	\begin{pmatrix}
		x_{11} & x_{12} \\
		x_{21} & x_{22}
	\end{pmatrix} \, , \qquad \qquad
	\Gamma_2 = 
	\begin{pmatrix}
		y_{11} & y_{12} \\
		y_{21} & y_{22}
	\end{pmatrix}\, ,
\label{param_G_n=2}
\end{equation}
where all parameters are complex.
Furthermore, using the result in \eq{eq:GCPReduced},
we choose the quark basis where the GCP symmetry matrices
take the simple form 
\begin{equation}
	X_L = 
	\begin{pmatrix}
		\phm c_\alpha & \phm s_\alpha \\
		-s_\alpha & \phm c_\alpha
	\end{pmatrix} \, , \qquad\qquad 
	X_{n_R} = 
	\begin{pmatrix}
		\phm c_\beta &\phm  s_\beta \\
		-s_\beta & \phm c_\beta
	\end{pmatrix}\, ,
\end{equation}
where $c_\alpha \equiv\cos\alpha$, $c_\beta\equiv\cos\beta$, $s_\alpha \equiv \sin\alpha$, and $s_\beta\equiv\sin\beta$, with $0\leq \alpha,\beta\leq\half\pi$ [as noted below \eq{eq:GCPReduced}].
The GCP2 symmetry equations given by  \eqst{GCP_invariant1}{GCP_invariant4}  are obtained by setting 
\be \label{XGCP2}
X=\begin{pmatrix} \phm 0 & \phm 1 \\ -1 & \phm 0\end{pmatrix}\,,
\ee
in light of \eqs{eq:GCPSymmetry1}{eq:CP2}.  
We can then rewrite \eq{GCP_invariant3} as
\be
	X_L \Gamma_1^* + \Gamma_2 X_{n_R} = 0 \, ,\qquad\quad
	X_L \Gamma^*_2 - \Gamma_1 X_{n_R} = 0 \,.
\label{LuisA}
\ee
Since $X_L$ and $X_{n_R}$ are real orthogonal matrices, we can eliminate $\Gamma_2$ to obtain
\be \label{XL2}
X_L^2\Gamma_1+\Gamma_1 X_{n_R}^2=0\,.
\ee

Substituting the parametrization of \eq{param_G_n=2} into \eq{LuisA},
we obtain 8 complex linear equations in 8 complex variables, which one can write in matrix form as
\be \label{Amatrix}
A\boldsymbol{x}=\boldsymbol{0}\,,\qquad \text{where $\boldsymbol{x}=\bigl(x_{11}^*,x_{12}^*,x_{21}^*,x_{22}^*,y_{11},y_{12},y_{21},y_{22}\bigr)^{\T}$},
\ee
where $A$ is a real $8\times 8$ matrix that can be written in block matrix form as
\be \label{Adef}
A=\begin{pmatrix} \phm c_\alpha\id & \,\,\,s_\alpha\id &\ \phm X_{n_R}^{\T} & \,\,\,\boldsymbol{0} \\
-s_\alpha\id & \,\,\, c_\alpha\id & \boldsymbol{0} & \,\,\,  X_{n_R}^{\T}\\
-X_{n_R}^{\T} & \,\,\, \boldsymbol{0} & \phm c_\alpha \id & \,\,\, s_\alpha\id \\
\boldsymbol{0} & \,\,\, -X_{n_R}^{\T} &   -s_\alpha\id & \,\,\, c_\alpha\id\end{pmatrix},
\ee
where $\id$ is the $2\times 2$ identity matrix and $\boldsymbol{0}$ is the eight-component zero vector in \eq{Amatrix} and the $2\times 2$ zero matrix in \eq{Adef}.
Since $\boldsymbol{x}\neq 0$, it follows that $\det A=0$, which will constrain the values of $\alpha$ and $\beta$ as we now demonstrate.
In Appendix~\ref{app:det}, we provide an explicit evaluation of $\det A$, which yields
\be \label{detAvalue}
\det A=16(c^2_\alpha c^2_\beta-s^2_\alpha s^2_\beta)^2=16\cos^2(\alpha+\beta)\cos^2(\alpha-\beta)=0\,,
\ee
where we have set $\det A=0$ as indicated below \eq{Adef}.  Since $0\leq\alpha,\beta\leq\half\pi$, it then follows that
\begin{equation}
	\alpha+\beta=\half\pi \,.
\label{alphbetpi2}
\end{equation}

Using \eq{alphbetpi2} in evaluating \eq{XL2} then yields
\begin{equation}
	X^2_L(\half\pi-\beta)\Gamma_1+\Gamma_1X^2_{n_R}(\beta)=\begin{pmatrix}
		\phm x_{21}-x_{12} &\,\, \phm x_{11}+x_{22} \\
		-x_{11}-x_{22} &\,\, -x_{12}+x_{21}
	\end{pmatrix} \sin 2\beta
	= 0 \, ,
\end{equation}
which leads to three possible cases:
\begin{align}
	\text{I :} 
&\quad \text{$\alpha + \beta = \half\pi $ with $\alpha,\beta\neq 0\,,\,\half\pi$}
\quad\Longrightarrow\quad
x_{21}=x_{12} \,,\,\, x_{22}=-x_{11}\,,	\ \label{eq:CP2case1}  \\
	\text{II :} 
&\quad \alpha = \half\pi\,,\, \beta = 0\,,	\label{eq:CP2case2} \\
	\text{III :} 
&\quad \alpha = 0 \,,\, \beta =\half\pi\,. \label{eq:CP2case3}
\end{align}
After imposing the constraints corresponding to one of these three cases,
we are left with a single symmetry equation constraining the down-type Yukawa coupling matrices [cf.~\eq{LuisA}]:
\be \label{eq:CP2gamma2}
\Gamma_2^*=X_L^\dagger\Gamma_1 X_{n_R}\,.
\ee

\subsubsection{Case I}

In this case,
$\alpha$ and $\beta$ can take any value in the open interval $(0,\pi/2)$
that satisfy the equation $\alpha +\beta = \pi/2$.
Constraining the couplings with \eqs{eq:CP2case1}{eq:CP2gamma2},
we obtain the following $\Gamma$ matrices:
\begin{equation}
	\Gamma_1 = \begin{pmatrix}
		x_{11} & \phm x_{12} \\
		x_{12} & -x_{11}
	\end{pmatrix} \, , \qquad \qquad
	\Gamma_2 = \begin{pmatrix}
		-x_{12}^* &\phm  x_{11}^* \\
		\phm x_{11}^* &\phm  x_{12}^* 
	\end{pmatrix} \,.
\label{GCP2_n=2_Case_I}
\end{equation}

\subsubsection{Case II}

In the case $\alpha = \pi/2$ and $\beta = 0$,
there are no constraints on $\Gamma_1$.
Using  \eqs{eq:CP2case2}{eq:CP2gamma2}, we obtain the following~$\Gamma$ matrices:
\begin{equation}
	\Gamma_1 = 
	\begin{pmatrix}
		x_{11} & \phm x_{12} \\
		x_{21} & \phm x_{22}
	\end{pmatrix} \, ,  \qquad \qquad
	\Gamma_2 = \begin{pmatrix}
		-x_{21}^* & -x_{22}^* \\
		\phm x_{11}^* &\phm  x_{12}^* 
	\end{pmatrix} \,.
\label{GCP2_n=2_Case_II}
\end{equation}

\subsubsection{Case III}

In the case $\alpha = 0$ and $\beta = \pi/2$,
there are no constraints on $\Gamma_1$.
Using  \eqs{eq:CP2case3}{eq:CP2gamma2}, we obtain the following~$\Gamma$ matrices:
\begin{equation}
	\Gamma_1 = 
	\begin{pmatrix}
		x_{11} & \phm x_{12} \\
		x_{21} &\phm  x_{22}
	\end{pmatrix}  \, ,  \qquad \qquad
	\Gamma_2 = 
	\begin{pmatrix}
		-x_{12}^* & \phm x_{11}^* \\
		-x_{22}^* & \phm x_{21}^*
	\end{pmatrix} \, .
\label{GCP2_n=2_Case_III}
\end{equation}

\subsubsection{Cabibbo angle and summary of cases}
\label{subsubsec:Cabibbo_no}

In contrast to Case I where $\Gamma_1$ and $\Gamma_2$ are fixed by two independent complex parameters,
Cases II and III are governed by four complex parameters.  This distinction will be significant when we compare the Yukawa sectors of the two-generation models constrained by 
the GCP2 and the $\ZPi$ symmetry, respectively, in Section~\ref{compare}.  Moreover, it is straightforward
to compute $\det H_d$ for all three cases:   
\ba
 \text{Case I:} &&\quad \det H_d=|(v_1 x_{11}-v_2 x_{12}^*)^2+(v_1 x_{12}+v_2 x_{11}^*)^2|^2\,, \nonumber \\[4pt]
 \text{Case II:} &&\quad \det H_d=|(v_1 x_{11}-v_2 x_{21}^*)(v_1 x_{22}+v_2 x_{12}^*)-(v_1 x_{12}-v_2 x_{22}^*)(v_1 x_{21}+v_2 x_{11}^*)|^2\,, \nonumber \\[4pt]
 \text{Case III:} &&\quad \det H_d=|(v_1 x_{11}-v_2 x_{12}^*)(v_1 x_{22}+v_2 x_{21}^*)-(v_1 x_{12}+v_2 x_{11}^*)(v_1 x_{21}-v_2 x_{22}^*)|^2\,.
\ea
In all three cases, $\det H_d$ is nonzero for a generic choice of parameters,\footnote{For example, in Case 1, $\det H_d=0$ only if $v_1/v_2=x_{11}/x_{12}^*$ and $\Re(v_1/v_2)=0$.}  which implies that the two down-type quark masses are generically nonzero.   
The same analysis yields the up-type quark Yukawa matrices $\Delta_1$ and $\Delta_2$ with precisely the same textures as the ones obtained for $\Gamma_1$ and $\Gamma_2$ above, which implies that the two up-type quark masses are also generically nonzero.  Even though $X_{p_R}$ and $X_{nR}$ are initially unrelated, the application of the GCP2 symmetry equations analogous to \eq{LuisA} to the up-type Yukawa coupling matrices,
\be
	X_L \Delta_1^* + \Delta_2 X_{p_R} = 0 \, ,\qquad\quad
	X_L \Delta^*_2 - \Delta_1 X_{p_R} = 0 \,.
\ee
yields the same three cases as in \eqst{eq:CP2case1}{eq:CP2case3}, with the angle $\beta$ replaced by a new angle $\gamma$ that is likewise constrained by the value of the angle $\alpha$.
It follows that the textures of the $\Gamma$ and $\Delta$ matrices must match, implying that there are three distinct classes of GCP2 models in total, which we shall denote below by I/I, II/II, and III/III to indicate the corresponding textures of the $\Gamma$ and $\Delta$ matrices, respectively.

In the case of two quark generations,
one can build a basis invariant quantity measuring the Cabibbo angle $\theta_c$
using
\begin{equation}
	J_c \equiv \textrm{det}\bigl\{[H_u,H_d]\bigr\}=
\det\bigl[V^\dagger D_u^2 V D_d^2-D_d^2 V^\dagger D_u^2 V\bigr]
= (m_d^2-m_s^2)^2(m_c^2-m_{u}^2)^2  \cos^2\theta_c \sin^2\theta_c \, ,
\label{JC}
\end{equation} 
where $V$ is now the $2 \times 2$ Cabibbo mixing matrix
\be
V= 	\begin{pmatrix}
	\phm \cos{\theta_c} &\phm \sin{\theta_c} \\
	-\sin{\theta_c} & \phm \cos{\theta_c}
	\end{pmatrix}  \, .
\ee
In model I/I, $H_d$ and $H_u$ are given by
\be \label{HdHu2gen}
H_d=\begin{pmatrix} \phm A & \phm B\\ -B & \phm A\end{pmatrix}\,,\qquad\quad H_u=\begin{pmatrix} \phm C & \phm D\\ -D & \phm C\end{pmatrix}\,,
\ee
where
\ba
A&\equiv &|v_1 x_{11}-v_2 x_{12}^*|^2+|v_1 x_{12}+v_2 x_{11}^*|^2=\bigl(|v_1|^2+|v_2|^2\bigr)\bigl(|x_{11}|^2+|x_{12}|^2\bigr)\,, \nonumber \\[4pt]
B&\equiv & (v_1 x_{11}-v_2 x_{12}^*)(v_1^* x_{12}^* + v_2^* x_{11})-(v_1 x_{12}+v_2 x_{11}^*)(v_1^* x_{11}^*-v_2^* x_{12})\,.
\ea
and $C$ and $D$ are obtained from $A$ and $B$ by replacing the corresponding elements of the $\Gamma$ matrices with those of the $\Delta$ matrices.
The form of $H_d$ and $H_u$ given in \eq{HdHu2gen} immediately yield $[H_u,H_d]=0$.   Thus $J_c=0$ which implies that 
$\sin 2\theta_c=0$, which is 
experimentally excluded.   In contrast, in models II/II and III/III, the four matrix elements of $H_d$ and $H_u$, respectively,  are independent quantities so that $[H_u,H_d]\neq 0$.  In particular, $\det\bigl\{[H_u,H_d]\bigr\}\neq 0$,
or equivalently $J_c\neq 0$.  Thus, we are left with two possible classes of two-generation models that extend the GCP2 symmetry to the Yukawa sector, each of which is
governed by four complex parameters in the down-type quark and up-type quark sectors, respectively.

\subsection{$\ZPi$ for two generations}
\label{twogenZPi}

Having found that GCP2 can be extended to the fermions with two generations,
we now want to check whether an extension also exists for $\ZPi$, and if such an
extension corresponds to GCP2 in another choice of scalar field and quark field basis.
Following the analysis of the three generation model of Section~\ref{sec:ZPi_n=3},
we begin by determining all possible extensions of the $\mathbb{Z}_2$ symmetry defined in \eq{eq:HF1} to the Yukawa sector.
To do this, we will employ a simplified version of the method used in Ref.~\cite{Ferreira:2010ir}.

\subsubsection{Extensions of $\mathbb{Z}_2$ to the Yukawa sector: the basics}
\label{Extensions22}

As noted in Ref.~\cite{Ferreira:2010ir},
due to the unitarity of the quark symmetry matrices,
one can always choose a basis such that $\mathbb{Z}_2$ symmetry matrices, $S^{(\mathbb{Z}_2)}_L$ and $S^{(\mathbb{Z}_2)}_{nR}$ are diagonal.
In this basis, the symmetry matrices can be written as 
\begin{equation}
	S^{(\mathbb{Z}_2)} = 
	\begin{pmatrix}
		1 & \phm 0 \\ 
		0 & -1
	\end{pmatrix} \, , \qquad \quad
	S^{(\mathbb{Z}_2)}_L = 
	\begin{pmatrix}
		e^{i\alpha_1} & 0 \\ 
		0 & e^{i\alpha_2}
	\end{pmatrix} \, , \qquad\quad
	S^{(\mathbb{Z}_2)}_{nR} = 
	\begin{pmatrix}
		e^{i\beta_1} & 0 \\ 
		0 & e^{i\beta_2}
	\end{pmatrix} \, .
	\label{eq:ParamsSymZ2}
\end{equation}
This implies that the $\mathbb{Z}_2$ symmetry equation given by \eq{eq:S_on_Gamma} can be written in the simple form
\begin{equation} \label{zeetwosym}
	(\Gamma_1)_{ab} = (\Gamma_1)_{ab} e^{i\theta_{ab}} \, ,
\quad\qquad
(\Gamma_2)_{ab} = (\Gamma_2)_{ab} e^{i(\theta_{ab}-\pi)} \, ,
\end{equation}
where $a,b\in\{1,2\}$ are left-handed quark generation indices and we have adopted the notation such that 
\begin{equation}
	\theta_{ab} \equiv  \alpha_a - \beta_b \,.
	\end{equation}
From \eq{zeetwosym}, we can readily see that there are
only three possibilities:
\begin{enumerate}
	\item  $\theta_{ab} = 0$, then $(\Gamma_1)_{ab}$ can take any value, and $(\Gamma_2)_{ab} = 0$
	\item  $\theta_{ab} = \pi$, then $(\Gamma_2)_{ab}$ can take any value, and $(\Gamma_1)_{ab} = 0$
	\item  $\theta_{ab} \neq 0,\pi$, then $(\Gamma_1)_{ab} = (\Gamma_2)_{ab} =0 $ 
\end{enumerate}
where all conditions on $\theta_{ab}$ are taken to be mod~$2\pi$.

\subsubsection{Extensions of $\mathbb{Z}_2$ to the Yukawa sector: left space constraints \label{sec:LSpaceConstr}}

Consider the following combinations of Yukawa matrices,
\ba
\Gamma_i \Gamma_j^\dagger
&=& (-1)^{(1-\delta_{ij})} S_L \Gamma_i \Gamma_j^\dagger  S_L^\dagger \, ,
\label{Z2gammaij} \\[3pt]
\Delta_i \Delta_j^\dagger
&=& (-1)^{(1-\delta_{ij})} S_L \Delta_i \Delta_j^\dagger  S_L^\dagger  \, ,\label{Z2deltaij}
\ea
where we have used \eq{eq:S_on_Gamma} with $S$ given in \eq{eq:ParamsSymZ2}.
We can see that, contrary to the right-handed symmetries that
act on either the up or the down quarks,
the left-handed symmetries affect both,
and are thus very constraining.  For example \eq{Z2gammaij} yields the following constraint,
\begin{equation} \label{gamgamconstraint}
	(\Gamma_i \Gamma_j^\dagger )_{ab}
= (\Gamma_i \Gamma_j^\dagger )_{ab} \, e^{i(\alpha_a-\alpha_b)+i \pi (1-\delta_{ij})}\,,
\end{equation}
after inserting the expression for $S_L$ given in 
\eq{eq:ParamsSymZ2}.
In particular, 
the phase on the right-hand side of \eq{gamgamconstraint}
is given by
\begin{equation}
\alpha_a-\alpha_b+ \pi(1-\delta_{ij})=	\begin{cases}
	\begin{pmatrix}
		0 & \alpha_1 - \alpha_2 \\ 
		\alpha_2-\alpha_1 & 0 
 	\end{pmatrix}_{ab}\,, & \text{for $i=j$}, \\\\
 	\begin{pmatrix}
 		\pi & \pi + \alpha_1-\alpha_2 \\
 		\pi + \alpha_2 - \alpha_1 & \pi 
 	\end{pmatrix}_{ab}\,, & \text{for $i\neq j$},
	\end{cases}
\end{equation}
where $a$,$b\in\{1,2\}$ label the elements of the $2\times 2$ matrices above.
Thus, there are three possibilities:
\begin{enumerate}
\item If $\alpha_1 - \alpha_2 = 0 $,
then $\Gamma_1\Gamma_2^\dagger = \Gamma_2\Gamma_1^\dagger = 0$,
with $\Gamma_1\Gamma_1^\dagger$, and $\Gamma_2\Gamma_2^\dagger $ unconstrained. 
\item If  $\alpha_1 - \alpha_2 = \pi$,
then $\Gamma_1\Gamma_1^\dagger$ and $\Gamma_2\Gamma_2^\dagger $ are diagonal,
whereas $\Gamma_1\Gamma_2^\dagger$ and $\Gamma_2\Gamma_1^\dagger$ are off-diagonal.
\item If  $\alpha_1 - \alpha_2 \neq 0,\pi$,
then $\Gamma_1\Gamma_1^\dagger$, $\Gamma_2\Gamma_2^\dagger$ are diagonal,
whereas $\Gamma_1\Gamma_2^\dagger=\Gamma_2\Gamma_1^\dagger = 0$.
This implies that  $H_d$ is diagonal.  A similar analysis of the up-type quark sector yields a diagonal $H_u$ in this case.
Hence, $\det\bigl\{[H_u,H_d]\bigr\}=0$ which yields $J_c=0$.  It follows that $\sin 2\theta_c=0$, which is experimentally excluded.
\end{enumerate}
In case 1 above where $\alpha_1 = \alpha_2$, the matrix $\theta_{ab}$ is given by
\begin{equation}
	\theta = \begin{pmatrix}
		\theta_{11} & \phm\theta_{12} \\
		\theta_{11} & \phm\theta_{12}
	\end{pmatrix} \, ,
\end{equation}
In case 2 above where $\alpha_2 = \pi + \alpha_1$, 
the matrix $\theta_{ab}$ is given by 
\begin{equation}
	\theta = \begin{pmatrix}
		\theta_{11} &\phm  \theta_{12} \\
		\theta_{11} + \pi & \phm  \theta_{12} + \pi
	\end{pmatrix} \,,
\end{equation}
where the elements of the matrix $\theta$ are evaluated mod~$2\pi$.
Combining the results obtained in this section
with those found below \eq{zeetwosym},
we obtain all the possible extensions of $\mathbb{Z}_2$ in
Table~\ref{tab:Z2Models_n=2}, where x stands for an arbitrary complex number. 

\begin{table}[H]
\centering
	\begin{tabular}{|c|c|c|c|}
		\hline
		\multicolumn{4}{|c|}{$\alpha_1=\alpha_2$}\\
		\hline
		$\theta_{11}$ & $\theta_{12}$ & $\Gamma_1$ & $\Gamma_2$ \\
	 	\hline
	 	0 & 0 & $\YukZerosTwo{\text{x}}{\text{x}}{\text{x}}{\text{x}}$
& $\YukZerosTwo{0}{0}{0}{0}$\\ 
	 	\hline	 
	 	$\pi$ & 0 &$\YukZerosTwo{0}{\text{x}}{0}{\text{x}}$
& $\YukZerosTwo{\text{x}}{0}{\text{x}}{0}$\\  
	 	\hline
	 	0 & $\pi$ & $\YukZerosTwo{\text{x}}{0}{\text{x}}{0}$
& $\YukZerosTwo{0}{\text{x}}{0}{\text{x}}$\\ 
	 	\hline
	 	$\pi$ & $\pi$ & $\YukZerosTwo{0}{0}{0}{0}$
& $\YukZerosTwo{\text{x}}{\text{x}}{\text{x}}{\text{x}}$\\ 
	 	\hline	 
	 \end{tabular}
	 \quad \quad
	 \begin{tabular}{|c|c|c|c|}
	 	\hline
		\multicolumn{4}{|c|}{$\alpha_2= \pi + \alpha_1$}\\
		\hline
		$\theta_{11}$ & $\theta_{12}$ &$\Gamma_1$ & $\Gamma_2$ \\
	 	\hline
	 	0 & 0 & $\YukZerosTwo{\text{x}}{\text{x}}{0}{0}$
& $\YukZerosTwo{0}{0}{\text{x}}{\text{x}}$\\ 
	 	\hline	 
	 	$\pi$ & 0 &$\YukZerosTwo{0}{\text{x}}{\text{x}}{0}$
& $\YukZerosTwo{\text{x}}{0}{0}{\text{x}}$\\  
	 	\hline
	 	0 & $\pi$ & $\YukZerosTwo{\text{x}}{0}{0}{\text{x}}$
& $\YukZerosTwo{0}{\text{x}}{\text{x}}{0}$\\ 
	 	\hline
	 	$\pi$ & $\pi$ & $\YukZerosTwo{0}{0}{\text{x}}{\text{x}}$
& $\YukZerosTwo{\text{x}}{\text{x}}{0}{0}$\\ 
	 	\hline
	 \end{tabular}
	\caption{Extensions of the $\mathbb{Z}_2$ symmetry to the Yukawa sector consisting of two quark generations.}
	\label{tab:Z2Models_n=2}
\end{table}

The cases in Table~\ref{tab:Z2Models_n=2} that are related by the interchange of $\Gamma_1$ and $\Gamma_2$ are physically equivalent, as this transformation simply corresponds to a change of the scalar field basis in which $\Phi_1$ and $\Phi_2$ are interchanged.
We can therefore choose the form of the Yukawa matrices $\Gamma_1$ and $\Gamma_2$ for the four inequivalent cases as exhibited below:
\ba
	\text{\textbf{Case 0}:}&\qquad
\Gamma_1 = \YukZerosTwo{\text{x}}{\text{x}}{\text{x}}{\text{x}} \, , \qquad\quad
\Gamma_2 = \YukZerosTwo{0}{0}{0}{0}\,, \label{eq:CasesZ2_2fam0} \\[2pt]
	\text{\textbf{Case 1}:}&\qquad
\Gamma_1 = \YukZerosTwo{\text{x}}{\text{x}}{0}{0} \, , \qquad\quad
\Gamma_2 = \YukZerosTwo{0}{0}{\text{x}}{\text{x}} \,,\label{eq:CasesZ2_2fam1}\\[2pt]
	\text{\textbf{Case 2}:}&\qquad
\Gamma_1 = \YukZerosTwo{\text{x}}{0}{\text{x}}{0} \, , \qquad\quad
\Gamma_2 = \YukZerosTwo{0}{\text{x}}{0}{\text{x}}\,, \label{eq:CasesZ2_2fam2} \\[2pt]
	\text{\textbf{Case 3}:}&\qquad
\Gamma_1 = \YukZerosTwo{\text{x}}{0}{0}{\text{x}} \, , \qquad\quad
\Gamma_2 = \YukZerosTwo{0}{\text{x}}{\text{x}}{0}\,.\label{eq:CasesZ2_2fam3}
\ea

These cases correspond, respectively, to the following $\mathbb{Z}_2$ symmetry matrices in a scalar field basis where $S^{(\mathbb{Z}_2)}=\sigma_Z$:
\ba
&&\text{\textbf{Case 0} [$\theta_{11}=\theta_{12}=0$, $\alpha_1=\alpha_2$]:}\phantom{xxii}\qquad\quad\, S^{(\mathbb{Z}_2)}_L=e^{i\alpha_1}\mathbbm{1} \,,\qquad\,\,\,\,\, S^{(\mathbb{Z}_2)}_{n_R}=e^{i\alpha_1}\mathbbm{1} \,, \label{Z22case0}\\[2pt]
&&\text{\textbf{Case 1} [$\theta_{11}=\theta_{12}=0$, $\alpha_2=\pi+\alpha_1$]:}\phantom{xii}\qquad S^{(\mathbb{Z}_2)}_L=e^{i\alpha_1}\sigma_Z\,,\qquad \,\,\! S^{(\mathbb{Z}_2)}_{n_R}=e^{i\alpha_1} \mathbbm{1} \,,\label{Z22case1} \\[2pt]
&&\text{\textbf{Case 2} [$\theta_{11}=0$, $\theta_{12}=\pi$, $\alpha_1=\alpha_2$]:}\phantom{xxii}\qquad S^{(\mathbb{Z}_2)}_L=e^{i\alpha_1}\mathbbm{1} \,,\qquad\,\,\,\,\,S^{(\mathbb{Z}_2)}_{n_R}=e^{i\alpha_1}\sigma_Z \,,\label{Z22case2} \\[2pt]
&&\text{\textbf{Case 3} [$\theta_{11}=0$, $\theta_{12}=\pi$,  $\alpha_2=\pi+\alpha_1$]:}\qquad\,\, S^{(\mathbb{Z}_2)}_L=e^{i\alpha_1}\sigma_Z\,,\qquad S^{(\mathbb{Z}_2)}_{n_R}=e^{i\alpha_1}\sigma_Z\,,\label{Z22case3}
\ea
where
\be \label{sigma_Z}
 \sigma_Z\equiv 
\begin{pmatrix}
		1 & \phm 0 \\ 0 & -1
	\end{pmatrix}\,.
	\ee
Without loss of generality, we may simply set the remaining global phase $\alpha_1=0$, as it plays no role in constraining 
the forms of $\Gamma_1$ and $\Gamma_2$.

\subsubsection{Extensions of $\mathbb{Z}_2$ to the two generation quark sector: compatibility with $\Pi_2$}
\label{subsubsec:ZPi_n=2}

If the scalar potential exhibits a $\ZPi$ symmetry, then the scalar potential is also invariant under the interchange of $\Phi_1$ and $\Phi_2$, corresponding to 
the $\Pi_2$ symmetry matrix $S^{(\Pi_2)}=\sigma_\Pi$, where 
\be \label{sigmaPimatrix}
\sigma_\Pi\equiv\begin{pmatrix} 0 & \phm 1 \\ 1 & \phm 0\end{pmatrix}.
\ee
When extending the $\mathbb{Z}_2$ symmetry to $\ZPi$ in the Yukawa sector, one must satisfy the additional constraint given by
\eq{eq:Pi2_On_Gamma}, which we repeat here in a more explicit notation,
\be \label{Pi2Gamma12}
\Gamma_1 = S^{(\Pi_2)}_L \Gamma_2 \bigl[S_{nR}^{(\Pi_2)}\bigr]^\dagger\, ,
\qquad\qquad \Gamma_2 = S^{(\Pi_2)}_L \Gamma_1 \bigl[S_{nR}^{(\Pi_2)}\bigr]^\dagger\, .
\ee
These constraints immediately eliminate Case 0 [\eqs{eq:CasesZ2_2fam0}{Z22case0}], since the imposition of \eq{Pi2Gamma12} would yield $\Gamma_1=\Gamma_2=0$,
corresponding to vanishing couplings of the down-type quarks to the Higgs doublet fields.

To determine all possible viable models with the $\ZPi$ symmetry extended to the Yukawa sector, one must 
determine all possible forms for the symmetry matrices $S^{(\Pi_2)}_L$ and $S_{nR}^{(\Pi_2)}$.  We proceed as follows.
We first consider the simpler problem of extending the $\Pi_2$ to the Yukawa sector.   But a $\Pi_2$-symmetric scalar potential is equivalent to a $\mathbb{Z}_2$-symmetric scalar potential in another scalar field basis as shown in \eqst{eq:Pi2}{UZUPi}.  Thus, as a first step we perform a scalar field basis transformation so that the $\Pi_2$ symmetry matrix is $S^{(\Pi_2)}=\sigma_Z$.
We can now extend the $\Pi_2$ symmetry to the Yukawa sector by using the results of Section~\ref{Extensions22}.   We will then end up with three potential cases for the choice of $S_L$ and $S_{n_R}$ given by \eqst{Z22case1}{Z22case3}, where we shall again set the global phase to zero with no loss of generality.
We can now transform back to the basis in which the $\Pi_2$ symmetry matrix is $S^{(\Pi_2)}=\sigma_\Pi$.
That is, we choose $U$ given by \eq{ULuis} and
\be \label{SymMatPi2}
S_L^{(\Pi_2)}=U_L S_L U^\dagger_L\,,\qquad\qquad  S_{n_R}^{(\Pi_2)}=U_{n_R} S_L U^\dagger_{n_R}\,,
\ee
where $S_L$ and $S_{n_R}$ are taken to be the symmetry matrices corresponding to one of the following three cases:
\be \label{Schoices}
(S_L\,,\,S_{n_R}) \in\Bigl\{\bigl( \sigma_Z\,,\,  \mathbbm{1}\bigr)\,,\,\,\,\bigl( \mathbbm{1}\,,\,  \sigma_Z\bigr)\,,\,\,\, \bigl( \sigma_Z\,,\,   \sigma_Z\bigr)\Bigr\}.
\ee

Having obtained the symmetry matrices for the $\Pi_2$ symmetry extended to the Yukawa sector, we can now consider the constraints on $\Gamma_1$ and $\Gamma_2$ obtained in \eqst{eq:CasesZ2_2fam1}{eq:CasesZ2_2fam3} under the $\ZPi$ symmetry extended to the Yukawa sector.  We simply insert the 
results of \eq{SymMatPi2} into \eq{Pi2Gamma12} to obtain
\ba
\Gamma_1(U_{n_R} S_{nR} U^\dagger_{n_R})& =& (U_L S_L U_L^\dagger)\Gamma_2 \, ,\label{Pi2Gamma12again1}
\\[4pt]  \Gamma_2 (U_{n_R} S_{nR}U^\dagger_{n_R})&=&  (U_L S_L U_L^\dagger)\Gamma_1 \,,\label{Pi2Gamma12again2}
\ea
where the three possible cases for $\{\Gamma_1\,,\,\Gamma_2\}$ are given in \eqst{eq:CasesZ2_2fam1}{eq:CasesZ2_2fam3}. 
Note that for any of the three sets of choices of $S_L$ and $S_{n_R}$ given in \eq{Schoices}, the two equations above are equivalent.

As a result of this analysis, we can consider nine possible models, corresponding to the three possible choices for $\Gamma_1$ and $\Gamma_2$ exhibited in \eqst{eq:CasesZ2_2fam1}{eq:CasesZ2_2fam3} and the three possible sets of $S_L$ and $S_{n_R}$ listed in \eq{Schoices}, which we shall henceforth denote by sets 1, 2, and 3.   The resulting model corresponding to Case $n$ for the choice of $\Gamma_1$ and $\Gamma_2$ and the $m$th set of possible choices for $S_L$ and $S_{n_R}$ will be denoted in the following by model $(n$-$m)$.   Thus, there are nine possible models to consider.

At this stage, we have yet to fix the unitary matrices $U_L$ and $U_{n_R}$.  
We shall employ the parametrization
\begin{equation} \label{unitary}
	U_\sigma = e^{i\phi_\sigma} \begin{pmatrix}
		\phm e^{i\alpha_\sigma} \cos\theta_\sigma & \phm e^{-i\beta_\sigma} \sin\theta_\sigma \\
		- e^{i\beta_\sigma} \sin\theta_\sigma &\phm  e^{-i\alpha_\sigma} \cos\theta_\sigma
	\end{pmatrix}\,,\quad \text{with $\sigma\in\{L,n_R\}$}.
\end{equation}
The global phase $\phi_\sigma$ has no effect on the transformation
of the symmetry matrices, so we may set $\phi_\sigma=0$ without loss of generality.  In this convention, $U_\sigma\in{\rm SU}(2)$, and the entire SU(2) group manifold can
be covered by taking the ranges of the remaining parameters to be $0\leq\theta_\sigma\leq\half\pi$ and $0\leq \alpha_\sigma$, $\beta_\sigma<2\pi$.

As an example, consider model (1-3), where $\Gamma_1$ and $\Gamma_2$, which take the form given by \eq{eq:CasesZ2_2fam1},
can be parametrized as
\be
\Gamma_1 = \YukZerosTwo{x_{11}}{x_{12}}{0}{0} \, , \qquad\quad
\Gamma_2 = \YukZerosTwo{0}{0}{x_{21}}{x_{22}} \,,
\ee
and $S_L=S_{n_R}= \sigma_Z$, as specified in the third set of \eq{Schoices}. Plugging these choices along with \eq{unitary} into \eq{Pi2Gamma12again1} yields
\ba
\cos 2\theta_L&=& 0\,, \label{c2th}\\[4pt] 
x_{21}e^{i(\alpha_L-\beta_L)} \sin 2\theta_L&=& x_{12}e^{-i(\alpha_R-\beta_R)}\sin 2\theta_R - x_{11}\cos 2\theta_R\,, \label{first}\\[4pt]
x_{22}e^{i(\alpha_L-\beta_L)}\sin 2\theta_L &=& x_{11} e^{i(\alpha_R-\beta_R)}\sin 2\theta_R+x_{12}\cos 2\theta_R\,,\label{second}
\ea
where we have simplified the $R$ subscript in writing $\alpha_R\equiv \alpha_{n_R}$, $\beta_R\equiv\beta_{n_R}$, and $\theta_R\equiv\theta_{n_R}$.  Since $0\leq\theta_L\leq\half\pi$, it follows that  $\theta_L=\pi/4$.  Then, \eq{SymMatPi2} yields
\ba
S_L^{(\Pi_2)}&=&U_L\sigma_Z U_L^\dagger =\begin{pmatrix} 0 &  -e^{i(\alpha_L-\beta_L)} \\  -e^{-i(\alpha_L-\beta_L)}  &  0\end{pmatrix}, \\[4pt]
S_{n_R}^{(\Pi_2)}&=&U_{n_R}\sigma_Z U_{n_R}^\dagger =\begin{pmatrix}  \cos 2\theta_R & -e^{i(\alpha_R-\beta_R)}\sin 2\theta_R \\ -e^{-i(\alpha_R-\beta_R)}\sin 2\theta_R & -\cos 2\theta_R \end{pmatrix}.
\ea
The $\ZPi$ symmetry constraints do not fix the remaining free parameter, $\alpha_L$, $\beta_L$, $\alpha_R$, $\beta_R$, and $\theta_R$.  Indeed, one is free to transform to a different 
quark field basis as long as the $\mathbb{Z}_2$ symmetry matrices $S_L^{(\mathbb{Z}_2)}=\sigma_Z$ and $S_{n_R}^{(\mathbb{Z}_2)}=\id$ are unchanged.  
In light of \eqs{Sprime1}{Sprime2}, we shall transform
\be \label{SLSRtrans}
S_L^{(\Pi_2)}\to U_L^\prime S_L^{(\Pi_2)} U_L^{\prime\dagger}\,,\qquad\quad 
S_{n_R}^{(\Pi_2)}\to U^\prime_{n_R} S_{n_R}^{(\Pi_2)} U_{n_R}^{\prime\dagger}\,,
\ee
where $U_L^\prime={\diag}\bigl(e^{i(\gamma+\delta)}\,,\,e^{i(\gamma-\delta)}\bigr)$ is the most general $2\times 2$ unitary matrix that leaves  $S_L^{(\mathbb{Z}_2)}$ unchanged and
$U_{n_R}^\prime$ is an arbitrary $2\times 2$ unitary matrix that (trivially) leaves $S_{n_R}^{(\mathbb{Z}_2)}$ unchanged.  With this freedom, it is convenient to 
choose $\gamma$, $\delta$, and the matrix elements of $U^\prime_{n_R}$ such that \eq{SLSRtrans} yields:
\be \label{parmchoice}
\alpha_L-\beta_L=\alpha_R-\beta_R=\pi\,,\qquad \theta_R=\tfrac14\pi\,.
\ee
Using these results and \eq{c2th} to simplify \eqs{first}{second}, we end up with $x_{21}=x_{12}$ and $x_{22}=x_{11}$.   That is, model (1-3) is equivalent to a model in which the Yukawa coupling matrices and the $\Pi_2$ symmetry matrices are given by:
\be
	\Gamma_1= \begin{pmatrix} x_{11} & \phm x_{12} \\ 0 &\phm 0 \end{pmatrix} \,,\qquad 
	\Gamma_2= \begin{pmatrix} 0 &\phm  0 \\ x_{12} &\phm  x_{11} \end{pmatrix} \,,\qquad
	S_L^{(\Pi_2)} =  \sigma_\Pi  \,,\qquad
	S_{nR}^{(\Pi_2)} =  \sigma_\Pi\,,
\ee
where 
$\sigma_\Pi$ is defined in \eq{sigmaPimatrix}.

 Moreover, it is straightforward to compute $H_d=(v_1\Gamma_1+v_2\Gamma_2) (v_1\Gamma_1+v_2\Gamma_2)^\dagger$ and its trace and determinant, which yield
 \ba
  \Tr H_d &=&  \bigl[|x_{11}|^2+|x_{12}|^2\bigr]v^2\,, \label{trace1}\\[4pt]
\det H_d&=&|v_1|^2|v_2|^2|x^2_{11}-x^2_{12}|^2\,,\label{det1}
\ea
where $v$ is defined in \eq{vdef}.
Note that $\det H_d$ is nonzero for a generic choice of parameters, which implies that the two down-type quark masses are generically nonzero.

In Appendix~\ref{app:details}, we analyze the remaining eight models.  Some of these models can be immediately excluded as they contain either a massless down-type quark or else vanishing down-type Yukawa coupling matrices.  
Furthermore, we find that model (3,3) actually represents a class of models that are parametrized by the angles $(\theta_L,\theta_R)$.  Two of these models, corresponding to
$(\theta_L,\theta_R)=(0,\tfrac14\pi)$ and $(\tfrac14\pi,0)$, which shall denote by (3-3)$_0$ and (3-3)$_1$, respectively,  are phenomenologically viable.   In the remaining models, collectively denoted by  (3-3)$_X$, all down-type quarks are massive.  However, when the up-type quark Yukawa couplings are taken into account, the resulting models predict a vanishing Cabibbo angle (as shown in Appendix~\ref{app:details}) and hence are phenomenologically excluded.

\subsubsection*{Summary of viable two-generation $\ZPi$-symmetric models}

In addition to model $(1,3)$ analyzed above, we show in Appendix~\ref{app:details} that in models 
(2-3), (3-1), (3-2),
(3-3)$_0$,  (3-3)$_1$, and (3-3)$_X$ all down-type quarks are massive.  As noted above, the class of models denoted by (3-3)$_X$ is phenomenologically excluded.
The symmetry matrices and the corresponding Yukawa coupling matrices of the remaining models are listed in Table~\ref{tab:symmetries_Z2PI2_2fam} below.
\begin{table}[H]
\centering
\begin{tabular}{|c|c|c|c|c|c|c|} \hline
	model		& $S_L^{(\mathbb{Z}_2)}$ & $S_{nR}^{(\mathbb{Z}_2)}$ & $S_L^{(\Pi_2)}$ & $S_{nR}^{(\Pi_2)}$
& $\Gamma_1$ & $\Gamma_2$ \TBstrut \\
		\hline
(1-3) & $\sigma_Z$ & $\mathbbm{1}$ & $\sigma_\Pi$ & $\sigma_\Pi$
& $\begin{pmatrix} x_{11}&x_{12}\\0&0 \end{pmatrix}$
& $\begin{pmatrix} 0&0\\x_{12}&x_{11}\ \end{pmatrix}$ \Ttstrut\Bstrut \\ 
			\hline	
(2-3) &  $\mathbbm{1}$ & $\sigma_Z$ & $\sigma_\Pi$ & $\sigma_\Pi$
& $\begin{pmatrix} x_{11}&\phm 0\\ x_{21}&\phm 0  \end{pmatrix}$
& $\begin{pmatrix} 0&\phm x_{21}\\ 0&\phm x_{11}\ \end{pmatrix}$ \TBstrut \\ 
			\hline	
(3-1) & $\sigma_Z$ & $\sigma_Z$ & $\sigma_\Pi$ & $\mathbbm{1}$
& $\begin{pmatrix} x_{11} & 0 \\ 0 & x_{22} \end{pmatrix}$
& $\begin{pmatrix} 0 & x_{22} \\ x_{11} & 0 \end{pmatrix}$ \TBstrut \\  
			\hline
(3-2) & $\sigma_Z$ & $\sigma_Z$ & $\mathbbm{1}$ & $\sigma_\Pi$
& $\begin{pmatrix} x_{11} & 0 \\ 0 & x_{22} \end{pmatrix}$
& $\begin{pmatrix} 0 & x_{11} \\ x_{22} & 0 \end{pmatrix}$ \TBstrut \\  
			\hline
(3-3)$_0$ & $\sigma_Z$ & $\sigma_Z$ & $\sigma_Z$ & $\sigma_\Pi$
& $\begin{pmatrix} x_{11} & 0 \\ 0 & x_{22} \end{pmatrix}$
& $\begin{pmatrix} 0 & x_{11} \\ -x_{22} & 0 \end{pmatrix}$  \TBstrut\\  
			\hline
(3-3)$_1$  & $\sigma_Z$ & $\sigma_Z$ & $\sigma_\Pi$ & $\sigma_Z$
& $\begin{pmatrix} x_{11} & 0 \\ 0 & x_{22} \end{pmatrix}$
& $\begin{pmatrix} 0 & -x_{22} \\ x_{11} & 0 \end{pmatrix}$ \TBstrut \\
\hline
\end{tabular}
\caption{Symmetry matrices for viable two-generation $\mathbb{Z}_2\otimes\Pi_2$-symmetric models 
in a scalar field basis where $S^{(\mathbb{Z}_2)}=\sigma_Z$ and  $S^{(\Pi_2)}=\sigma_\Pi$, 
with the corresponding forms for the $\mathbb{Z}_2$ and $\Pi_2$ symmetry matrices of the down-type Yukawa sector and the corresponding Yukawa coupling matrices. \\[-15pt]}
\label{tab:symmetries_Z2PI2_2fam}
\end{table}

\begin{table}[t]
	\centering
	\begin{tabular}{|c|c|c|c|c|c|c|}
	\hline
		 down/up sector models & $S_L^{(\mathbb{Z}_2)}$ & $S_{nR}^{(\mathbb{Z}_2)}$ & $S_{pR}^{(\mathbb{Z}_2)}$ & $S_L^{(\Pi_2)}$ & $S_{nR}^{(\Pi_2)}$ & $S_{pR}^{(\Pi_2)}$ \\
		\hline\hline
		(1-3) / (1-3) & $\sigma_Z$ & $\mathbbm{1}$ & $\mathbbm{1}$ & $\sigma_\Pi$ & $\sigma_\Pi$ & $\sigma_\Pi$  \\ 
		\hline
		(1-3) / (3-1) & $\sigma_Z$ & $\mathbbm{1}$ & $\sigma_Z$ & $\sigma_\Pi$ & $\sigma_\Pi$ & $\mathbbm{1}$  \\ 
		\hline
		(1-3) / (3-3)$_1$ & $\sigma_Z$ & $\mathbbm{1}$ & $\sigma_Z$ & $\sigma_\Pi$ & $\sigma_\Pi$ & $\sigma_Z$  \\ 
		\hline
		(2-3) / (2-3) & $\mathbbm{1}$ & $\sigma_Z$ & $\sigma_Z$ & $\sigma_\Pi$ & $\sigma_\Pi$ & $\sigma_\Pi$  \\ 	
		\hline
		(3-3)$_0$/ (3-3)$_0$ & $\sigma_Z$ & $\sigma_Z$ & $\sigma_Z$ & $\sigma_Z$ & $\sigma_\Pi$ & $\sigma_\Pi$  \\ 
		 \hline
		(3-3)$_1$/ (1,3) & $\sigma_Z$ & $\sigma_Z$ & $\id$ & $\sigma_\Pi$ & $\sigma_Z$ & $\sigma_\Pi$  \\ 	
		\hline
		(3-3)$_1$/ (3,3)$_1$ & $\sigma_Z$ & $\sigma_Z$ & $\sigma_Z$ & $\sigma_\Pi$ & $\sigma_Z$ & $\sigma_Z$  \\ 
		\hline\hline	 \rowcolor[gray]{0.95}	
		(3-1) / (1-3) & $\sigma_Z$ & $\sigma_Z$ & $\id$ & $\sigma_\Pi$ & $\mathbbm{1}$ & $\sigma_\Pi$  \\ 
		\hline \rowcolor[gray]{0.95}
		(3-1) / (3-1) & $\sigma_Z$ & $\sigma_Z$ & $\sigma_Z$ & $\sigma_\Pi$ & $\mathbbm{1}$ & $\mathbbm{1}$  \\ 
		\hline	 \rowcolor[gray]{0.95}
		(3-1) / (3-3)$_1$ & $\sigma_Z$ & $\sigma_Z$ & $\sigma_Z$ & $\sigma_\Pi$ & $\mathbbm{1}$ & $\sigma_Z$  \\ 
		\hline  \rowcolor[gray]{0.95}
		(3-2) / (3-2) & $\sigma_Z$ & $\sigma_Z$ & $\sigma_Z$ & $\mathbbm{1}$ & $\sigma_\Pi$ & $\sigma_\Pi$  \\
		\hline  \rowcolor[gray]{0.95}
		(3-3)$_1$/ (3,1) & $\sigma_Z$ & $\sigma_Z$ & $\sigma_Z$ & $\sigma_\Pi$ & $\sigma_Z$ & $\id$  \\ 
		\hline
	\end{tabular}
	\caption{$\mathbb{Z}_2$ and $\Pi_2$ symmetry matrices for each viable Yukawa sector model that is compatible with the $\mathbb{Z}_2\otimes\Pi_2$ symmetry of the 2HDM scalar potential in a scalar field basis where $S^{(\mathbb{Z}_2)}\!=\sigma_Z$ and  $S^{(\Pi_2)}\!=\sigma_\Pi$.
Of the 12 models listed above, the first seven models are inequivalent with respect to basis changes.  Each of the last five models (shaded in gray) can be shown to be equivalent
to one of the first seven models listed above via a particular change in the scalar field and quark field basis, as shown in Appendix~\ref{app:equiv}.  The corresponding equivalent models
are given in \eqss{equiv1}{equiv2}{equiv3}.}
	\label{tab:completemodels}
\end{table}

To determine the viability of the possible two-generation models, one must now consider the corresponding results for the up-type Yukawa coupling matrices $\Delta_1$ and $\Delta_2$.  In the analysis of  the possible forms for $\Delta_1$ and $\Delta_2$ consistent with the $\ZPi$ symmetry, one must use the same $S_L^{(\mathbb{Z}_2)}$ and  $S_L^{(\Pi_2)}$ employed in the analysis
of the down-type Yukawa sector.   One is still free to fix $S_{p_R}^{(\mathbb{Z}_2)}$ and  $S_{p_R}^{(\Pi_2)}$ consistent with the symmetry requirements.  The end result is a table identical with
Table~\ref{tab:symmetries_Z2PI2_2fam}, with the possible forms for $\Delta_1$ and $\Delta_2$ coinciding with those of $\Gamma_1$ and $\Gamma_2$ but with different nonzero matrix elements (which we shall denote by $y_{ij})$.   Consequently, none of the allowed choices for $\Delta_1$ and $\Delta_2$ yield massless up-type quarks for a generic choice of the parameters.
The resulting Yukawa sector models are specified by a pair of model types that share the same $S_L^{(\mathbb{Z}_2)}$ and  $S_L^{(\Pi_2)}$, which are listed above in Table~\ref{tab:completemodels}.  It is now straightforward to check that for all models listed in Table~\ref{tab:completemodels}, $\det\bigl\{[H_u,H_d]\bigr\}\neq 0$, or equivalently $J_c\neq 0$ [cf.~\eq{JC}].  Hence, all 
the models of Table~\ref{tab:completemodels} possess a nonzero Cabibbo angle, as required by experimental data. 

The models exhibited in Table~\ref{tab:completemodels}
are not all inequivalent, as some of the listed models are related by a change in the Higgs field and the quark field basis.  In particular, we show in Appendix~\ref{app:equiv} that there are five sets of model pairs, specified in \eqss{equiv1}{equiv2}{equiv3}, where the two models that make up a given pair are related by an appropriate set of basis transformations. 
We may take the first seven models listed in Table~\ref{tab:completemodels}
to constitute the list of inequivalent models.   Each of the remaining five models (which are shaded in gray) is shown in Appendix~\ref{app:equiv} to be equivalent to 
one of the (unshaded) seven inequivalent models in Table~\ref{tab:completemodels}.

\subsection{(Non-)Correspondence between $\mathbb{Z}_2\otimes \Pi_2$ and GCP2 with two quark generations}
\label{compare}

The question now arises: are the Yukawa-extended two-generation GCP2-symmetric models of Section~\ref{sec:GCPfortwo}
equivalent (i.e., the same model but expressed in different choices of the
Higgs field and quark field basis) to the corresponding $\mathbb{Z}_2\otimes \Pi_2$-symmetric
models of Section~\ref{twogenZPi}?  In Section~\ref{sec:GCPfortwo}, we classified the possible Yukawa-extended GCP2-symmetric models.
We found three distinct classes of models, denoted by I/I, II/II, and III/III, all of which exhibited nonzero up-type and down-type quark masses for
generic choices of the parameters.   In model class I/I, the corresponding down-type 
and up-type Yukawa coupling matrices 
exhibited forms that each depended on two independent complex parameters.   In light of these forms, we demonstrated that
the Cabibbo angle vanished.  In model classes II/II and III/III,  the corresponding down-type 
and up-type Yukawa coupling matrices 
exhibited forms that each depended on four independent complex parameters, which implied a nonvanishing Cabibbo angle.

In Section~\ref{twogenZPi}, we classified the possible Yukawa-extended $\mathbb{Z}_2\otimes \Pi_2$-symmetric
models.  In all cases but one, the corresponding down-type 
and up-type Yukawa coupling matrices 
exhibited forms that each depended on two independent complex parameters.  
Some of these models possessed at least one massless quark.  Among the class of models with nonzero up-type and down-type quark masses,
we showed the existence of seven inequivalent models, each of which allowed for a nonvanishing Cabibbo angle.
The one exceptional case (with Yukawa coupling matrices that each depend on only one independent parameter) was 
shown to have a vanishing Cabibbo angle in Appendix~\ref{app:details}.

Naively, one might have expected that the GCP2-symmetric models with down-type 
and up-type Yukawa coupling matrices that each depended on two independent complex parameters could be related via a basis transformation to
the corresponding $\mathbb{Z}_2\otimes \Pi_2$-symmetric
models with the same number of independent parameters.  However, the Cabibbo angle necessarily vanishes in the former, whereas
it is generically nonzero in the latter.  Thus, we conclude that in contrast to the behavior of these symmetries in the scalar sector of the
2HDM, GCP2 and $\mathbb{Z}_2\otimes \Pi_2$ are inequivalent symmetries when extended to the Yukawa sector.

Is it possible that the Yukawa-extended GCP2-symmetric models (models II/II and III/III),
 with down-type 
and up-type Yukawa coupling matrices that each depend on four independent complex parameters, could be related via a basis transformation to
the corresponding $\mathbb{Z}_2\otimes \Pi_2$-symmetric
models, which depend on half the number of independent complex parameters?  The answer is clearly negative.
Although the Cabibbo angle is nonvanishing in all these models, there are other physical observables that would distinguish
the Yukawa-extended GCP2-symmetric models from the corresponding $\mathbb{Z}_2\otimes \Pi_2$-symmetric models.
Hence, we have demonstrated that the extension of the GCP2 and the $\mathbb{Z}_2\otimes \Pi_2$ symmetries from the scalar sector
(where the corresponding scalar potentials are related by a change in the scalar field basis) to the Yukawa sector effectively 
``removes the degeneracy'' and yields inequivalent models.

\section{Conclusions}
\label{sec:concl}

In this paper we have explored 
 the curious connection
between the 2HDM scalar potential obtained by imposing invariance under the 
Higgs-flavor symmetry, $\ZPi$ and the scalar potential obtained by imposing invariance under 
the generalized CP symmetry, GCP2.  As first noticed in Ref.~\cite{Davidson:2005cw},
the resulting scalar potentials after imposing $\ZPi$ and GCP2 were related by a transformation
of the scalar field basis, and thus could be considered as physically equivalent.  
The objective of this paper was to extend these symmetries to the Yukawa sector, 
to see whether the extended $\ZPi$-symmetric 2HDM was still equivalent to the
extended GCP2-symmetric 2HDM, or whether the extension to the Yukawa sector
removes the degeneracy between the two models.
In Ref.~\cite{Ferreira:2010bm}, Ferreira and Silva proved that one could not extend 
the GCP2 symmetry to the Yukawa sector in a way that was consistent with 
nonzero quark masses and a CKM mixing angle that were consistent
with experimental observations.
The more difficult case of extending the $\ZPi$ symmetry to the three-generation Yukawa sector
is addressed in this paper for the first time.   We find that, similar to the results obtained
for the GCP2 symmetry in Ref.~\cite{Ferreira:2010bm}, there is no extension of the $\ZPi$ symmetry
that is consistent with experimental observations.

In analogy with the connection between $\ZPi$ and GCP2, there is also a similar relation between
the 2HDM scalar potential obtained by imposing invariance under the 
Higgs-flavor symmetry, $\UPi$ and the scalar potential obtained by imposing invariance under 
the generalized CP symmetry, GCP3.
It was also observed in Ref.~\cite{Ferreira:2009wh} that the corresponding scalar potentials 
were related by a transformation of the scalar field basis.
In Ref.~\cite{Ferreira:2010bm}, it was shown that there
is a unique extension of the GCP3 symmetry to the Yukawa sector that yields a model
with nonzero quark masses and a nonvanishing CKM angle.
This provided the possibility of a realistic fully GCP3-symmetric 2HDM, although further 
analysis presented in Ref.~\cite{Ferreira:2010bm} showed that the model was unable
to yield a CKM mixing matrix that was fully compatible with experimental data.

In this paper, we examined for the first time all possible extensions of the $\UPi$ symmetry to the Yukawa sector.
We found that there is again a unique model with nonzero quark masses and a nonvanishing CKM angle.
Moreover, we showed that the corresponding three-generation Yukawa-extended GCP3-symmetric and $\UPi$-symmetric
2HDM are related by a simultaneous transformation of the Higgs field basis and the quark field basis. 

It was tempting to conclude that the results described above imply that the physical equivalence
of the models obtained by imposing a  $\ZPi$ ($\UPi$) and GCP2 (GCP3) symmetry
was a general feature of the 2HDM.
In this paper, we have also proved that this conclusion is not generally correct by focusing on 
a 2HDM toy model with two quark generations.
We have classified all such two-generation GCP2-symmetric and $\ZPi$-symmetric models, 
where the corresponding symmetries have been extended to the Yukawa sector.
We have found inequivalent GCP2-symmetric and $\ZPi$-symmetric models
that possess nonzero quark masses and a nonzero Cabibbo angle.
For example, the corresponding down-type Yukawa coupling matrices of the
GCP2-symmetric models generically depend on four complex parameters, whereas
those of the $\ZPi$-symmetric models generically depend on two complex parameters.  Indeed, there
are no scalar field and quark field basis transformations that can relate the 
phenomenologically viable GCP2-symmetric and $\ZPi$-symmetric models.  That is,
the degeneracy between these two symmetry classes has been removed.  

One can perform a similar classification of two-generation GCP3-symmetric and $\UPi$-symmetric models, 
where the corresponding symmetries have been extended to the Yukawa sector.
We again find inequivalent GCP3-symmetric and $\UPi$-symmetric models
that possess nonzero quark masses and a nonzero Cabibbo angle.
Details of this analysis, which mirrors the calculations presented in Section~\ref{sec:Z2PivsGCP2_n=2},
can be found in Ref.~\cite{CarroloThesis}.  Once again, the degeneracy between these two symmetry classes has been removed.  
We conclude that the physical equivalence
of the models obtained by imposing a $\ZPi$ ($\UPi$) and GCP2 (GCP3)
symmetry is an accidental feature of the 2HDM with three quark generations.

\section*{Acknowledgments}
H.E.H. is grateful for the hospitality and support during his visits to the Instituto Superior
T\'{e}cnico, Universidade de Lisboa, where their work was initiated.  This work is supported in part
by the Portuguese Funda\c{c}\~{a}o
para a Ci\^{e}ncia e Tecnologia\/ (FCT) under Contracts
2024.01362.CERN (\texttt{https://doi.org/10.54499/2024.01362.CERN}), UIDB/00777/2020, and UIDP/00777/2020\,;
these projects are partially funded through POCTI (FEDER),
COMPETE, QREN, and the EU.
The work of H.E.H. is supported in part by the U.S. Department of Energy Grant No.~\uppercase{DE-SC}0010107.
The work of L.L. was funded during the final stage of this project by the Deutsche Forschungsgemeinschaft (DFG, German Research Foundation) within the DFG Research Training Group 
RTG2994---project number 517344132.
The work of S.C. was partially funded by the European Union (ERC, UNIVERSE PLUS, 101118787). Views and opinions expressed are however those of the author(s) only and do not necessarily reflect those of the European Union or the European Research Council Executive Agency. Neither the European Union nor the granting authority can be held responsible for them.


\appendix

\section{Extensions of U(1) and $\mathbb{Z}_2$ symmetries of the 2HDM scalar potential to the Yukawa sector}
\label{appEuler:derivation}

In this appendix, we will list all the extensions of  $\mathbb{Z}_2$ and U(1)
symmetries of the 2HDM scalar potential to the Yukawa sector that were obtained in Ref.~\cite{Ferreira:2010ir}.
We shall employ the notation of Ref.~\cite{Ferreira:2010ir}, where the $\text{x}$ represents the freedom to choose any complex
number for that matrix element.

\subsection{Extensions of $\mathbb{Z}_2$ to the Yukawa sector}

In the list of independent forms obtained in Ref.~\cite{Ferreira:2010ir} for the down-type Yukawa coupling matrices $\Gamma_1$ and $\Gamma_2$ that are compatible with the $\mathbb{Z}_2$ symmetry of the 2HDM scalar potential when extended to the Yukawa sector, extensions that were 
equivalent with respect to the permutations of quark flavors were excluded from the list.  In contrast, extensions that 
are related by a permutation of the scalar doublets were not removed.  In Table~\ref{tab:Z2Models}, we exhibit the list as 
presented in Ref.~\cite{Ferreira:2010ir}

\begin{table}[H]
\centering
	\begin{tabular}{|c|c|c|c|}
		\hline
		Eqs. & $\Gamma_1$ & $\Gamma_2$ \\
	 	\hline
	 	66 & $\YukZeros{\text{x}}{\text{x}}{\text{x}}{\text{x}}{\text{x}}{\text{x}}{\text{x}}{\text{x}}{\text{x}}$
& $\YukZeros{0}{0}{0}{0}{0}{0}{0}{0}{0}$\\ 
	 	\hline	 
	 	67 & $\YukZeros{\text{x}}{\text{x}}{0}{\text{x}}{\text{x}}{0}{\text{x}}{\text{x}}{0}$
& $\YukZeros{0}{0}{\text{x}}{0}{0}{\text{x}}{0}{0}{\text{x}}$\\ 
	 	\hline
	 	68 & $\YukZeros{\text{x}}{0}{0}{\text{x}}{0}{0}{\text{x}}{0}{0}$
& $\YukZeros{0}{\text{x}}{\text{x}}{0}{\text{x}}{\text{x}}{0}{\text{x}}{\text{x}}$\\ 
	 	\hline
	 	69 & $\YukZeros{0}{0}{0}{0}{0}{0}{0}{0}{0}$
& $\YukZeros{\text{x}}{\text{x}}{\text{x}}{\text{x}}{\text{x}}{\text{x}}{\text{x}}{\text{x}}{\text{x}}$\\ 
	 	\hline	 
	 \end{tabular}
	 \quad \quad
	 \begin{tabular}{|c|c|c|}
	 \hline
		Eqs. & $\Gamma_1$ & $\Gamma_2$\\
	 	\hline
		71 & $\YukZeros{\text{x}}{\text{x}}{\text{x}}{\text{x}}{\text{x}}{\text{x}}{0}{0}{0}$
& $\YukZeros{0}{0}{0}{0}{0}{0}{\text{x}}{\text{x}}{\text{x}}$\\ 
	 	\hline
	 	73 & $\YukZeros{\text{x}}{\text{x}}{0}{\text{x}}{\text{x}}{0}{0}{0}{\text{x}}$
& $\YukZeros{0}{0}{\text{x}}{0}{0}{\text{x}}{\text{x}}{\text{x}}{0}$\\ 
	 	\hline
	 	75 & $\YukZeros{\text{x}}{0}{0}{\text{x}}{0}{0}{0}{\text{x}}{\text{x}}$
& $\YukZeros{0}{\text{x}}{\text{x}}{0}{\text{x}}{\text{x}}{\text{x}}{0}{0}$\\
	 	\hline
	 	79 & $\YukZeros{0}{0}{0}{0}{0}{0}{\text{x}}{\text{x}}{\text{x}}$
& $\YukZeros{\text{x}}{\text{x}}{\text{x}}{\text{x}}{\text{x}}{\text{x}}{0}{0}{0}$\\ 
	 	\hline
	 \end{tabular}
	\caption{Independent forms for the down-type Yukawa coupling matrices $\Gamma_1$ and $\Gamma_2$ that are compatible with the $\mathbb{Z}_2$ symmetry of the 2HDM scalar potential when extended to the Yukawa sector, labeled by the corresponding equation numbers of Ref.~\cite{Ferreira:2010ir}.}
	\label{tab:Z2Models}
\end{table}

\subsection{Extensions of U(1) to the Yukawa sector}

A list of the independent forms obtained in Ref.~\cite{Ferreira:2010ir} for the down-type Yukawa coupling matrices $\Gamma_1$ and $\Gamma_2$ that are compatible with a global U(1)  symmetry of the 2HDM scalar potential when extended to the Yukawa sector is given in Tables~\ref{tab:1U1},
\ref{tab:2U1}, \ref{tab:3U1}, and \ref{tab:4U1} below.

\begin{table}[H]
\centering
	\begin{tabular}{|c|c|c|}
		\hline
		Eq. & $\Gamma_1$ & $\Gamma_2$\\
	 	\hline
	 	57 & $\YukZeros{0}{0}{0}{0}{0}{\text{x}}{\text{x}}{\text{x}}{0}$
& $\YukZeros{\text{x}}{\text{x}}{0}{0}{0}{0}{0}{0}{\text{x}}$\\ 
	 	\hline	 
	 	58 & $\YukZeros{0}{0}{0}{0}{0}{0}{\text{x}}{\text{x}}{0}$
& $\YukZeros{\text{x}}{\text{x}}{0}{0}{0}{\text{x}}{0}{0}{0}$\\ 
	 	\hline
	 	59 & $\YukZeros{0}{0}{\text{x}}{\text{x}}{\text{x}}{0}{0}{0}{0}$
& $\YukZeros{0}{0}{0}{0}{0}{0}{\text{x}}{\text{x}}{0}$\\ 
	 	\hline
	 	60 & $\YukZeros{0}{0}{0}{\text{x}}{\text{x}}{0}{0}{0}{\text{x}}$
& $\YukZeros{0}{0}{\text{x}}{0}{0}{0}{\text{x}}{\text{x}}{0}$\\ 
	 	\hline
	\end{tabular}
	\quad \quad
	\begin{tabular}{|c|c|c|}
		\hline
		Eq. & $\Gamma_1$ & $\Gamma_2$\\
	 	\hline
		61 & $\YukZeros{\text{x}}{0}{0}{0}{\text{x}}{0}{0}{0}{\text{x}}$
& $\YukZeros{0}{0}{\text{x}}{0}{0}{0}{0}{\text{x}}{0}$\\ 
	 	\hline
	 	62 & $\YukZeros{\text{x}}{0}{0}{0}{0}{0}{0}{0}{\text{x}}$
& $\YukZeros{0}{0}{\text{x}}{0}{\text{x}}{0}{0}{0}{0}$\\ 
	 	\hline
	 	63 & $\YukZeros{\text{x}}{0}{0}{0}{\text{x}}{0}{0}{0}{0}$
& $\YukZeros{0}{0}{0}{0}{0}{\text{x}}{0}{\text{x}}{0}$\\ 
	 	\hline
	 	64 & $\YukZeros{0}{0}{0}{0}{\text{x}}{0}{\text{x}}{0}{0}$
& $\YukZeros{\text{x}}{0}{0}{0}{0}{\text{x}}{0}{\text{x}}{0}$\\ 
	 	\hline
	\end{tabular}
	\caption{Independent forms for the down-type Yukawa coupling matrices $\Gamma_1$ and $\Gamma_2$ that are compatible with the global U(1) symmetry of the 2HDM scalar potential when extended to the Yukawa sector, labeled by the corresponding equation numbers of Ref.~\cite{Ferreira:2010ir}---Part 1.}
	\label{tab:1U1}
\end{table}
\begin{table}[H]
\centering
	\begin{tabular}{|c|c|c|}
		\hline
		Eq. & $\Gamma_1$ & $\Gamma_2$\\
	 	\hline
	 	66& $\YukZeros{\text{x}}{\text{x}}{\text{x}}{\text{x}}{\text{x}}{\text{x}}{\text{x}}{\text{x}}{\text{x}}$
& $\YukZeros{0}{0}{0}{0}{0}{0}{0}{0}{0}$\\ 
	 	\hline
	 	67& $\YukZeros{\text{x}}{\text{x}}{0}{\text{x}}{\text{x}}{0}{\text{x}}{\text{x}}{0}$
& $\YukZeros{0}{0}{\text{x}}{0}{0}{\text{x}}{0}{0}{\text{x}}$\\ 
	 	\hline
	 	68& $\YukZeros{\text{x}}{0}{0}{\text{x}}{0}{0}{\text{x}}{0}{0}$
& $\YukZeros{0}{\text{x}}{\text{x}}{0}{\text{x}}{\text{x}}{0}{\text{x}}{\text{x}}$\\ 
	 	\hline
	 	69& $\YukZeros{0}{0}{0}{0}{0}{0}{0}{0}{0}$
& $\YukZeros{\text{x}}{\text{x}}{\text{x}}{\text{x}}{\text{x}}{\text{x}}{\text{x}}{\text{x}}{\text{x}}$\\ 
	 	\hline
	\end{tabular}
	\quad \quad 
	\begin{tabular}{|c|c|c|}
		\hline
	 	Eq & $\Gamma_1$ & $\Gamma_2$\\
	 	\hline
	 	71& $\YukZeros{\text{x}}{\text{x}}{\text{x}}{\text{x}}{\text{x}}{\text{x}}{0}{0}{0}$
& $\YukZeros{0}{0}{0}{0}{0}{0}{\text{x}}{\text{x}}{\text{x}}$\\ 
	 	\hline
	 	72 & $\YukZeros{\text{x}}{\text{x}}{0}{\text{x}}{\text{x}}{0}{0}{0}{0}$
& $\YukZeros{0}{0}{\text{x}}{0}{0}{\text{x}}{\text{x}}{\text{x}}{0}$\\ 
	 	\hline
	 	74 & $\YukZeros{\text{x}}{0}{0}{\text{x}}{0}{0}{0}{0}{0}$
& $\YukZeros{0}{\text{x}}{\text{x}}{0}{\text{x}}{\text{x}}{\text{x}}{0}{0}$\\ 
	 	\hline
	 	76 & $\YukZeros{\text{x}}{\text{x}}{0}{\text{x}}{\text{x}}{0}{0}{0}{\text{x}}$
& $\YukZeros{0}{0}{0}{0}{0}{0}{\text{x}}{\text{x}}{0}$\\ 
	 	\hline
	\end{tabular}
	\caption{Independent forms for the down-type Yukawa coupling matrices $\Gamma_1$ and $\Gamma_2$ that are compatible with the global U(1) symmetry of the 2HDM scalar potential when extended to the Yukawa sector, labeled by the corresponding equation numbers of Ref.~\cite{Ferreira:2010ir}---Part 2.}
	\label{tab:2U1}
\end{table}
\begin{table}[H]
\centering
	\begin{tabular}{|c|c|c|}
		\hline
	 	Eq & $\Gamma_1$ & $\Gamma_2$\\
	 	\hline
	 	77 & $\YukZeros{\text{x}}{0}{0}{\text{x}}{0}{0}{0}{0}{\text{x}}$
& $\YukZeros{0}{\text{x}}{0}{0}{\text{x}}{0}{\text{x}}{0}{0}$\\ 
	 	\hline
	 	78 & $\YukZeros{0}{0}{0}{0}{0}{0}{0}{0}{\text{x}}$
& $\YukZeros{\text{x}}{\text{x}}{0}{\text{x}}{\text{x}}{0}{0}{0}{0}$\\ 
	 	\hline
	 	79 & $\YukZeros{0}{0}{0}{0}{0}{0}{\text{x}}{\text{x}}{\text{x}}$
& $\YukZeros{\text{x}}{\text{x}}{\text{x}}{\text{x}}{\text{x}}{\text{x}}{0}{0}{0}$\\
	 	\hline
	 	81 & $\YukZeros{\text{x}}{\text{x}}{0}{\text{x}}{\text{x}}{0}{0}{0}{\text{x}}$
& $\YukZeros{0}{0}{\text{x}}{0}{0}{\text{x}}{0}{0}{0}$\\ 
	 	\hline
	 \end{tabular}
	 \quad \quad 
	 \begin{tabular}{|c|c|c|}
	 	\hline
	 	Eq & $\Gamma_1$ & $\Gamma_2$\\
	 	\hline
	 	82 & $\YukZeros{\text{x}}{0}{0}{\text{x}}{0}{0}{0}{\text{x}}{\text{x}}$
& $\YukZeros{0}{\text{x}}{\text{x}}{0}{\text{x}}{\text{x}}{0}{0}{0}$\\ 
	 	\hline
	 	83 & $\YukZeros{\text{x}}{\text{x}}{0}{\text{x}}{\text{x}}{0}{0}{0}{0}$
& $\YukZeros{0}{0}{0}{0}{0}{0}{0}{0}{\text{x}}$\\ 
	 	\hline
	 	84 & $\YukZeros{\text{x}}{0}{0}{\text{x}}{0}{0}{0}{\text{x}}{0}$
& $\YukZeros{0}{\text{x}}{0}{0}{\text{x}}{0}{0}{0}{\text{x}}$\\ 
	 	\hline
	 	85 & $\YukZeros{0}{0}{0}{0}{0}{0}{\text{x}}{\text{x}}{\text{x}}$
& $\YukZeros{\text{x}}{\text{x}}{\text{x}}{\text{x}}{\text{x}}{\text{x}}{0}{0}{0}$\\ 
	 	\hline
	 \end{tabular}
	\caption{Independent forms for the down-type Yukawa coupling matrices $\Gamma_1$ and $\Gamma_2$ that are compatible with the global U(1) symmetry of the 2HDM scalar potential when extended to the Yukawa sector, labeled by the corresponding equation numbers of Ref.~\cite{Ferreira:2010ir}---Part 3.}
	\label{tab:3U1}
\end{table}

\begin{table}[H]
\centering
	\begin{tabular}{|c|c|c|}
		\hline
	 	Eq & $\Gamma_1$ & $\Gamma_2$\\
		\hline
	 	86 & $\YukZeros{0}{0}{0}{0}{0}{0}{\text{x}}{\text{x}}{0}$
& $\YukZeros{\text{x}}{\text{x}}{0}{\text{x}}{\text{x}}{0}{0}{0}{\text{x}}$\\ 
	 	\hline
	 	89 & $\YukZeros{\text{x}}{\text{x}}{0}{0}{0}{\text{x}}{0}{0}{0}$
& $\YukZeros{0}{0}{0}{0}{0}{\text{x}}{\text{x}}{\text{x}}{0}$\\ 
	 	\hline
	 	90 & $\YukZeros{\text{x}}{\text{x}}{0}{0}{0}{0}{0}{0}{\text{x}}$
& $\YukZeros{0}{0}{0}{0}{0}{\text{x}}{\text{x}}{\text{x}}{0}$\\
	 	\hline
	 	91 & $\YukZeros{\text{x}}{0}{0}{0}{\text{x}}{\text{x}}{0}{0}{0}$
& $\YukZeros{0}{\text{x}}{\text{x}}{0}{0}{0}{\text{x}}{0}{0}$\\ 
	 	\hline
	 \end{tabular}
	 \quad 
	 \begin{tabular}{|c|c|c|}
		\hline
	 	Eq & $\Gamma_1$ & $\Gamma_2$\\
	 	\hline
	 	92 & $\YukZeros{\text{x}}{0}{0}{0}{\text{x}}{0}{0}{0}{\text{x}}$
& $\YukZeros{0}{\text{x}}{0}{0}{0}{\text{x}}{\text{x}}{0}{0}$\\ 
	 	\hline
	 	93 & $\YukZeros{\text{x}}{0}{0}{0}{0}{0}{0}{\text{x}}{\text{x}}$
& $\YukZeros{0}{0}{0}{0}{\text{x}}{\text{x}}{\text{x}}{0}{0}$\\ 
	 	\hline
	 	94 & $\YukZeros{0}{0}{0}{\text{x}}{\text{x}}{0}{0}{0}{\text{x}}$
& $\YukZeros{\text{x}}{\text{x}}{0}{0}{0}{\text{x}}{0}{0}{0}$\\ 
	 	\hline
	 	95 & $\YukZeros{0}{0}{0}{\text{x}}{0}{0}{0}{\text{x}}{\text{x}}$
& $\YukZeros{\text{x}}{0}{0}{0}{\text{x}}{\text{x}}{0}{0}{0}$\\ 
	 	\hline
	 \end{tabular}
	\caption{Independent forms for the down-type Yukawa coupling matrices $\Gamma_1$ and $\Gamma_2$ that are compatible with the global U(1) symmetry of the 2HDM scalar potential when extended to the Yukawa sector, labeled by the corresponding equation numbers of Ref.~\cite{Ferreira:2010ir}---Part 4.}
	\label{tab:4U1}
\end{table}

\section{Evaluation of a determinant}
\label{app:det}

In this appendix, we provide an explicit evaluation of the determinant of the $8\times 8$ matrix given in \eq{Adef}.   We begin by noticing that one can express
 $A$ in $2\times 2$ block matrix form consisting of $4\times 4$ matrix blocks that can be written in terms of Kronecker products of
$2\times 2$ matrices,
\be \label{app:Adef}
A=\begin{pmatrix} \phm c_\alpha\id & \,\,\,s_\alpha\id &\ \phm X_{n_R}^{\T} & \,\,\,\boldsymbol{0} \\
-s_\alpha\id & \,\,\, c_\alpha\id & \boldsymbol{0} & \,\,\,  X_{n_R}^{\T}\\
-X_{n_R}^{\T} & \,\,\, \boldsymbol{0} & \phm c_\alpha \id & \,\,\, s_\alpha\id \\
\boldsymbol{0} & \,\,\, -X_{n_R}^{\T} &   -s_\alpha\id & \,\,\, c_\alpha\id\end{pmatrix}=\begin{pmatrix} X_L\otimes\id &\,\,\, \id\otimes X^{\T}_{n_R} \\   -\id\otimes X^{\T}_{n_R} & X_L\otimes\id\end{pmatrix}.
\ee
where $\id$ is the $2\times 2$ identity matrix and $\boldsymbol{0}$ is the $2\times 2$ zero matrix in \eq{app:Adef}.

Using the well-known formula for the determinant of a $2\times 2$ block matrix (e.g., see Ref.~\cite{Johnston:2021}),
\be \label{app:schur}
\det \begin{pmatrix} M & N \\ P & Q\end{pmatrix}=\det M \det(Q-PM^{-1}N)\,,
\ee
where $Q-PM^{-1}N$ is the Schur complement of $M$ (under the assumption that $M$ is invertible), 
and noting that $\det(X_L\otimes\id)=1$ and $X_L^{-1}=X_L^{\T}$, we obtain
\ba \label{app:detA1}
\det A= \det\bigl[X_L\otimes\id+(\id\otimes X^{\T}_{nR})(X_L^{\T}\otimes\id)(\id\otimes X^{\T}_{nR})\bigr]\,.
\ea
We can manipulate \eq{app:detA1} into a more useful form by using the properties of the Kronecker product of two matrices,
\ba
\det A&=& \det\bigl[X_L\otimes\id+(\id\otimes X^{\T}_{nR})(X_L^{\T}\otimes X^{\T}_{nR})\bigr]
= \det\Bigl\{(X_L\otimes\id)\bigl[\id_4+(X_L^{\T}\otimes X^{\T}_{nR})^2\bigr]\Bigr\}  \nonumber \\
&=& \det\bigl[\id_4+(X_L^{\T}\otimes X^{\T}_{nR})^2\bigr]\,,
\ea
after using $(X_L\otimes\id)(X_L^{\T}\otimes\id) = \id_4$, where $\id_4$ is the $4\times 4$ identity matrix.  Noting that $X_L\otimes X_{n_R}$ is an orthogonal $4\times 4$ matrix with unit determinant, it follows that
$\id_4=(X_L\otimes X_{n_R})(X^{\T}_L\otimes X^{\T}_{n_R})$ and
\be
\det A=\det\bigl[(X^{\T}_L\otimes X^{\T}_{nR})(X_L\otimes X_{nR}+X^{\T}_L\otimes X^{\T}_{nR})\bigr]=\det\bigl[X_L\otimes X_{nR}+X^{\T}_L\otimes X^{\T}_{nR}\bigr]\,.
\ee
Explicitly, we have
\be \label{app:IJ}
X_L\otimes X_{nR}+X^{\T}_L\otimes X^{\T}_{nR} = \begin{pmatrix} \phm c_\alpha(X_{n_R}+X_{n_R}^{\T}) & \quad s_\alpha (X_{n_R}-X_{n_R}^{\T}) \\ -s_\alpha (X_{n_R}-X_{n_R}^{\T})  & \quad
c_\alpha(X_{n_R}+X_{n_R}^{\T}) \end{pmatrix} =\begin{pmatrix} \phm 2c_\alpha c_\beta\id & \,\,\, 2 s_\alpha s_\beta \boldsymbol{J} \\ -2 s_\alpha s_\beta \boldsymbol{J} & \,\,\, 2c_\alpha c_\beta\id\end{pmatrix} \,,
\ee
where $\boldsymbol{J}\equiv\left(\begin{smallmatrix} \phm 0 & \phm 1 \\ -1 & \phm 0\end{smallmatrix}\right)$.  Using \eq{app:schur} to evaluate the determinant of \eq{app:IJ} with $\boldsymbol{J}^2=-\id$, one quickly obtains
\be
\det A=16(c^2_\alpha c^2_\beta-s^2_\alpha s^2_\beta)^2\,,
\ee
as indicated in \eq{detAvalue}.

\section{Extensions of $\mathbb{Z}_2 \otimes \Pi_2$ to the Yukawa sector of the 2HDM with two quark generations}
\label{app:details}

In Section~\ref{subsubsec:ZPi_n=2}, we identified nine classes of models that are compatible with the extension of the $\ZPi$ symmetry to the two-generation Yukawa sector.
These models were obtained by determining the unitary matrices $U_L$ and $U_{n_R}$ that satisfy 
\be \label{app:Pi2Gamma12}
\Gamma_1(U_{n_R} S_{nR} U^\dagger_{n_R}) = (U_L S_L U_L^\dagger)\Gamma_2\,,
\ee
where the Yukawa coupling matrices $\{\Gamma_1\,,\,\Gamma_2\}$ have the form given by Cases 1, 2, or 3 specified in \eqst{eq:CasesZ2_2fam1}{eq:CasesZ2_2fam3}
and the choices of the symmetry matrices $S_L$ and $S_{n_R}$ correspond to one of the following three choices exhibited in \eq{Schoices}, which we repeat here for the convenience of the reader:
\be \label{app:Schoices}
(S_L\,,\,S_{n_R}) \in \Bigl\{\bigl( \sigma_Z\,,\,  \mathbbm{1}\bigr)\,,\,\,\,\bigl( \mathbbm{1}\,,\,  \sigma_Z\bigr)\,,\,\,\, \bigl( \sigma_Z\,,\,   \sigma_Z\bigr)\Bigr\},
\ee
where $\sigma_Z$ is defined in \eq{sigma_Z}.
This yields nine possible model types, denoted by $(n$-$m)$, corresponding to the Case~$n$ Yukawa coupling matrices and 
 the $m$th set of possible choices for $S_L$ and $S_{n_R}$ given in \eq{app:Schoices}. 

To solve \eq{app:Pi2Gamma12}, we parametrize the unitary matrices $U_L$ and $U_{n_R}$ as in \eq{unitary}
\begin{equation} \label{app:unitary}
	U_\sigma = \begin{pmatrix}
		\phm e^{i\alpha_\sigma} \cos\theta_\sigma & \phm e^{-i\beta_\sigma} \sin\theta_\sigma \\
		- e^{i\beta_\sigma} \sin\theta_\sigma &\phm  e^{-i\alpha_\sigma} \cos\theta_\sigma
	\end{pmatrix}\,,\quad \text{with $\sigma\in\{L,n_R\}$},
\end{equation}
where an arbitrary global phase has been set to zero, 
$0\leq\theta_\sigma\leq\half\pi$, and $0\leq \alpha_\sigma$, $\beta_\sigma<2\pi$.  In the analysis of the possible model types below, we shall make use of the following quantity:
\be \label{app:unitary2}
U_\sigma \sigma_Z U_\sigma^\dagger=\begin{pmatrix}  \cos 2\theta_\sigma &  -e^{i(\alpha_\sigma-\beta_\sigma)}\sin 2\theta_\sigma \\
 -e^{-i(\alpha_\sigma-\beta_\sigma)}\sin 2\theta_\sigma & -\cos 2\theta_\sigma\end{pmatrix}.
 \ee

\subsubsection*{Model (1-1)}
\begin{equation}
	\Gamma_1= \begin{pmatrix}
		x_{11} &  \phm x_{12} \\ 0 & \phm  0 \end{pmatrix} \,,\qquad 
	\Gamma_2= \begin{pmatrix}
		0 & \phm  0 \\ x_{21} & \phm  x_{22} \end{pmatrix} \,,\qquad S_L =  \sigma_Z \,,\qquad 
	S_{nR}=  \mathbbm{1} \,.
\end{equation}
Plugging these results into \eq{app:Pi2Gamma12} and using \eq{app:unitary2} yields
\ba
\cos 2\theta_L&=& 0\,, \\[3pt]
x_{11}&=&-x_{21}e^{i(\alpha_L-\beta_L)}\sin 2\theta_L\,, \\[3pt]
x_{12}&=&-x_{22}e^{i(\alpha_L-\beta_L)}\sin 2\theta_L\,.
\ea
It follows that $\theta_L=\pi/4$ and
\be
\Gamma_2=-e^{-i(\alpha_L-\beta_L)}
\begin{pmatrix} 0 &  \phm 0
		\\ x_{11} &  \phm  x_{12} \end{pmatrix} \,.
\ee
Finally, we compute
\be
H_d=(v_1\Gamma_1+v_2\Gamma_2)(v_1\Gamma_1+v_2\Gamma_2)^\dagger= \bigl[|x_{11}|^2+|x_{12}|^2\bigr]\begin{pmatrix} |v_1|^2 &  -v_1 v_2^*e^{i(\alpha_L-\beta_L)} \\
 -v^*_1 v_2 e^{-i(\alpha_L-\beta_L)} & |v_2|^2\end{pmatrix}\,.
 \ee
 Thus, $\det H_d=0$, corresponding to the existence of a massless down-type quark.  Thus, we discard this model.

\subsubsection*{Model (1-2)}
\begin{equation}
	\Gamma_1= \begin{pmatrix}
		x_{11} &  \phm  x_{12} \\ 0 & \phm   0 \end{pmatrix} \,,\qquad 
	\Gamma_2= \begin{pmatrix}
		0 & \phm  0 \\ x_{21} &  \phm  x_{22} \end{pmatrix} \,,\qquad S_L = \mathbbm{1} \,,\qquad 
	S_{nR}=   \sigma_Z \,.
\end{equation}
Plugging these results into \eq{app:Pi2Gamma12} and using \eq{app:unitary} yields $x_{21}=x_{22}=0$.   In light of \eq{app:Pi2Gamma12}, $\Gamma_1=\Gamma_2=0$, and the model is discarded. 

\subsubsection*{Model (1-3)}
\begin{equation}
	\Gamma_1= \begin{pmatrix}
		x_{11} &  \phm  x_{12} \\ 0 & 0 \end{pmatrix} \,,\qquad 
	\Gamma_2= \begin{pmatrix}
		0 & 0 \\ x_{21} & \phm  x_{22} \end{pmatrix} \,,\qquad S_L =S_{n_R}=  \sigma_Z \,.
\end{equation}
This case has been treated explicitly in Section~\ref{subsubsec:ZPi_n=2}.    We found that
 $x_{21}=x_{12}$ and $x_{22}=x_{11}$, and 
  \ba
  \Tr H_d &=&  \bigl[|x_{11}|^2+|x_{12}|^2\bigr]v^2\,,\\[4pt]
\det H_d&=&|v_1|^2|v_2|^2|x^2_{11}-x^2_{12}|^2\,,
\ea
where $v^2\equiv |v_1|^2+|v_2|^2$.
Thus, the two down-type quarks are generically nonzero.  
 In addition, the corresponding $\Pi_2$ symmetry matrices obtained from \eq{SymMatPi2} are 
\be
	S_L^{(\Pi_2)} =  \sigma_\Pi  \,,\qquad\quad
	S_{nR}^{(\Pi_2)} =  \sigma_\Pi\,,
\ee
where 
$\sigma_\Pi$ is defined in \eq{sigmaPimatrix}.

\subsubsection*{Model (2-1)}
We start in Case 2 of $\mathbb{Z}_2$:
\begin{equation}	
	\Gamma_1= \begin{pmatrix} x_{11} &\phm  0  \\ x_{21} &\phm  0  \end{pmatrix} \,,\qquad
	\Gamma_2= \begin{pmatrix} 0 &\phm   x_{12} \\ 0 &\phm   x_{22} \end{pmatrix} \,,\qquad S_L =  \sigma_Z  \,,\qquad 
	S_{nR}=  \mathbbm{1}   \,.
	\end{equation}
Plugging these results into \eq{app:Pi2Gamma12} and using \eq{app:unitary2} yields $x_{11}=x_{21}=0$.   In light of \eq{app:Pi2Gamma12}, $\Gamma_1=\Gamma_2=0$, and the model is discarded.

\subsubsection*{Model (2-2)}
\begin{equation}	
	\Gamma_1= \begin{pmatrix} x_{11} &\phm  0  \\ x_{21} &\phm  0  \end{pmatrix} \,,\qquad
	\Gamma_2= \begin{pmatrix} 0 &\phm   x_{12} \\ 0 &\phm   x_{22} \end{pmatrix} \,,\qquad S_L = \mathbbm{1} \,,\qquad 
	S_{nR}=   \sigma_Z \,.
	\end{equation}
Plugging these results into \eq{app:Pi2Gamma12} and using \eq{app:unitary2} yields 
\ba
\cos 2\theta_R&=& 0\,, \\[3pt]
x_{12}&=&-x_{11}e^{i(\alpha_R-\beta_R)}\sin 2\theta_R\,, \\[3pt]
x_{22}&=&-x_{21}e^{i(\alpha_R-\beta_R)}\sin 2\theta_R\,,
 \ea
where we have simplified the $R$ subscript in writing $\alpha_R\equiv \alpha_{n_R}$, $\beta_R\equiv\beta_{n_R}$, and $\theta_R\equiv\theta_{n_R}$. 
It follows that $\theta_R=\pi/4$ and
\be
\Gamma_2=-e^{i(\alpha_R-\beta_R)}
\begin{pmatrix} 0 & \phm x_{11}
		\\  0 & \phm x_{21} \end{pmatrix} \,.
\ee
Finally, we compute
\be
H_d=(v_1\Gamma_1+v_2\Gamma_2)(v_1\Gamma_1+v_2\Gamma_2)^\dagger=v^2\begin{pmatrix} |x_{11}|^2 & \phm x_{11}x_{21}^* \\[3pt]
x_{11}^* x_{21} & \phm |x_{21}|^2\end{pmatrix}\,.
 \ee
 Thus, $\det H_d=0$, corresponding to the existence of a massless down-type quark.  Thus, we discard this model.

\subsubsection*{Model (2-3)}
\begin{equation}\label{eq:Case23_Z2}
	\Gamma_1= \begin{pmatrix} x_{11} & \phm 0 \\ x_{21} &  \phm 0 \end{pmatrix} \,,\qquad
	\Gamma_2= \begin{pmatrix} 0 &  \phm x_{12} \\ 0 &  \phm  x_{22} \end{pmatrix} \,,\qquad S_L =S_{n_R}=  \sigma_Z \,.
	\end{equation}
Plugging these results into \eq{app:Pi2Gamma12} and using \eq{app:unitary2} yields 
\ba
\cos 2\theta_R&=& 0\,, \label{app:c2th}\\[4pt] 
x_{11}e^{i(\alpha_R-\beta_R)} \sin 2\theta_R&=& x_{22}e^{-i(\alpha_L-\beta_L)}\sin 2\theta_L - x_{12}\cos 2\theta_L\,, \label{app:first}\\[4pt]
x_{21}e^{i(\alpha_R-\beta_R)}\sin 2\theta_R &=& x_{12} e^{i(\alpha_L-\beta_L)}\sin 2\theta_L+x_{22}\cos 2\theta_L\,.\label{app:second}
\ea
It follows that  $\theta_R=\pi/4$.  Then, \eq{SymMatPi2} yields
\ba
S_L^{(\Pi_2)}&=&U_L\sigma_Z U_L^\dagger =\begin{pmatrix}  \cos 2\theta_L & -e^{i(\alpha_L-\beta_L)}\sin 2\theta_L \\ -e^{-i(\alpha_L-\beta_L)}\sin 2\theta_L & -\cos 2\theta_L \end{pmatrix}, \\[4pt]
S_{n_R}^{(\Pi_2)}&=&U_{n_R}\sigma_Z U_{n_R}^\dagger =\begin{pmatrix} 0 &  -e^{i(\alpha_R-\beta_R)} \\  -e^{-i(\alpha_R-\beta_R)}  &  0\end{pmatrix}.
\ea
The $\ZPi$ symmetry constraints do not fix the remaining free parameter, $\alpha_L$, $\beta_L$, $\alpha_R$, $\beta_R$, and $\theta_L$.  Indeed, one is free to transform to a different 
quark field basis as long as the $\mathbb{Z}_2$ symmetry matrices $S_L^{(\mathbb{Z}_2)}=e^{i\alpha_1}\sigma_Z$ and $S_{n_R}^{(\mathbb{Z}_2)}=e^{\alpha_1}\id$ are unchanged.  
In light of \eqs{Sprime1}{Sprime2}, we shall transform
\be \label{onemore}
S_L^{(\Pi_2)}\to U_L^\prime S_L^{(\Pi_2)} U_L^{\prime\dagger}\,,\qquad\quad 
S_{n_R}^{(\Pi_2)}\to U^\prime_{n_R} S_{n_R}^{(\Pi_2)} U_{n_R}^{\prime\dagger}\,,
\ee
where $U_L^\prime={\diag}\bigl(e^{i(\gamma+\delta)}\,,\,e^{i(\gamma-\delta)}\bigr)$ is the most general $2\times 2$ unitary matrix that leaves  $S_L^{(\mathbb{Z}_2)}$ unchanged and
$U_{n_R}^\prime$ is an arbitrary $2\times 2$ unitary matrix that (trivially) leaves $S_{n_R}^{(\mathbb{Z}_2)}$ unchanged.  With this freedom, it is convenient to set
\be \label{parmchoice2}
\alpha_L-\beta_L=\alpha_R-\beta_R=\pi\,,\qquad \theta_L=\tfrac14\pi\,.
\ee
With this choice, \eqst{app:c2th}{app:second} yield $x_{12}=x_{21}$ and $x_{22}=x_{11}$.   That is, model (2-3) corresponds to
\begin{equation}
	\Gamma_1= \begin{pmatrix} x_{11} & \phm 0 \\ x_{21} &  \phm 0 \end{pmatrix} \,,\qquad
	\Gamma_2= \begin{pmatrix} 0 &  \phm x_{21} \\ 0 & \phm  x_{11} \end{pmatrix} \,,\qquad
	S_L^{(\Pi_2)} = e^{i\xi_2} \sigma_\Pi  \,,\qquad
	S_{nR}^{(\Pi_2)} = e^{i\xi_2} \sigma_\Pi \,.
\end{equation}

It is straightforward to compute $H_d=(v_1\Gamma_1+v_2\Gamma_2) (v_1\Gamma_1+v_2\Gamma_2)^\dagger$ and its trace and determinant, which yield
 \ba
 \Tr H_d&=& \bigr[|x_{11}|^2+|x_{21}|^2\bigr]v^2\,,\\[4pt]
\det H_d&=&|v_1|^2|v_2|^2|x^2_{11}-x^2_{21}|^2\,.
\ea
Note that $\det H_d$ is nonzero for a generic choice of parameters, which implies that the two down-type quark masses are generically nonzero.

\subsubsection*{Model (3-1)}
\begin{equation}\label{eq:Case31_Z2}
	\Gamma_1= \begin{pmatrix} x_{11} & \phm  0 \\ 0 &  \phm  x_{22} \end{pmatrix} \,,\qquad  
	\Gamma_2= \begin{pmatrix} 0 &  \phm x_{12} \\ x_{21} &  \phm 0 \end{pmatrix} \,,\qquad S_L =  \sigma_Z \,,\qquad 
	S_{nR}=  \mathbbm{1} \,.
\end{equation}
Plugging these results into \eq{app:Pi2Gamma12} and using \eq{app:unitary2} yields
\ba
\cos 2\theta_L&=& 0\,, \\[3pt]
x_{11}&=&-x_{21}e^{i(\alpha_L-\beta_L)}\sin 2\theta_L\,, \\[3pt]
x_{22}&=&-x_{12}e^{-i(\alpha_L-\beta_L)}\sin 2\theta_L\,.
\ea
It follows that $\theta_L=\pi/4$ and
\be
\Gamma_2=
\begin{pmatrix} 0 &  -x_{22} e^{i(\alpha_L-\beta_L)}
		\\ -x_{11} e^{-i(\alpha_L-\beta_L)} & 0 \end{pmatrix} \,.
\ee
Then, \eq{SymMatPi2} yields
\be
S_L^{(\Pi_2)}=U_L\sigma_Z U_L^\dagger =\begin{pmatrix} 0 & -e^{i(\alpha_L-\beta_L)}\\ -e^{-i(\alpha_L-\beta_L)}& 0\end{pmatrix},\qquad\quad
S_{n_R}^{(\Pi_2)}=U_{n_R} U_{n_R}^\dagger =\id.
\ee
Once again, we are free to change the quark field basis using \eq{onemore}
where $U_L^\prime={\diag}\bigl(e^{i(\gamma+\delta)}\,,\,e^{i(\gamma-\delta)}\bigr)$ is the most general $2\times 2$ unitary matrix that leaves  $S_L^{(\mathbb{Z}_2)}$ unchanged and
$U_{n_R}^\prime$ is an arbitrary $2\times 2$ unitary matrix that (trivially) leaves $S_{n_R}^{(\mathbb{Z}_2)}$ unchanged.  With this freedom, it is convenient to set $\alpha_L-\beta_L=\pi$, 
which yields 
\begin{equation}
	\Gamma_1= \begin{pmatrix} x_{11} & \phm 0 \\    0 & \phm  x_{22} & \end{pmatrix} \,,\qquad
	\Gamma_2= \begin{pmatrix} 0 &  \phm x_{22} \\ x_{11} & \phm  0\end{pmatrix} \,,\qquad
	S_L^{(\Pi_2)} =  \sigma_\Pi  \,,\qquad
	S_{nR}^{(\Pi_2)} =  \id\,.
\end{equation}

It is straightforward to compute $H_d=(v_1\Gamma_1+v_2\Gamma_2) (v_1\Gamma_1+v_2\Gamma_2)^\dagger$ and its trace and determinant, which yield
 \ba
 \Tr H_d&=& \bigr[|x_{11}|^2+|x_{22}|^2\bigr]v^2\,,\label{trace2}\\[4pt]
\det H_d&=&|x_{11}|^2|x_{22}|^2 |v_1^2-v_2^2|^2\,.\label{det2}
\ea
Note that $\det H_d$ is nonzero for a generic choice of parameters, which implies that the two down-type quark masses are generically nonzero.

\subsubsection*{Model (3-2)}
\begin{equation}\label{eq:Case32_Z2}
	\Gamma_1= \begin{pmatrix} x_{11} & \phm 0 \\ 0 &\phm  x_{22} \end{pmatrix} \,,\qquad  
	\Gamma_2= \begin{pmatrix} 0 & \phm x_{12} \\ x_{21} & \phm  0 \end{pmatrix} \,,\qquad S_L = \mathbbm{1} \,,\qquad 
	S_{nR}=   \sigma_Z \,.
\end{equation}
Plugging these results into \eq{app:Pi2Gamma12} and using \eq{app:unitary2} yields
\ba
\cos 2\theta_R&=& 0\,, \\[3pt]
x_{12}&=&-x_{11}e^{i(\alpha_R-\beta_R)}\sin 2\theta_R\,, \\[3pt]
x_{21}&=&-x_{22}e^{-i(\alpha_R-\beta_R)}\sin 2\theta_R\,.
\ea
It follows that $\theta_R=\pi/4$ and
\be
\Gamma_2=
\begin{pmatrix} 0 &  -x_{11} e^{i(\alpha_R-\beta_R}
		\\ -x_{22} e^{-i(\alpha_R-\beta_R} & 0 \end{pmatrix} \,.
\ee
Then, \eq{SymMatPi2} yields
\be
S_L^{(\Pi_2)}=U_{n_L} U_{n_L}^\dagger =\id\,, \qquad\quad
S_{n_R}^{(\Pi_2)}=U_R\sigma_Z U_R^\dagger =\begin{pmatrix} 0 & -e^{i(\alpha_R-\beta_R)}\\ -e^{-i(\alpha_R-\beta_R)}& 0\end{pmatrix}.
\ee
We are free again to change the quark field basis using \eq{onemore}
where $U_R^\prime={\diag}\bigl(e^{i(\gamma+\delta)}\,,\,e^{i(\gamma-\delta)}\bigr)$ is the most general $2\times 2$ unitary matrix that leaves  $S_{n_R}^{(\mathbb{Z}_2)}$ unchanged and
$U_L^\prime$ is an arbitrary $2\times 2$ unitary matrix that (trivially) leaves $S_L^{(\mathbb{Z}_2)}$ unchanged.  With this freedom, it is convenient to set $\alpha_R-\beta_R=\pi$, 
which yields 
\begin{equation}
	\Gamma_1= \begin{pmatrix} x_{11} & \phm 0 \\    0 & \phm  x_{22} & \end{pmatrix} \,,\qquad
	\Gamma_2= \begin{pmatrix} 0 &  \phm x_{11} \\ x_{22} & \phm  0\end{pmatrix} \,,\qquad
	S_L^{(\Pi_2)} =  \id \,,\qquad
	S_{nR}^{(\Pi_2)} =  \sigma_\Pi \,.
\end{equation}

It is straightforward to compute $H_d=(v_1\Gamma_1+v_2\Gamma_2) (v_1\Gamma_1+v_2\Gamma_2)^\dagger$ and its trace and determinant, which yield
 \ba
 \Tr H_d&=& \bigr[|x_{11}|^2+|x_{22}|^2\bigr]v^2\,,\\[4pt]
\det H_d&=&|x_{11}|^2|x_{22}|^2 |v_1^2-v_2^2|^2\,.
\ea
Note that $\det H_d$ is nonzero for a generic choice of parameters, which implies that the two down-type quark masses are generically nonzero.

\subsubsection*{Models (3-3)}
We shall see below that there are a number of possible submodels within the class of Models (3-3).   This model class is defined by the following
Yukawa coupling matrices and symmetry matrices:
\begin{equation}\label{eq:Case33_Z2}
	\Gamma_1= \begin{pmatrix} x_{11} & \phm 0 \\ 0 & \phm x_{22}
	\end{pmatrix} \,,\qquad  
	\Gamma_2= \begin{pmatrix} 0 &\phm  x_{12} \\ x_{21} &\phm  0
	\end{pmatrix} \,,\qquad S_L =S_{n_R}=  \sigma_Z \,.
\end{equation}
Plugging these results into \eq{app:Pi2Gamma12} and using \eq{app:unitary2} yields
\ba
&& x_{11}\cos 2\theta_R+x_{21}e^{i(\alpha_L-\beta_L)}\sin 2\theta_L=0 \,,\label{sol1}\\[3pt]
&& x_{11}e^{i(\alpha_R-\beta_R)}\sin 2\theta_R+x_{12}\cos 2\theta_L=0\,, \label{sol2}\\[3pt]
&& x_{22}e^{-i(\alpha_R-\beta_R)}\sin 2\theta_R-x_{21}\cos 2\theta_L=0\,, \label{sol3}\\[3pt]
&& x_{22}\cos 2\theta_R-x_{12}e^{-i(\alpha_L-\beta_L)}\sin 2\theta_L=0 \,.\label{sol4}
\ea
This is a homogeneous system of four linear equations in the variables $x_{11}$, $x_{12}$, $x_{21}$, and $x_{22}$.   Nontrivial solutions exist only if
\be
\det\begin{pmatrix} \cos 2\theta_R & 0 & e^{i(\alpha_L-\beta_L)}\sin 2\theta_L & 0 \\ e^{i(\alpha_R-\beta_R)}\sin 2\theta_R & \cos 2\theta_L & 0 & 0 \\
0 & 0 &  -\cos 2\theta_L &e^{-i(\alpha_R-\beta_R)}\sin 2\theta_R \\ 0 & -e^{-i(\alpha_L-\beta_L)}\sin 2\theta_L & 0 & \cos 2\theta_R\end{pmatrix}=0\,,
\ee
which simplifies to
\be
\sin^2 2\theta_L\sin^2 2\theta_R-\cos^2 2\theta_L\cos^2 2\theta_R=0\,.
\ee
The solution to this equation is
\be \label{costwo}
\cos [2(\theta_L\pm\theta_R)]=0\,.
\ee
We now consider separately the following submodels.

\subsubsection*{Model (3,3)$_0$:  $\theta_L=0$ and $\theta_R=\pi/4$}

\Eqst{sol1}{sol4} yields $x_{12}=-x_{11}e^{i(\alpha_R-\beta_R)}$ and $x_{21}=x_{22}e^{-i(\alpha_R-\beta_R)}$.  
Using  \eq{SymMatPi2} yields
\be
S_L^{(\Pi_2)}=U_{L} \sigma_ZU_{L}^\dagger =\sigma_Z\,, \qquad\quad
S_{n_R}^{(\Pi_2)}=U_{n_R}\sigma_Z U_{n_R}^\dagger =\begin{pmatrix} 0 & -e^{i(\alpha_R-\beta_R)}\\ -e^{-i(\alpha_R-\beta_R)}& 0\end{pmatrix}.
\ee
As in previous cases, one can transform to another quark field basis where
$\alpha_R-\beta_R=\pi$, which yields
\begin{equation}
	\Gamma_1= \begin{pmatrix} x_{11} & 0 \\ 0 & \phm x_{22} \end{pmatrix} \,,\qquad 
	\Gamma_2= \begin{pmatrix} \phm 0 & \phm x_{11} \\ -x_{22} & \phm 0 \end{pmatrix} \,,\qquad
	S_L^{(\Pi_2)} =   \sigma_Z   \,,\qquad
	S_{nR}^{(\Pi_2)} =  \sigma_\Pi  \,.
\end{equation}

It is straightforward to compute $H_d=(v_1\Gamma_1+v_2\Gamma_2) (v_1\Gamma_1+v_2\Gamma_2)^\dagger$ and its trace and determinant, 
which yield\footnote{Note that in general $v^2\equiv |v_1|^2+|v_2|^2\neq  v_1^2+v_2^2$, since $v_1$ and $v_2$ are generically complex [cf.~\eq{vdef}].}
\ba
 \Tr H_d&=& \bigr[|x_{11}|^2+|x_{22}|^2\bigr]v^2\,,\\[4pt]
\det H_d&=&|x_{11}|^2|x_{22}|^2 |v_1^2+v_2^2|^2\,.
\ea
Note that $\det H_d$ is nonzero for a generic choice of parameters, which implies that the two down-type quark masses are generically nonzero.

\subsubsection*{Model (3,3)$_1$:  $\theta_L=\pi/4$ and $\theta_R=0$}

\Eqst{sol1}{sol4} yields $x_{21}=-x_{11}e^{-i(\alpha_L-\beta_L)}$ and $x_{12}=x_{22}e^{i(\alpha_L-\beta_L)}$.  
The computation is similar to the case of Model (3,3)$_0$.  Transforming to another quark field basis where $\alpha_L-\beta_L=\pi$, 
we end up with
\begin{equation}
	\Gamma_1= \begin{pmatrix} x_{11} &\phm  0 \\ 0 & \phm  x_{22} \end{pmatrix} \,,\qquad 
	\Gamma_2= \begin{pmatrix} 0 & -x_{22} \\ x_{11} & \phm 0 \end{pmatrix} \,,\qquad
	S_L^{(\Pi_2)} =   \sigma_\Pi     \,,\qquad
	S_{nR}^{(\Pi_2)} =  \sigma_Z \,.
\end{equation}
We again obtain:
 \ba
 \Tr H_d&=& \bigr[|x_{11}|^2+|x_{22}|^2\bigr]v^2\,, \label{trace3}\\[4pt]
\det H_d&=&|x_{11}|^2|x_{22}|^2 |v_1^2+v_2^2|^2\,,\label{det3}
\ea
corresponding to nonzero down-type quark masses for 
a generic choice of parameters.

\subsubsection*{Models (3-3)$_{\rm X}$:  $\theta_L$, $\theta_R\neq 0, \pi/4$}

The class of submodels under consideration here correspond to solutions of \eq{costwo} where neither $\theta_L$ nor $\theta_R$ is equal to 0 or $\pi/4$.
There are four possible cases, denoted by cases~(i)--(iv), which are defined in Table~\ref{models33}. 
Using \eqst{sol1}{sol4}, one easily obtains the constraints on the elements of the Yukawa coupling matrices in cases~(i)--(iv), which are listed in Table~\ref{models33}. It is convenient to write:
\be
x_{ij}=\xi_{ij}x_{12}\,,
\ee
where $\xi_{12}=1$ and $\xi_{ij}=\pm 1$ for $ij=12$, $21$ and $22$, where the signs are determined from the $x_{ij}$ listed in Table~\ref{models33}.

\begin{table}[H]
	\centering
	\begin{tabular}{|c|c|c|c|c|c|}\hline
		model  & $\theta_R$ & $\sin 2\theta_R$ & $\cos 2\theta_R$ & $x_{ij}$ relations & $\xi_{ij}$ \\
		\hline\hline
		$\theta_L=0$ &  $\tfrac14\pi$ & 1 & 0 & $x_{12}=x_{11}$ and $x_{21}=-x_{22}$ & \\
		\hline
		$\theta_L=\tfrac14\pi$ &  $0$ & 0 & 1 & $x_{21}=x_{11}$ and $x_{12}=-x_{22}$ &\\
		\hline
		case~(i) &  $\tfrac14\pi-\theta_L$ & $\phm\cos 2\theta_L$ & $\phm\sin 2\theta_L$ & $x_{11}=-x_{22}=x_{12}=x_{21}$ & $ \xi_{12}=\xi_{21}=1$, $\xi_{22}=-1$ \\ 
		\hline
		case~(ii) &  $\tfrac14\pi+\theta_L$ & $\phm\cos 2\theta_L$ & $-\sin 2\theta_L$ & $x_{11}=x_{22}=x_{12}=-x_{21}$ &  $\xi_{12}=\xi_{22}=1$, $\xi_{21}=-1$  \\ 
		\hline
		case~(iii) &  $\tfrac34\pi-\theta_L$ & $-\cos 2\theta_L$ & $-\sin 2\theta_L$ & $x_{11}=-x_{22}=-x_{12}=-x_{21}$  & $ \xi_{12}=\xi_{21}=\xi_{22}=-1$ \\ 
		\hline
		case~(iv) &  $-\tfrac14\pi+\theta_L$ & $-\cos 2\theta_L$ & $\phm\sin 2\theta_L$ & $x_{11}=x_{22}=-x_{12}=x_{21}$ &  $\xi_{21}=\xi_{22}=1$, $\xi_{12}=-1$ \\ 
		\hline	
	\end{tabular}
	\caption{Submodels within the class of Models (3-3).  Cases (i)--(iv) comprise models where neither $\theta_L$ nor $\theta_R$ is equal to $\pi/4$.
	Since $0\leq\theta_{L},\theta_R\leq \pi/2$,  it follows that $0<\theta_L<\pi/4$ in cases (i) and (ii) above and $\pi/4<\theta_L<\pi/2$
in cases (iii) and (iv) above.  The relations among the $x_{ij}$ are obtained from \eqst{sol1}{sol4} in a quark field basis where $\alpha_L-\beta_L=\alpha_R-\beta_R=\pi$.  For cases (i)--(iv), the $\xi_{ij}$ are defined such that $x_{ij}\equiv\xi_{ij} x_{11}$ with $\xi_{11}=1$.} 
	\label{models33}
\end{table}

As previously noted, one can simplify the analysis by a judicious choice of the quark field basis.  Following
\eqs{onemore}{parmchoice2}, we shall fix the phases $\alpha_L-\beta_L=\alpha_R-\beta_R=\pi$.   
With this choice, the Yukawa coupling matrices are given by
\be \label{LuisGamma}
	\Gamma_1= x_{11}\begin{pmatrix} 1 & \phm 0 \\ 0 &\phm \xi_{22}
	\end{pmatrix} \,,\qquad\quad 
	\Gamma_2= x_{11}\begin{pmatrix} 0 & \phm \xi_{12} \\  \xi_{21} & \phm 0
	\end{pmatrix},
\ee
and the $\Pi_2$ symmetry matrices, which are obtained from \eq{SymMatPi2}, are given by
\be \label{Luis333}
	S_L^{(\Pi_2)} =  
	\begin{pmatrix} \cos 2\theta_L & \phm \sin 2\theta_L
		\\ \sin 2\theta_L & -\cos 2\theta_L
	\end{pmatrix}     \,,\qquad\quad
	S_{nR}^{(\Pi_2)} = 
	\begin{pmatrix}  \xi_{21}\sin 2\theta_L &\phm \xi_{12}\cos 2\theta_L
		\\ \xi_{12}\cos 2\theta_L & -\xi_{21} \sin 2\theta_L
	\end{pmatrix} \,.
\ee

It is straightforward to compute $H_d=(v_1\Gamma_1+v_2\Gamma_2) (v_1\Gamma_1+v_2\Gamma_2)^\dagger$ and its trace and determinant:
\ba
H_d&=&(\Gamma_1 v_1+\Gamma_2 v_2)(\Gamma_1 v_1+\Gamma_2 v_2)^\dagger=|x_{11}^2|\begin{pmatrix}v^2 &\quad  2i\xi_{21}\Im(v_1 v_2^*) \\ -2i\xi_{21}\Im(v_1 v_2^*)  & \quad v^2 \end{pmatrix}, \\[5pt]
 \Tr H_d&=&2|x_{11}|^2v^2\,,\\[4pt]
\det H_d &=& |x_{11}|^4v^4\,,
\ea
where $\xi_{21}=\pm 1$, with the sign depending on the case chosen.  \pagebreak
As in the previous (3-3) submodels, the down-type quark masses are nonzero.   However, in contrast to Models (3,3)$_0$ and (3,3)$_1$, we note that the down-type quark masses are degenerate if 
$\Im(v_1 v_2^*)=0$.

The analysis of the up-type quark Yukawa coupling matrices yields the same textures for $\Delta_1$ and $\Delta_2$ exhibited in \eq{LuisGamma}, with the matrix element $x_{11}$ replaced by $y_{11}$ and sign factors  $\xi_{ij}$ replaced by $\xi^\prime_{ij}$.   
In addition, $S_L^{(\Pi_2)}$ is given by \eq{Luis333} and $S^{(\Pi_2)}_{p_R}$ is obtained from $S^{(\Pi_2)}_{n_R}$ by replacing $\xi_{ij}$ with $\xi^\prime_{ij}$.It therefore follows that
\be
H_u=(\Delta_1 v_1+\Delta_2 v_2)(\Delta_1 v_1+\Delta_2 v_2)^\dagger=|y_{11}|^2\begin{pmatrix}v^2 &\quad  2i\xi^\prime_{21}\Im(v_1 v_2^*) \\ -2i\xi^\prime_{21}\Im(v_1 v_2^*)  & \quad v^2 \end{pmatrix},
\ee
where $\xi^\prime_{21}=\pm 1$, with the sign depending on the case chosen.  Since $\det H_u=|y_{11}|^4 v^2$, it follows that the up-type quark masses are nonzero.  However, note that 
\be
\bigl[H_u\,,\,H_d\bigr]=0\,,
\ee
for either choice of $\xi_{21}\xi_{21}^\prime=\pm 1$.  In particular, $\det\bigl\{[H_u,H_d]\bigr\}= 0$
or equivalently $J_c=0$ [cf.~\eq{JC}], corresponding to a vanishing Cabibbo angle.  Thus, it follows that all models (3-3)$_X$ are phenomenologically excluded.

\section{Equivalent two-generation $\ZPi$-symmetric models}
\label{app:equiv}

In Table~\ref{tab:completemodels}, we classified all phenomenologically viable two-generation Yukawa-extended $\ZPi$-symmetric models (i.e., models with nonzero quark masses and a nonzero Cabibbo angle).  However, the models that appear in Table~\ref{tab:completemodels}, 
are not all inequivalent, as certain pairs of models are related by a particular change in the Higgs field and the quark field basis.   Suppose we transform to a new basis characterized by the basis transformation matrices 
\be \label{threeUs}
U=U_L=U_{n_R}=U_{p_R}=\frac{1}{\sqrt{2}} 
	\begin{pmatrix}
		1 & \phm 1 \\
		1 & -1
	\end{pmatrix} \,.
\ee
Using \eq{Gamtransform}, it follows that the model (1-3) down-type Yukawa coupling matrices listed in Table~\ref{tab:symmetries_Z2PI2_2fam} transform into
\be
\Gamma^\prime_1=\begin{pmatrix} x^\prime_{11} & 0 \\ 0 & x^\prime_{22} \end{pmatrix}\,,\qquad\qquad 
\Gamma^\prime_2=\begin{pmatrix} 0 & x^\prime_{22} \\ x^\prime_{11} & 0 \end{pmatrix}\,,
\ee
where 
\be
x_{11}^\prime=x_{21}^\prime=\frac{x_{11}+x_{12}}{\sqrt{2}}\,,\qquad\qquad
x_{12}^\prime=x_{22}^\prime=\frac{x_{11}-x_{12}}{\sqrt{2}}\,,
\ee
That is, the textures of $\Gamma_1^\prime$ and $\Gamma_2^\prime$ precisely match those of model (3-1) listed in Table~\ref{tab:symmetries_Z2PI2_2fam}.   Moreover, in light of \eq{vtrans},
the vevs are transformed into 
\be \label{vees}
  v^\prime_1=\frac{1}{\sqrt{2}}\bigl(v_1+v_2\bigr)\,,\qquad\qquad  v^\prime_2=\frac{1}{\sqrt{2}}\bigl(v_1-v_2\bigr)\,.
  \ee
It follows that the trace and determinant of $H_d$ given in \eqs{trace1}{det1} are transformed into  
  \ba
  \Tr H_d &=&  \bigl[|x^\prime_{11}|^2+|x^\prime_{12}|^2\bigr]v^2\,, \\[4pt]
\det H_d&=&|x^\prime_{11}|^2|x^\prime_{22}|^2\bigl|v_1^{\prime\,2}-v_2^{\prime\,2}\bigr|^2\,,
\ea
in agreement with the model (3-1) results obtained in \eqs{trace2}{det2}.  Finally, if we make use of \eqs{Sprime1}{Sprime2}, we obtain the corresponding symmetry matrices
\be \label{allthesyms}
S^{\prime(\mathbb{Z}_2)}=\sigma_\Pi\,,\qquad\
S_L^{\prime(\mathbb{Z}_2)}=\sigma_\Pi\,,\qquad S_{nR}^{\prime(\mathbb{Z}_2)}=\id\,,\qquad S^{\prime(\Pi_2)}=\sigma_Z\,,\qquad S_L^{\prime(\Pi_2)}=\sigma_Z\,,\qquad S_{nR}^{\prime(\Pi_2)}=\sigma_Z\,.
\ee
We cannot directly compare this result to Table~\ref{tab:symmetries_Z2PI2_2fam}, as the results of this table have assumed a
scalar field basis where $S^{(\mathbb{Z}_2)}=\sigma_Z$ and  $S^{(\Pi_2)}=\sigma_\Pi$.  But, we can overcome this impediment simply by interchanging the roles of the $\mathbb{Z}_2$ and $\Pi_2$ symmetries, in which case \eq{allthesyms} becomes
\be \label{allthesyms2}
S^{\prime(\mathbb{Z}_2)}=\sigma_Z\,,\qquad\
S_L^{\prime(\mathbb{Z}_2)}=\sigma_Z\,,\qquad S_{nR}^{\prime(\mathbb{Z}_2)}=\sigma_Z\,,\qquad S^{\prime(\Pi_2)}=\sigma_\Pi\\,,\qquad S_L^{\prime(\Pi_2)}=\sigma_\Pi\,,\qquad S_{nR}^{\prime(\Pi_2)}=\id\,,
\ee
which precisely matches the symmetry matrices of model (3-1) listed in Table~\ref{tab:symmetries_Z2PI2_2fam}.  Hence, we conclude that models (1-3) and (3-1) are simply related by a basis \pagebreak change and hence can be regarded as equivalent.  

Note that the basis transformation matrices given in \eq{threeUs} are equal to their inverses.  Thus,
if we start with the model (3-1) down-type Yukawa coupling matrices and apply the same procedure as outlined above, one ends up with corresponding Yukawa coupling matrices of model (1-3).
Likewise, applying the same basis transformation matrices followed by interchanging the identification of the $\mathbb{Z}_2$ and $\Pi_2$ symmetries also converts down-type Yukawa coupling matrices of model (2-3) into those of model (3-2) and vice versa.

The same arguments also apply to the up-type Yukawa coupling matrices.  
In particular, the textures of $\Delta_1$ and $\Delta_2$ mirror those of $\Gamma_1$ and $\Gamma_2$ given in Table~\ref{tab:symmetries_Z2PI2_2fam}. 
For the basis transformation characterized by the matrices specified in \eq{threeUs}, 
we can use the same treatment that was employed above in the analysis of the down-type Yukawa coupling matrices to conclude that the two models in each of the following model pairs that appear in Table~\ref{tab:completemodels}, 
\be \label{equiv1}
\text{$\{$(1-3)\! / \!(1-3); (3-1)\! / \!(3-1)$\}$,  \qquad 
$\{$(1-3)\!  / \!(3-1); (3-1)\!  / \!(1-3)$\}$,\qquad  $\{$(2-3)\! / \!(2-3); (3-2)\! / \!(3-2)$\}$,}
\ee
are equivalent  (i.e., they correspond to the same model with a different choice of basis).  
However, the above analysis also shows that, e.g., the models (1-3)\!  / \!(3-1) and (1-3)\!  / \!(1-3) are \textit{not} equivalent, since one must employ the \textit{same} basis transformation matrices to both the down-type and the up-type Yukawa coupling matrices.

One can also consider a change of basis for which $U_{p_R}\neq U_{n_R}$.  
Suppose we take $U$, $U_L$, and $U_{p_R}$ given in \eq{threeUs} and $U_{n_R}=\sigma_Z$.   
Using
\eq{Gamtransform}, it follows that model (3,3)$_1$ is the unique model of Table~\ref{tab:symmetries_Z2PI2_2fam} such that $\Gamma^\prime_i=\Gamma_i$ in the transformed basis.   The transformed symmetry matrices of model (3-3)$_1$ are obtained using \eqs{Sprime1}{Sprime2},
\be \label{allthesyms3}
S^{\prime(\mathbb{Z}_2)}=\sigma_\Pi\,,\qquad\
S_L^{\prime(\mathbb{Z}_2)}=\sigma_\Pi\,,\qquad S_{nR}^{\prime(\mathbb{Z}_2)}=\sigma_Z\,,\qquad S^{\prime(\Pi_2)}=\sigma_Z\,,\qquad S_L^{\prime(\Pi_2)}=\sigma_Z\,,\qquad S_{nR}^{\prime(\Pi_2)}=\sigma_Z\,.
\ee
Finally, after interchanging the identification of the $\mathbb{Z}_2$ and $\Pi_2$ symmetries, we reproduce the original symmetry matrices of model (3-3)$_1$.   Consequently,  it follows that 
the two models in the following model pair are equivalent:
\be \label{equiv2}
\text{$\{$(3-3)$_1$\! / \!(1-3);  (3-3)$_1$\! / \!(3-1)$\}$}.
\ee
 
 Similarly, suppose we take $U$, $U_L$, and $U_{n_R}$ given in \eq{threeUs} and $U_{p_R}=\sigma_Z$.  Using the same analysis as the one just given, where \eq{Deltransform} is now employed, it follows that $\Delta_i^\prime=\Delta_i$ in the transformed basis.  Hence,
the two models in the following model pair are equivalent:
\be \label{equiv3}
\text{$\{$(1-3)\! / \!(3-3)$_1$;  (3-1)\! / \!(3-3)$_1\}$}.
\ee
 This completes the search for equivalent models.  Taking \eqss{equiv1}{equiv2}{equiv3} into account, there
 are seven inequivalent models for the Yukawa sector in Table~\ref{tab:completemodels}.

\bibliographystyle{JHEP}
\bibliography{ref}

\providecommand{\href}[2]{#2}\begingroup\raggedright\begin{thebibliography}{10}

\bibitem{ATLAS:2012yve}
{\bf ATLAS} Collaboration, G.~Aad et~al., {\it {Observation of a new particle
  in the search for the Standard Model Higgs boson with the ATLAS detector at
  the LHC}},  {\em Phys. Lett. B} {\bf 716} (2012) 1,
  [\href{http://arxiv.org/abs/1207.7214}{{\tt arXiv:1207.7214}}].

\bibitem{CMS:2012qbp}
{\bf CMS} Collaboration, S.~Chatrchyan et~al., {\it {Observation of a New Boson
  at a Mass of 125 GeV with the CMS Experiment at the LHC}},  {\em Phys. Lett.
  B} {\bf 716} (2012) 30, [\href{http://arxiv.org/abs/1207.7235}{{\tt
  arXiv:1207.7235}}].

\bibitem{ATLAS:2022vkf}
{\bf ATLAS} Collaboration, G.~Aad et~al., {\it {A detailed map of Higgs boson
  interactions by the ATLAS experiment ten years after the discovery}},  {\em
  Nature} {\bf 607} (2022), no.~7917 52--59,
  [\href{http://arxiv.org/abs/2207.00092}{{\tt arXiv:2207.00092}}]. [Erratum:
  Nature 612, (2022) 7941, E24].

\bibitem{CMS:2022dwd}
{\bf CMS} Collaboration, A.~Tumasyan et~al., {\it {A portrait of the Higgs
  boson by the CMS experiment ten years after the discovery}},  {\em Nature}
  {\bf 607} (2022), no.~7917 60--68,
  [\href{http://arxiv.org/abs/2207.00043}{{\tt arXiv:2207.00043}}]. [Erratum:
  Nature 623 (2023) 7985, E4].

\bibitem{Branco:2011iw}
G.~C. Branco, P.~M. Ferreira, L.~Lavoura, M.~N. Rebelo, M.~Sher, and J.~P.
  Silva, {\it {Theory and phenomenology of two-Higgs-doublet models}},  {\em
  Phys. Rept.} {\bf 516} (2012) 1, [\href{http://arxiv.org/abs/1106.0034}{{\tt
  arXiv:1106.0034}}].

\bibitem{Lee:1973iz}
T.~D. Lee, {\it {A Theory of Spontaneous T Violation}},  {\em Phys. Rev. D}
  {\bf 8} (1973) 1226.

\bibitem{Cohen:1993nk}
A.~G. Cohen, D.~B. Kaplan, and A.~E. Nelson, {\it {Progress in electroweak
  baryogenesis}},  {\em Ann. Rev. Nucl. Part. Sci.} {\bf 43} (1993) 27,
  [\href{http://arxiv.org/abs/hep-ph/9302210}{{\tt hep-ph/9302210}}].

\bibitem{Deshpande:1977rw}
N.~G. Deshpande and E.~Ma, {\it {Pattern of Symmetry Breaking with Two Higgs
  Doublets}},  {\em Phys. Rev. D} {\bf 18} (1978) 2574.

\bibitem{Barbieri:2006dq}
R.~Barbieri, L.~J. Hall, and V.~S. Rychkov, {\it {Improved naturalness with a
  heavy Higgs: An Alternative road to LHC physics}},  {\em Phys. Rev. D} {\bf
  74} (2006) 015007, [\href{http://arxiv.org/abs/hep-ph/0603188}{{\tt
  hep-ph/0603188}}].

\bibitem{Glashow:1976nt}
S.~L. Glashow and S.~Weinberg, {\it {Natural Conservation Laws for Neutral
  Currents}},  {\em Phys. Rev. D} {\bf 15} (1977) 1958.

\bibitem{Paschos:1976ay}
E.~A. Paschos, {\it {Diagonal Neutral Currents}},  {\em Phys. Rev. D} {\bf 15}
  (1977) 1966.

\bibitem{Fritzsch:1977za}
H.~Fritzsch, {\it {Calculating the Cabibbo Angle}},  {\em Phys. Lett. B} {\bf
  70} (1977) 436.

\bibitem{Ivanov:2007de}
I.~P. Ivanov, {\it {Minkowski space structure of the Higgs potential in 2HDM.
  II. Minima, symmetries, and topology}},  {\em Phys. Rev. D} {\bf 77} (2008)
  015017, [\href{http://arxiv.org/abs/0710.3490}{{\tt arXiv:0710.3490}}].

\bibitem{Ferreira:2009wh}
P.~M. Ferreira, H.~E. Haber, and J.~P. Silva, {\it {Generalized CP symmetries
  and special regions of parameter space in the two-Higgs-doublet model}},
  {\em Phys. Rev. D} {\bf 79} (2009) 116004,
  [\href{http://arxiv.org/abs/0902.1537}{{\tt arXiv:0902.1537}}].

\bibitem{Davidson:2005cw}
S.~Davidson and H.~E. Haber, {\it {Basis-independent methods for the
  two-Higgs-doublet model}},  {\em Phys. Rev. D} {\bf 72} (2005) 035004,
  [\href{http://arxiv.org/abs/hep-ph/0504050}{{\tt hep-ph/0504050}}]. [Erratum:
  Phys.~Rev.~D {\bf 72} (2005) 099902].

\bibitem{Ferreira:2010yh}
P.~M. Ferreira, H.~E. Haber, M.~Maniatis, O.~Nachtmann, and J.~P. Silva, {\it
  {Geometric picture of generalized-CP and Higgs-family transformations in the
  two-Higgs-doublet model}},  {\em Int. J. Mod. Phys. A} {\bf 26} (2011) 769,
  [\href{http://arxiv.org/abs/1010.0935}{{\tt arXiv:1010.0935}}].

\bibitem{Battye:2011jj}
R.~A. Battye, G.~D. Brawn, and A.~Pilaftsis, {\it {Vacuum Topology of the Two
  Higgs Doublet Model}},  {\em JHEP} {\bf 08} (2011) 020,
  [\href{http://arxiv.org/abs/1106.3482}{{\tt arXiv:1106.3482}}].

\bibitem{Pilaftsis:2011ed}
A.~Pilaftsis, {\it {On the Classification of Accidental Symmetries of the Two
  Higgs Doublet Model Potential}},  {\em Phys. Lett. B} {\bf 706} (2012) 465,
  [\href{http://arxiv.org/abs/1109.3787}{{\tt arXiv:1109.3787}}].

\bibitem{Haber:2021zva}
H.~E. Haber and J.~P. Silva, {\it {Exceptional regions of the 2HDM parameter
  space}},  {\em Phys. Rev. D} {\bf 103} (2021) 115012,
  [\href{http://arxiv.org/abs/2102.07136}{{\tt arXiv:2102.07136}}]. [Erratum:
  Phys.~Rev.~D {\bf 105} (2022) 119902].

\bibitem{Botella:1994cs}
F.~J. Botella and J.~P. Silva, {\it {Jarlskog--like invariants for theories
  with scalars and fermions}},  {\em Phys. Rev. D} {\bf 51} (1995) 3870,
  [\href{http://arxiv.org/abs/hep-ph/9411288}{{\tt hep-ph/9411288}}].

\bibitem{Trautner:2018ipq}
A.~Trautner, {\it {Systematic construction of basis invariants in the 2HDM}},
  {\em JHEP} {\bf 05} (2019) 208, [\href{http://arxiv.org/abs/1812.02614}{{\tt
  arXiv:1812.02614}}].

\bibitem{Bento:2020jei}
M.~P. Bento, R.~Boto, J.~P. Silva, and A.~Trautner, {\it {A fully basis
  invariant Symmetry Map of the 2HDM}},  {\em JHEP} {\bf 21} (2020) 229,
  [\href{http://arxiv.org/abs/2009.01264}{{\tt arXiv:2009.01264}}].

\bibitem{Doring:2024kdg}
C.~D\"oring and A.~Trautner, {\it {Symmetries from outer automorphisms and
  unorthodox group extensions}},  \href{http://arxiv.org/abs/2410.11052}{{\tt
  arXiv:2410.11052}}.

\bibitem{Ferreira:2010ir}
P.~M. Ferreira and J.~P. Silva, {\it {Abelian symmetries in the
  two-Higgs-doublet model with fermions}},  {\em Phys. Rev. D} {\bf 83} (2011)
  065026, [\href{http://arxiv.org/abs/1012.2874}{{\tt arXiv:1012.2874}}].

\bibitem{Ivanov:2011ae}
I.~P. Ivanov, V.~Keus, and E.~Vdovin, {\it {Abelian symmetries in
  multi-Higgs-doublet models}},  {\em J. Phys. A} {\bf 45} (2012) 215201,
  [\href{http://arxiv.org/abs/1112.1660}{{\tt arXiv:1112.1660}}].

\bibitem{Ivanov:2013bka}
I.~P. Ivanov and C.~C. Nishi, {\it {Abelian symmetries of the N-Higgs-doublet
  model with Yukawa interactions}},  {\em JHEP} {\bf 11} (2013) 069,
  [\href{http://arxiv.org/abs/1309.3682}{{\tt arXiv:1309.3682}}].

\bibitem{Ferreira:2010bm}
P.~M. Ferreira and J.~P. Silva, {\it {A Two-Higgs Doublet Model With Remarkable
  CP Properties}},  {\em Eur. Phys. J. C} {\bf 69} (2010) 45,
  [\href{http://arxiv.org/abs/1001.0574}{{\tt arXiv:1001.0574}}].

\bibitem{Bree:2024edl}
I.~Bree, D.~D. Correia, and J.~P. Silva, {\it {Generalized CP symmetries in
  three-Higgs-doublet models}},  {\em Phys. Rev. D} {\bf 110} (2024) 035028,
  [\href{http://arxiv.org/abs/2407.09615}{{\tt arXiv:2407.09615}}].

\bibitem{Boto:2020wyf}
R.~Boto, T.~V. Fernandes, H.~E. Haber, J.~C. Rom\~ao, and J.~P. Silva, {\it
  {Basis-independent treatment of the complex 2HDM}},  {\em Phys. Rev. D} {\bf
  101} (2020) 055023, [\href{http://arxiv.org/abs/2001.01430}{{\tt
  arXiv:2001.01430}}].

\bibitem{Branco:1999fs}
G.~C. Branco, L.~Lavoura, and J.~P. Silva, {\em {CP Violation}}.
\newblock Oxford University Press, Oxford, UK, 1999.

\bibitem{Branco:1987mj}
G.~C. Branco and L.~Lavoura, {\it {Rephasing Invariant Parametrization of the
  Quark Mixing Matrix}},  {\em Phys. Lett. B} {\bf 208} (1988) 123.

\bibitem{Bree:2023ojl}
I.~Bree, S.~Carrolo, J.~C. Romao, and J.~P. Silva, {\it {A viable $A_4$ 3HDM
  theory of quark mass matrices}},  {\em Eur. Phys. J. C} {\bf 83} (2023) 292,
  [\href{http://arxiv.org/abs/2301.04676}{{\tt arXiv:2301.04676}}].

\bibitem{Jarlskog:1985ht}
C.~Jarlskog, {\it {Commutator of the Quark Mass Matrices in the Standard
  Electroweak Model and a Measure of Maximal CP Nonconservation}},  {\em Phys.
  Rev. Lett.} {\bf 55} (1985) 1039.

\bibitem{Jarlskog:1985cw}
C.~Jarlskog, {\it {A Basis Independent Formulation of the Connection Between
  Quark Mass Matrices, CP Violation and Experiment}},  {\em Z. Phys. C} {\bf
  29} (1985) 491.

\bibitem{Dunietz:1985uy}
I.~Dunietz, O.~W. Greenberg, and D.-d. Wu, {\it {A Priori Definition of Maximal
  CP Violation}},  {\em Phys. Rev. Lett.} {\bf 55} (1985) 2935.

\bibitem{Silva:2004gz}
J.~P. Silva, {\it {Phenomenological aspects of CP violation}},  in {\em
  {Central European School in Particle Physics}}, 10, 2004.
\newblock \href{http://arxiv.org/abs/hep-ph/0410351}{{\tt hep-ph/0410351}}.

\bibitem{Bernabeu:1986fc}
J.~Bernabeu, G.~C. Branco, and M.~Gronau, {\it {CP Restrictions on Quark Mass
  Matrices}},  {\em Phys. Lett. B} {\bf 169} (1986) 243.

\bibitem{Gunion:2005ja}
J.~F. Gunion and H.~E. Haber, {\it {Conditions for CP-violation in the general
  two-Higgs-doublet model}},  {\em Phys. Rev. D} {\bf 72} (2005) 095002,
  [\href{http://arxiv.org/abs/hep-ph/0506227}{{\tt hep-ph/0506227}}].

\bibitem{Ecker:1987qp}
G.~Ecker, W.~Grimus, and H.~Neufeld, {\it {A Standard Form for Generalized {CP}
  Transformations}},  {\em J. Phys. A} {\bf 20} (1987) L807.

\bibitem{Ivanov:2005hg}
I.~P. Ivanov, {\it {Two-Higgs-doublet model from the group-theoretic
  perspective}},  {\em Phys. Lett. B} {\bf 632} (2006) 360,
  [\href{http://arxiv.org/abs/hep-ph/0507132}{{\tt hep-ph/0507132}}].

\bibitem{Ivanov:2006yq}
I.~P. Ivanov, {\it {Minkowski space structure of the Higgs potential in 2HDM}},
   {\em Phys. Rev. D} {\bf 75} (2007) 035001,
  [\href{http://arxiv.org/abs/hep-ph/0609018}{{\tt hep-ph/0609018}}]. [Erratum:
  Phys.~Rev.~D {\bf 76} (2007) 039902].

\bibitem{Peccei:1977ur}
R.~D. Peccei and H.~R. Quinn, {\it {Constraints Imposed by CP Conservation in
  the Presence of Instantons}},  {\em Phys. Rev. D} {\bf 16} (1977) 1791.

\bibitem{Ferreira:2008zy}
P.~M. Ferreira and J.~P. Silva, {\it {Discrete and continuous symmetries in
  multi-Higgs-doublet models}},  {\em Phys. Rev. D} {\bf 78} (2008) 116007,
  [\href{http://arxiv.org/abs/0809.2788}{{\tt arXiv:0809.2788}}].

\bibitem{Artin}
M.~Artin, {\em {Algebra}}.
\newblock Pearson Education, Inc., Upper Saddle River, New Jersey, 2nd~ed.,
  2017.

\bibitem{Arhrib:2016rlj}
A.~Arhrib, R.~Benbrik, S.~J.~D. King, B.~Manaut, S.~Moretti, and C.~S. Un, {\it
  {Phenomenology of 2HDM with vectorlike quarks}},  {\em Phys. Rev. D} {\bf 97}
  (2018) 095015, [\href{http://arxiv.org/abs/1607.08517}{{\tt
  arXiv:1607.08517}}].

\bibitem{Draper:2016cag}
P.~Draper, H.~E. Haber, and J.~T. Ruderman, {\it {Partially Natural Two Higgs
  Doublet Models}},  {\em JHEP} {\bf 06} (2016) 124,
  [\href{http://arxiv.org/abs/1605.03237}{{\tt arXiv:1605.03237}}].

\bibitem{Draper:2020tyq}
P.~Draper, A.~Ekstedt, and H.~E. Haber, {\it {A natural mechanism for
  approximate Higgs alignment in the 2HDM}},  {\em JHEP} {\bf 05} (2021) 235,
  [\href{http://arxiv.org/abs/2011.13159}{{\tt arXiv:2011.13159}}].

\bibitem{CarroloThesis}
S.~Carrolo, {\it {Building Consistent Multi Higgs Models}}, . Master's thesis
  (2022), Instituto Superior Tecnico, Universidade de Lisboa,
  \texttt{https://fenix.tecnico.ulisboa.pt/downloadFile/1970719973968612/Tese\_sergioCarrolo\_90192.pdf}.

\bibitem{Johnston:2021}
N.~Johnston, {\em {Advanced Linear and Matrix Algebra}}.
\newblock Springer Nature Switzerland AG, Cham, Switzerland, 2021.

\end{thebibliography}\endgroup

\end{document}